\documentclass[a4paper,UKenglish,cleveref,autoref,thm-restate]{lipics-v2021}
\hideLIPIcs

\AtBeginDocument{\nolinenumbers} %
\microtypesetup{protrusion=false,expansion=false} %
\usepackage{xcolor}

\usepackage{mathtools}
\usepackage{booktabs}
\usepackage{longtable}

\usepackage{tikz}
\usetikzlibrary{arrows.meta,positioning,automata,calc,shapes.geometric}
\usepackage{xspace}
\usepackage{algorithm}       %
\usepackage{algpseudocode}
\usepackage{float}
\usepackage{comment}
\usepackage{pgfplots}
\pgfplotsset{compat=1.18}
\definecolor{citationblue}{RGB}{40,75,115}
\definecolor{citationteal}{RGB}{35,90,95}
\definecolor{citationgrayblue}{RGB}{70,85,105}
\hypersetup{
  colorlinks=true,
  citecolor=magenta,
  linkcolor=citationteal,
  urlcolor=citationblue
}
\title{Stochastic Games with Distributional Entropy Objectives}
\author{Shankaranarayanan Krishna}{IIT Bombay, India}{}{}{}
\author{Durgam Latha}{IIT Bombay, India}{}{}{}
\author{Dion Reji}{IIT Bombay, India}{}{}{}
\author{Ashutosh Trivedi}{University of Colorado Boulder, USA}{}{}{}
\authorrunning{S. Krishna, D. Latha, D. Reji, and A. Trivedi}
\Copyright{Shankaranarayanan Krishna, Durgam Latha, Dion Reji, and Ashutosh Trivedi}
\category{}
\relatedversion{} %
\EventEditors{}
\EventNoEds{1}
\EventLongTitle{}
\EventShortTitle{}
\EventAcronym{}
\EventYear{}
\EventDate{}
\EventLocation{}
\EventLogo{}
\SeriesVolume{}
\ArticleNo{}
\newcommand{\States}{S}
\newcommand{\Acts}{A}

\newcommand{\Dist}{\mathcal{D}}
\newcommand{\DistS}{\Dist(\States)}
\newcommand{\Trans}{\Delta}

\newcommand{\Hist}{\mathsf{Hist}}              %

\newcommand{\StratMin}{\Pi_{\min}}
\newcommand{\StratMax}{\Pi_{\max}}
\newcommand{\stratmin}{\pi_{\min}}
\newcommand{\stratmax}{\pi_{\max}}
\newcommand{\stratminat}[2]{\pi_{\min,#2}^{#1}}
\newcommand{\stratmaxat}[2]{\pi_{\max,#2}^{#1}}
\newcommand{\strpair}{(\stratmin,\stratmax)}
\newcommand{\distr}{\mu}
\newcommand{\distro}{\mu_0}

\newcommand{\RandDistr}{\boldsymbol{\mu}}
\newcommand{\DistOp}{\Upsilon}
\newcommand{\TransReal}{\Xi}
\newcommand{\PathSpace}{\Omega}
\newcommand{\PathSigma}{\mathcal{F}}
\newcommand{\PathProb}[4]{\mathbb{P}_{#1}^{#2,#3,#4}}
\newcommand{\PathExpect}[4]{\mathbb{E}_{#1}^{#2,#3,#4}}
\newcommand{\Ent}{\mathbf{H}}

\newcommand{\DiscPay}{\mathsf{Pay}}

\newcommand{\NP}{\mathsf{NP}}

\newcommand{\AP}{\mathsf{AP}}
\newcommand{\PSPACE}{\mathsf{PSPACE}}

\newcommand{\Supp}{\mathsf{supp}}

\newcommand{\Real}{\mathbb{R}}

\newcommand{\Discount}{\gamma}

\newcommand{\defeq}{\coloneqq}
\newcommand{\MaxPl}{\textsc{Max}\xspace}
\newcommand{\MinPl}{\textsc{Min}\xspace}

\newcommand{\GapVal}{\mathsf{GapVal}}
\newcommand{\ValDec}{\mathsf{ValDec}}

\newcommand{\Sem}{\star}

\newcommand{\dl}[1]{}

\usepackage{wrapfig}

\newtheorem{construction}[theorem]{Construction}
\usepackage[most]{tcolorbox}
\definecolor{problemblue}{RGB}{40,75,115}
\newcounter{verificationproblem}[section]

\crefname{verificationproblem}{problem}{problems}
\Crefname{verificationproblem}{Problem}{Problems}
\newenvironment{problembox}[1][]
  {%
    \begin{tcolorbox}[
      enhanced,
      colback=problemblue!3,
      colframe=problemblue!70!black,
      boxrule=0.5pt,
      arc=1.5pt,
      outer arc=1.5pt,
      left=8pt,
      right=8pt,
      top=6pt,
      bottom=6pt,
      before skip=10pt,
      after skip=10pt,
      fonttitle=\bfseries,
      coltitle=black,
      title={Problem: #1},
      attach boxed title to top left={
        xshift=8pt,
        yshift*=-\tcboxedtitleheight/2
      },
      boxed title style={
        colback=white,
        colframe=problemblue!70!black,
        boxrule=0.5pt,
        arc=1pt,
        left=5pt,
        right=5pt,
        top=2pt,
        bottom=2pt
      }
    ]%
  }
  {\end{tcolorbox}}
\begin{document}
\maketitle
\begin{abstract}
We introduce \emph{stochastic games with distributional entropy objectives}, two-player zero-sum games
whose configurations are probability distributions over the states of a
finite arena and whose stage reward is Shannon entropy. Over an infinite
discounted horizon, \MinPl{} seeks to concentrate the distribution, while
\MaxPl{} seeks to disperse it. Because entropy is nonlinear, it matters
whether strategic and transition randomness are averaged before entropy is
evaluated or sampled first. We formalise three resulting semantics:
averaged-distribution, sampled-profile, and sampled-transition.

This semantic choice fundamentally changes the game-theoretic structure.
Averaged-distribution games need not be determined, even on a four-state
deterministic arena, whereas both sampled semantics are determined and admit
optimal stationary randomised strategies. We give contractive Bellman
characterisations and convergent finite-horizon approximations for the upper
and lower averaged values and the two sampled values. Moreover, even the
lower value under averaged-distribution semantics dominates the value under
either sampled semantics, while the sampled-profile and sampled-transition
values are incomparable in general.

For fixed rational discount and positive rational promise gap, the averaged-distribution (upper and lower value) gap problems and both sampled gap problems lie in $\PSPACE$; for
a single entropy minimiser, the averaged and sampled-profile problems lie in
$\NP$. Finally, under all three semantics, entropy optimisation is
$\NP$-hard to approximate within a suitable fixed additive constant, already
for deterministic one-player games, where pure memoryless strategies are optimal.
\end{abstract}

\section{Introduction}
\label{sec:introduction}

Stochastic games are a canonical model of quantitative adversarial decision making under uncertainty~\cite{shapley1953stochastic}. In the classical finite discounted setting, the current configuration is a single state of an arena. The players choose actions, the arena makes a stochastic transition, and the resulting rewards are accumulated in expectation. Such games have a value and admit optimal stationary strategies. Their values are characterised as the unique fixed point of a contractive Shapley operator, giving rise to value- and policy-iteration algorithms. Underlying this theory is the affine manner in which strategic randomisation and stochastic transitions enter the expected payoff.

\medskip\noindent\textbf{Probability distributions as configurations.}
In many systems, however, the current configuration is naturally represented by a distribution rather than a single state. Consider a distributed computing system in which a fixed population of jobs moves among a collection of servers. The system operator controls routing at some servers, while an adversary controls compromised servers. The operator seeks to spread the workload across the system, whereas the adversary seeks to concentrate it at a few servers and create bottlenecks. The resulting configuration is described by a distribution $\mu$ over the servers, where $\mu(s)$ is the fraction of jobs at server $s$. Its Shannon entropy,
$\Ent(\mu)=-\sum_s \mu(s)\log_2\mu(s)$,
measures workload dispersion: high entropy corresponds to a broadly distributed load, while low entropy indicates concentration.
Analogous configurations arise in transportation networks, where mass represents vehicles distributed across locations, and in epidemiological and ecological models, where it represents populations distributed across regions or compartments. A complementary interpretation arises in security, surveillance, and active sensing, where the distribution represents uncertainty about a hidden system state. In quantitative opacity and information-flow analysis, for example, a defender may seek to keep an observer's belief dispersed across several possible states, while the observer seeks to concentrate it sufficiently to infer a secret~\cite{jacob2016overview,lafortune2018history,clark2005quantitative,smith2009foundations}. Thus, across these settings, opposing objectives may favour either dispersion or concentration.

Motivated by these examples, we introduce \emph{stochastic games with distributional entropy objectives}, two-player zero-sum games played over a finite arena whose configurations are probability distributions over its states. Shannon entropy of the distribution serves as the stage reward, and the payoff is its infinite discounted sum. We discount because this makes the Bellman operators contractions and the values well defined for every continuous stage reward; it also weights the early rounds more, so the objective is about how dispersed the distribution becomes soon. A mean-payoff variant is a natural alternative that we leave open (Section~\ref{sec:conclusion}). The entropy-minimising player \MinPl{} seeks to concentrate the distribution, while the entropy-maximising player \MaxPl{} seeks to disperse it.

\medskip\noindent\textbf{Three roles for probability.}
Once distributions themselves become game configurations, probability plays three conceptually different roles: the current distribution may represent either population mass or uncertainty over states, the players may randomise their decisions, and the transitions of the arena may be stochastic. This raises a modelling question that does not arise for state-based games: when forming the next distribution, which sources of randomness describe \emph{fractional flow}, and which describe an event that is \emph{actually realised}?

One possible interpretation is that all probabilities describe fractional flow. If a routing decision selects two routes with probabilities $0.3$ and $0.7$, then the corresponding fractions of the population follow those routes. A stochastic transition with probabilities $0.2$ and $0.8$ similarly divides the population in those proportions. Once the players' mixed decisions are fixed, strategic and transition randomness therefore produce a single successor distribution. We call this \emph{averaged-distribution semantics}.

A second possibility is that strategic randomisation selects a control regime rather than dividing the population. At the beginning of a time interval, for example, each player may randomly select one complete routing plan. The selected plans are realised, but the population subsequently moves according to their transition rates. Thus, action profiles are sampled, while stochastic transitions continue to represent fractional population flow. We call this \emph{sampled-profile semantics}.

Finally, transition probabilities may themselves describe uncertainty about a common environmental condition. A transition probability may, for example, specify whether a communication link is available during the interval. The link is either available or unavailable, and all jobs at a given source server experience the same realised outcome. In this interpretation, both the control regime and the stochastic environment are realised before entropy is evaluated on the resulting distribution. We call this \emph{sampled-transition semantics}.

These interpretations are indistinguishable when rewards are affine in the distribution, since affine functions commute with averaging. Entropy behaves differently: it is strictly concave, so mixing distinct distributions produces greater entropy than the corresponding average of their entropies. Formally, if $\widetilde{\mu}$ is a random successor distribution, Jensen's inequality gives
$\Ent\!\left(\mathbb{E}[\widetilde{\mu}]\right)
\geq
\mathbb{E}\!\left[\Ent(\widetilde{\mu})\right]$,
with strict inequality whenever $\widetilde{\mu}$ is not almost surely constant. Thus, averaging random outcomes before evaluating entropy need not agree with first realising those outcomes and then averaging their entropies. The three semantics accordingly capture different interpretations of probability: distributional mass or belief, randomised decisions, and uncertainty over realised worlds.

This leads to the central theoretical question of the paper: which parts of the classical theory of discounted stochastic games survive when configurations are probability distributions and rewards are nonlinear functions of those distributions? Our results show that the answer depends sharply on how randomness is interpreted.

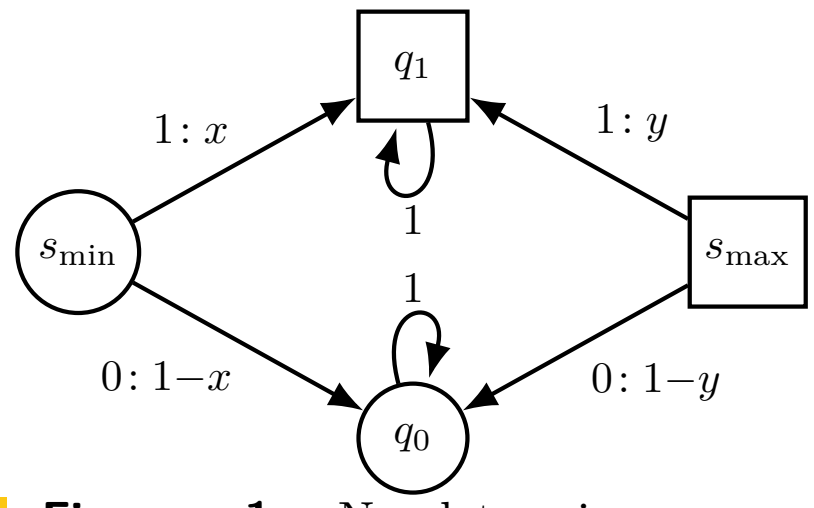
\begin{wrapfigure}{r}{0.43\textwidth}
\vspace{-1\baselineskip}
\centering
\begin{tikzpicture}[>=Latex, thick, scale=0.82,
                    every node/.style={transform shape}]
  \tikzstyle{minstate}=[state]
  \tikzstyle{maxstate}=[state, rectangle]

  \node[minstate] (smin) at (0,0) {$s_{\min}$};
  \node[maxstate] (smax) at (5.4,0) {$s_{\max}$};
  \node[minstate] (q0) at (2.7,-1.5) {$q_0$};
  \node[maxstate] (q1) at (2.7,1.5) {$q_1$};

  \path[->]
  (smin) edge node[midway, below left] {$0\!:1{-}x$} (q0)
  (smin) edge node[midway, above left] {$1\!:x$} (q1)
  (smax) edge node[midway, above right] {$1\!:y$} (q1)
  (smax) edge node[midway, below right] {$0\!:1{-}y$} (q0)
  (q0) edge [loop above] node {$1$} (q0)
  (q1) edge [loop below] node {$1$} (q1);
\end{tikzpicture}
\vspace{-0.8\baselineskip}
\caption{Nondeterminacy
under averaged-distribution semantics.}
\label{fig:nondet-gadget}
\vspace{1em}
\end{wrapfigure}

\noindent\textbf{Determinacy: averaging versus sampling.}
Under averaged-distribution semantics, the continuation value is applied to
the distribution produced jointly by the players' mixed choices. Because this
dependence is nonlinear, the classical minimax argument no longer applies.
More strongly, determinacy itself can fail, even in a four-state arena with
deterministic transitions. In the game shown in
\cref{fig:nondet-gadget}, the initial distribution places equal mass on 
 $s_{\min}$ controlled by \MinPl{} and $s_{\max}$ controlled by \MaxPl{}. 
Each player
routes its mass to one
 of two absorbing states. If $x$ and $y$ are the
probabilities with which \MinPl{} and \MaxPl{}, respectively, choose the
second absorbing state, then the normalised discounted payoff is
\[
G_\gamma(x,y)
=(1-\gamma)+\gamma h\!\left(\frac{x+y}{2}\right),
\]
where $h$ is binary entropy, $h(z)=-z\log_2z-(1-z)\log_2(1-z)$. The initial distribution is uniform on $\{s_{\min},s_{\max}\}$, so the first stage has entropy $1$, and its discounted weight gives the first term $(1-\gamma)\cdot1$. From the second stage on, the mass is $\bigl(1-\frac{x+y}2,\frac{x+y}2\bigr)$ on the two absorbing states and stays fixed there, so every later stage has entropy $h\bigl(\frac{x+y}2\bigr)$, and together they contribute the second term $\gamma\,h\bigl(\frac{x+y}2\bigr)$. The two orders of optimisation give
\[
\inf_{x\in[0,1]}\sup_{y\in[0,1]}G_\gamma(x,y)
=1
>
(1-\gamma)+\gamma h\!\left(\frac14\right)
=
\sup_{y\in[0,1]}\inf_{x\in[0,1]}G_\gamma(x,y).
\]
Thus, the upper and lower values differ: averaging the players' mixed choices
before applying entropy can destroy determinacy. 
Averaging can destroy determinacy without destroying Bellman structure:
the two ordered Bellman operators remain contractions, separately
characterising the upper and lower values and yielding convergent
finite-horizon approximations.

The situation changes under the two sampled semantics. Once a pair of pure action profiles has been selected, it determines a scalar continuation payoff: the value at the resulting successor distribution under sampled-profile semantics, and the expected value over realised transitions under sampled-transition semantics. Randomisation over pure-profile pairs therefore produces a payoff that is bilinear in the players' mixed strategies. The one-step interaction becomes a finite zero-sum matrix game, allowing minimax to be applied at every distribution. We prove that both sampled semantics are determined and that both players have optimal stationary randomised strategies, even against opponents whose strategies depend on the complete history of observed distributions.

We compare the values induced by the three interpretations. Even the lower
value under averaged-distribution semantics dominates the value under
either sampled semantics (\cref{thm:semantic-value-dominance}). The
sampled-profile and sampled-transition values, however, are incomparable
(\cref{ex:sp-st-reversal}). With a single entropy-minimising player,
averaged-distribution and sampled-profile semantics coincide, and their
common value dominates that under sampled-transition semantics
(\cref{thm:semantic-value-dominance}).

\medskip\noindent\textbf{Algorithms and complexity.}
The Bellman characterisations yield convergent finite-horizon approximations.
Under averaged-distribution semantics, separate ordered recursions approximate
the upper and lower values; under either sampled semantics, each Bellman
update is a finite zero-sum matrix game.

For every fixed rational discount and positive rational promise gap, the
averaged-distribution (upper and lower value) gap problems and both sampled gap problems lie in
$\PSPACE$. Averaged-distribution semantics uses a rational grid for mixed
actions; sampled-profile semantics uses sparse near-optimal mixtures over
pure profiles; and sampled-transition semantics additionally enumerates
transition realisations sequentially. Depth-first evaluation keeps the space
polynomial.

For a single entropy-minimising player, the averaged-distribution and
sampled-profile problems improve to $\NP$, since pure profiles suffice over
the truncated horizon; sampled-transition remains in $\PSPACE$. Conversely,
entropy optimisation is $\NP$-hard under all three semantics, already for
deterministic one-player games, where pure memoryless strategies are optimal, and remains
$\NP$-hard to approximate within a suitable fixed additive constant. This
contrasts with classical discounted MDPs, which are solvable in polynomial time~\cite{puterman1994markov}.

\Cref{tab:disc} summarises the upper bounds. The averaged-distribution and
sampled-profile one-player cases are tight; the remaining cases lie between
$\NP$-hardness and $\PSPACE$. The lower bounds follow
immediately from the one-player construction of \cref{thm:set-cover-hard}.

\begin{table}[t]
\centering
\caption{Upper bounds for the discounted gap problem for rational
$\Discount$ and $\varepsilon_0>0$;
the general-game averaged-distribution entry holds for both $V^+$ and $V^-$.}
\label{tab:disc}
\small
\renewcommand{\arraystretch}{1}
\setlength{\tabcolsep}{5pt}
\begin{tabularx}{\textwidth}{
@{}l
>{\centering\arraybackslash}X
>{\centering\arraybackslash}X@{}
}
\toprule
Semantics & \textbf{One player} & \textbf{General games} \\
\midrule
Averaged distribution
& $\NP$ (\cref{cor:1p-av-np})
& $\PSPACE$ (\cref{thm:etr-ub}) \\
Sampled profile
& $\NP$ (\cref{cor:1p-sp-np})
& $\PSPACE$ (\cref{thm:sp-pspace}) \\
Sampled transition
& $\PSPACE$ (\cref{cor:1p-st-pspace})
& $\PSPACE$ (\cref{thm:st-pspace}) \\
\bottomrule
\end{tabularx}
\end{table}

\medskip\noindent\textbf{Related Work.}
Opacity enforcement in discrete-event systems modifies system dynamics to prevent an observer from inferring a secret predicate
\cite{jacob2016overview,lafortune2018history,wu2014synthesis}, while quantitative information flow studies how much uncertainty about a secret is retained or revealed through observations
\cite{clark2005quantitative,kopf2007information,smith2009foundations,alvim2012measuring}.
Moving-target defence can similarly be viewed as maintaining uncertainty against an entropy-minimising observer. Our setting instead treats this uncertainty as the state of a sequential two-player game in which both players influence the entropy of the evolving distribution. Entropy-based objectives have also been studied in single-agent exploration and control
\cite{hazan2019provably}, whereas we consider a zero-sum interaction between entropy minimisation and maximisation. Cyber-physical security games likewise model adversarial interaction
\cite{huang2019stochastic}, but typically use state- or action-based rewards rather than Shannon entropy of the full state distribution.
Entropy has appeared in stochastic games through entropic-risk objectives
\cite{baier2023entropic}, where an exponential transformation is applied to accumulated conventional reward.

Games called entropy games in \cite{asarin2016entropy,akian2017operator,akian2019operator} have a payoff given by the growth rate of the number of behaviours, equivalently of matrix products (in the operator view of \cite{akian2017operator,akian2019operator}, a mean-payoff game with relative-entropy payments over action simplices); there is no state distribution, and the reward is not the Shannon entropy of one. Maximising the entropy (rate) of the path distribution of Markov chains, interval Markov chains and MDPs \cite{biondi2014maximizing,chen2014complexity,savas2018entropy} concerns a single controller and the entropy of the induced distribution on paths, whereas we study a two-player zero-sum game on the entropy of the state distribution at each step. MDPs and Markov chains have been analysed as transformers of probability distributions, with properties given by linear predicates on the distribution, temporal and monadic logics of probability over such predicates, and safety of the distribution inside a polytope \cite{korthikanti2010reasoning,agrawal2015approximate,beauquier2002logic,akshay2018distribution}, but not with a nonlinear quantitative payoff on the distribution. A related distinction between chance (sampled) and mass (averaged) readings of probabilities is studied by Tsai et al.~\cite{tsai2025chance} for a single agent and threshold reachability, with configurations read as mass. Belief-dependent rewards, including entropy of the belief, appear in $\rho$-POMDPs \cite{araya2010pomdp}; there a single agent controls the belief, while here two players with opposed aims control the distribution, and the averaged and sampled semantics distinguish how randomisation reaches it.

Finally, \cite{akshay2026entropyobjectivesmarkovdecision} study entropy objectives for a single controller in a standard MDP: they minimise the worst-case (over time, after a warm-up) entropy of the state distribution induced by a policy, allow history-dependent policies, and show that memoryless policies do not suffice for their objective. For memoryless policies their update is our one-player averaged-distribution update, but they have no analogue of the sampled semantics, no adversary, and a different objective (worst case over time rather than a discounted sum). In the one-player case both hardness proofs use a one-step deterministic gadget; theirs is a randomized reduction from bounded-occurrence MAX-3SAT \cite{trevisan2001nonapproximability}, so it needs $\mathrm{NP}\neq\mathrm{RP}$, whereas our \cref{thm:set-cover-hard} is a deterministic reduction from the gap of \cite{petrank1994gap} and needs only $\mathrm{P}\neq\mathrm{NP}$, so it also yields their hardness result under the weaker assumption.

\medskip\noindent\textbf{Contributions.}
Our main contributions are:
\begin{enumerate}
\item \emph{Model and semantics.} We introduce stochastic games with distributional entropy objectives, in which the stage reward is the Shannon entropy of the state distribution and two players with opposed aims control that distribution. We formalise three semantics (averaged distribution, sampled profile, sampled transition), which differ in how randomisation reaches the distribution, and show that they give different values: we establish dominance and incomparability relations among them.
\item \emph{Determinacy and Bellman characterisations.} We give contractive Bellman characterisations of the upper and lower averaged values and the two sampled values. Averaged-distribution semantics may be nondetermined even on a four-state deterministic arena, whereas both sampled semantics are determined and admit optimal stationary randomised strategies.
\item \emph{Upper bounds.} For fixed rational discount and positive rational promise gap, the averaged-distribution (upper and lower value) gap problems and both sampled gap problems lie in $\PSPACE$; for a single entropy-minimising player, the averaged and sampled-profile problems lie in $\NP$.
\item \emph{Hardness.} Under all three semantics, for every discount there is a constant $\varepsilon_0>0$ such that the gap problem is $\NP$-hard already for deterministic one-player games, where pure memoryless strategies are optimal. The reduction is deterministic, so the hardness holds unless $\mathrm{P}=\mathrm{NP}$.
\end{enumerate}

\section{Problem Definition}
\label{sec:preliminaries}
\label{sec:model}
For a finite set $X$, let
$\Dist(X)\defeq\{\mu:X\to[0,1]\mid\sum_{x\in X}\mu(x)=1\}$ denote the set of probability distributions over $X$.
For $\mu \in \Dist(X)$, let $\Supp(\mu)\defeq\{x:\mu(x)>0\}$,
and write $\delta_x$ for the Dirac distribution concentrated at $x$. 
For $\mu\in\Dist(X)$ and $g:X\to\mathbb R$, the expectation of $g$ under $\mu$ is
$\mathbb E_{x\sim\mu}[g(x)]\defeq\sum_x\mu(x)g(x)$.
For vector-valued $g$, expectation is taken componentwise.

\medskip\noindent\textbf{Entropy.}
The Shannon entropy of
$\mu\in\Dist(X)$ is
\[
\Ent(\mu)\defeq-\sum_{x\in\Supp(\mu)}\mu(x)\log_2\mu(x),
\qquad 0\log_2 0\defeq0.
\]
For $|X|=n$, $0\le\Ent(\mu)\le\log_2n$; $\Ent$ is strictly concave and
continuous on $\Dist(X)$. For $X=\{0,1\}$, write
$h(z)\defeq-z\log_2z-(1-z)\log_2(1-z)$; its restriction to
$[0,\tfrac12]$ is a strictly increasing bijection onto $[0,1]$.
See \cref{prop:entropy-props}, Appendix \ref{app:prelim} for these standard facts.

\medskip\noindent\textbf{Our game setting.}
A \emph{stochastic game with distributional entropy objectives} is a finite turn-based stochastic arena
$\mathcal E=(\States,\States_{\min},\States_{\max},\Acts,\Trans,\distro)$,
where $\States=\States_{\min}\uplus\States_{\max}$, each $s\in\States$
has enabled actions $\Acts(s)\subseteq\Acts$,
$\Trans:\States\times\Acts\to\Dist(\States)$, and
initial distribution $\distro\in\DistS$. The game state is a distribution
$\distr\in\DistS$, so mass may lie simultaneously on states controlled
by both players.

A \emph{pure profile} $a\in\mathcal P_{\min}$ assigns an enabled action
$a(s)$ to every $s\in\States_{\min}$ (likewise
$b\in\mathcal P_{\max}$). Writing $\rho_{a,b}(s)=a(s)$ on
$\States_{\min}$ and $b(s)$ on $\States_{\max}$, define
\[
\Phi_{a,b}(\distr)(s')
\defeq
\sum_{s\in\States}\distr(s)\Trans(s,\rho_{a,b}(s),s').
\]
A \emph{statewise mixed action} for \MinPl{} is
$x:\States_{\min}\to\Dist(\Acts)$ with $x(s)$ supported on $\Acts(s)$;
write $\mathcal X$ for all such $x$, and analogously $\mathcal Y$ for
\MaxPl{}. With $\pi_{x,y}(s)=x(s)$ on $\States_{\min}$ and $y(s)$ on
$\States_{\max}$,
\[
\DistOp_{x,y}(\distr)(s')
\defeq
\sum_{s\in\States}\distr(s)
\sum_{\alpha\in\Acts(s)}
\pi_{x,y}(s)(\alpha)\Trans(s,\alpha,s').
\]
Thus $\Phi_{a,b}=\DistOp_{x,y}$ for $x(s) = \delta_{a(s)}$ and $y(s) = \delta_{b(s)}$. Mixed actions induce
product profile laws
$\widehat x(a)\defeq\prod_{s\in\States_{\min}}x(s)(a(s))$
and analogously $\widehat y$.

\medskip\noindent\textbf{Three semantics.}
In words, the semantics differ in which randomness is averaged into the distribution and which is sampled before entropy is evaluated:
\begin{center}
\begin{tabular}{lll}
\toprule
& players' randomisation & transition randomisation\\
\midrule
$\mathsf{av}$ & averaged & averaged\\
$\mathsf{sp}$ & sampled (a pure profile) & averaged\\
$\mathsf{st}$ & sampled (a pure profile) & sampled (a successor per state)\\
\bottomrule
\end{tabular}
\end{center}
A fourth combination, which we do not consider, averages the players' mixed actions but samples transitions: it splits the mass at each state across actions and lets each part move to its own sampled successor. Throughout, the game starts from the distribution $\distro$ itself, with mass $\distro(s)$ at each state $s$; starting from a single state drawn from $\distro$ is the case of a Dirac start. 
Strategies observe the sequence of distributions. Under $\mathsf{av}$ nothing else is revealed; under $\mathsf{sp}$ the sampled profile is also revealed through the next distribution; under $\mathsf{st}$ the realised successors are revealed as well.

A history is $h_t=(\distr_0,\ldots,\distr_t)\in\DistS^{t+1}$.
Under averaged-distribution semantics ($\mathsf{av}$), strategies are
Borel maps from histories to $\mathcal X$ and $\mathcal Y$ denoted by $\pi^{\min}_{\mathsf{av},t},\pi^{\max}_{\mathsf{av},t}$; under
sampled-profile and sampled-transition semantics
($\Sem \in \{\mathsf{sp},\mathsf{st}\}$), they map histories to
$\Dist(\mathcal P_{\min})$ and $\Dist(\mathcal P_{\max})$ denoted by $\pi^{\min}_{\Sem,t},\pi^{\max}_{\Sem,t}$.
Write $\StratMin^\Sem,\StratMax^\Sem$ for the resulting strategy spaces.
For $\TransReal\defeq\States^\States$ and $\xi \in \TransReal$, define
\[
P_{a,b}(\xi)\defeq
\prod_{s\in\States}\Trans(s,\rho_{a,b}(s),\xi(s)),
\qquad
\Psi_\xi(\distr)(s')
\defeq
\sum_{s:\xi(s)=s'}\distr(s).
\]
Given a history $h_t$, let $x_t=\pi^{\min}_{\mathsf{av},t}(h_t),y_t = \pi^{\max}_{\mathsf{av},t}(h_t)$ under $\mathsf{av}$ and
$\sigma_t = \pi^{\min}_{\Sem,t}(h_t), \tau_t = \pi^{\max}_{\Sem,t}(h_t)$ under $\Sem \in \{\mathsf{sp},\mathsf{st}\}$, the three distribution updates are
\begin{align*}
\mathsf{av}\ \text{(averaged distribution)}:\quad&
\distr_{t+1}=\DistOp_{x_t,y_t}(\distr_t),\\
\mathsf{sp}\ \text{(sampled profile)}:\quad&
\distr_{t+1}=\Phi_{a_t,b_t}(\distr_t),
\qquad (a_t,b_t)\sim\sigma_t\times\tau_t,\\
\mathsf{st}\ \text{(sampled transition)}:\quad&
\distr_{t+1}=\Psi_{\xi_t}(\distr_t),
\qquad
\xi_t\sim P_{a_t,b_t}\ \text{conditional on }(a_t,b_t).
\end{align*}
Thus $\mathsf{av}$ averages both strategic and transition randomness,
$\mathsf{sp}$ samples the profile but averages transitions, and
$\mathsf{st}$ samples both. Moreover,
$\mathbb E_{\xi\sim P_{a,b}}[\Psi_\xi(\distr)]=\Phi_{a,b}(\distr)$.

A strategy is \emph{stationary} if it depends only on the current
distribution. It is \emph{pure} under $\mathsf{av}$ if its mixed actions
are point-valued, and under $\mathsf{sp},\mathsf{st}$ if it selects a
single pure profile.

These updates induce Borel kernels on $\DistS$ and hence, by the
Ionescu-Tulcea theorem~\cite{kallenberg2002foundations}, unique path
measures and expectations
$\PathProb{\distr}{\Sem}{\stratmin}{\stratmax}$ and
$\PathExpect{\distr}{\Sem}{\stratmin}{\stratmax}$.
Measurability details are deferred to Appendix \ref{app:sec:model-full}.
For fixed $x,y$, couple
$(\mathsf A,\mathsf B)\sim\widehat x\times\widehat y$ and
$\boldsymbol\xi\sim P_{\mathsf A,\mathsf B}$ conditionally. Then
\[
\DistOp_{x,y}(\distr)
=
\mathbb E[\Phi_{\mathsf A,\mathsf B}(\distr)]
=
\mathbb E[\Psi_{\boldsymbol\xi}(\distr)],
\]
and Jensen's inequality gives
\[
\Ent(\DistOp_{x,y}(\distr))
\ge
\mathbb E[\Ent(\Phi_{\mathsf A,\mathsf B}(\distr))]
\ge
\mathbb E[\Ent(\Psi_{\boldsymbol\xi}(\distr))].
\]
In words, $\mathsf{av}$ evaluates entropy on the averaged successor, $\mathsf{sp}$ on a successor that already fixes the pure profile, and $\mathsf{st}$ on one that also fixes each state's realised successor. By Jensen, for fixed mixed actions each sampling step can only lower the expected one-step entropy; this does not order the values, since $V_{\mathsf{sp}}$ and $V_{\mathsf{st}}$ are incomparable (\cref{ex:sp-st-reversal}).

This is the one-step source of the differing semantics
(\cref{lem:mean-profile-successor,cor:entropy-ordering}); full proofs are
in Appendix \ref{app:sec:model-full}. The space $\DistS$ is compact and
convex, and all three successor maps are affine and continuous
(\cref{prop:compact}).

\medskip\noindent\textbf{Determinacy and Decision Problems.}
Fix $\Discount\in(0,1)$. For a path
$\omega=(\RandDistr_t)_{t\ge0}$, define
\begin{equation}
\label{eq:discounted-path-payoff}
\DiscPay_{\Discount}(\omega)
\defeq
(1-\Discount)\sum_{t=0}^{\infty}\Discount^t\Ent(\RandDistr_t).
\end{equation}
Since $0\le\Ent\le\log_2|\States|$, this payoff is bounded. For
$\Sem\in\{\mathsf{av},\mathsf{sp},\mathsf{st}\}$, define
\begin{equation}
\label{eq:objective-path-expectation}
\mathsf{Obj}^{\Sem}_{\Discount}
(\stratmin,\stratmax;\distr)
\defeq
\PathExpect{\distr}{\Sem}{\stratmin}{\stratmax}
[\DiscPay_{\Discount}].
\end{equation}
By Tonelli's theorem ~\cite[Theorem 2.37]{folland1999real}, this equals
$(1-\Discount)\sum_{t\ge0}\Discount^t
\PathExpect{\distr}{\Sem}{\stratmin}{\stratmax}
[\Ent(\RandDistr_t)]$.
Define the upper and lower values
\[
V^{+}_{\Sem}(\distr)
\defeq
\inf_{\stratmin}\sup_{\stratmax}
\mathsf{Obj}^{\Sem}_{\Discount}(\stratmin,\stratmax;\distr),
\qquad
V^{-}_{\Sem}(\distr)
\defeq
\sup_{\stratmax}\inf_{\stratmin}
\mathsf{Obj}^{\Sem}_{\Discount}(\stratmin,\stratmax;\distr).
\]
By the minimax inequality, we have 
$V^-_\Sem(\distr)\le V^+_\Sem(\distr)$. 

\begin{problembox}[Determinacy and Decision Problems]
\label{prob:determinacy}
\label{prob:instance}
Fix $\Sem\in\{\mathsf{av},\mathsf{sp},\mathsf{st}\}$ and rational
$\Discount\in(0,1)$. The game is \emph{determined from $\distr$ under
$\Sem$} if $V^-_\Sem(\distr)=V^+_\Sem(\distr)$.
An instance consists of a stochastic game with distributional entropy objectives, having rational transitions and
initial distribution $\distro$, together with a rational threshold $v$.
For rational $\varepsilon_0>0$,
$\GapVal^{\diamond}_{\Sem,\varepsilon_0}$, $\diamond\in\{+,-\}$, is the
promise problem of distinguishing
\[
V^{\diamond}_{\Sem}(\distro)\le v-\varepsilon_0
\qquad\text{from}\qquad
V^{\diamond}_{\Sem}(\distro)\ge v+\varepsilon_0 .
\]
The exact problem $\ValDec^{\diamond}_{\Sem}$ asks whether
$V^{\diamond}_{\Sem}(\distro)\le v$. When $\Sem$ is determined we omit
$\diamond$; unqualified $\ValDec$ and $\GapVal_{\varepsilon_0}$ denote
the  upper-value problems.
\end{problembox}

A promise gap instance guarantees that the value is not within $\varepsilon_0$ of the threshold $v$; an algorithm must answer correctly only on such instances, which lets it absorb truncation, discretisation and numerical errors below $\varepsilon_0$. The ``discounted gap problem'' of \cref{tab:disc} is $\GapVal^{\diamond}_{\Sem,\varepsilon_0}$ for a fixed rational discount $\Discount$ and gap $\varepsilon_0$; the averaged entry covers both $\diamond\in\{+,-\}$.

\medskip\noindent\textbf{Why discounting.} The payoff $(1-\Discount)\sum_t\Discount^t\Ent(\distr_t)$ is the expected entropy at a random stopping time $T$ with $\Pr[T=t]=(1-\Discount)\Discount^t$, that is, the game ends after each round with probability $1-\Discount$. It models sessions of unknown, bounded length, where dispersing or concentrating mass matters while the session lasts, and it makes all Bellman operators $\Discount$-contractions. A mean-payoff objective is natural when long-run behaviour matters, but the geometric-tail arguments used here do not apply to it, and we leave it open (\cref{sec:conclusion}).

\section{Bellman Characterisations}
\label{sec:bellman}

We first give a Bellman characterisation of the four relevant value
functions. Let $C(\DistS)$ denote the Banach space of continuous
real-valued functions on $\DistS$, equipped with the supremum norm
$\|f\|_\infty\defeq\sup_{\distr\in\DistS}|f(\distr)|$.

\begin{restatable}[Bellman characterisations]{theorem}{bellmancharacterisations}
\label{thm:bellman}
For $f\in C(\DistS)$, define the averaged Bellman operators by
\begin{align}
(\mathcal T_{\mathsf{av}}^+f)(\distr)
&\defeq
(1-\Discount)\Ent(\distr)
+\Discount\min_{x\in\mathcal X}\max_{y\in\mathcal Y}
f(\DistOp_{x,y}(\distr)),
\label{eq:bellman-av-upper}\\
(\mathcal T_{\mathsf{av}}^-f)(\distr)
&\defeq
(1-\Discount)\Ent(\distr)
+\Discount\max_{y\in\mathcal Y}\min_{x\in\mathcal X}
f(\DistOp_{x,y}(\distr)).
\label{eq:bellman-av-lower}
\end{align}
For $\Sem\in\{\mathsf{sp},\mathsf{st}\}$, define
\begin{equation}
(\mathcal T_\Sem f)(\distr)
\defeq
(1-\Discount)\Ent(\distr)
+\Discount\,\operatorname{val}(M_f^\Sem(\distr)),
\label{eq:bellman-sampled}
\end{equation}
where $M_f^\Sem(\distr)$ is the finite matrix indexed by
$a\in\mathcal P_{\min}$ and $b\in\mathcal P_{\max}$, with entries
\begin{align}
M_f^{\mathsf{sp}}(\distr)_{a,b}
&\defeq f(\Phi_{a,b}(\distr)),
\label{eq:bellman-matrix-sp}\\
M_f^{\mathsf{st}}(\distr)_{a,b}
&\defeq
\sum_{\xi\in\TransReal}
P_{a,b}(\xi)\,f(\Psi_\xi(\distr)).
\label{eq:bellman-matrix-st}
\end{align}
Here $\operatorname{val}$ is the value of the zero-sum matrix game
in which \MinPl{} chooses rows and \MaxPl{} chooses columns.
Each of these operators maps $C(\DistS)$ into itself and is a
$\Discount$-contraction:
\[
\|\mathcal T f-\mathcal T g\|_\infty
\le \Discount\|f-g\|_\infty.
\]
The unique fixed points of $\mathcal T_{\mathsf{av}}^+$ and
$\mathcal T_{\mathsf{av}}^-$ are respectively
$V_{\mathsf{av}}^+$ and $V_{\mathsf{av}}^-$.
For $\Sem\in\{\mathsf{sp},\mathsf{st}\}$, the unique fixed point of
$\mathcal T_\Sem$ equals both $V_\Sem^-$ and $V_\Sem^+$; we write
$V_\Sem$ for this common value.

For each operator $\mathcal T$ and corresponding value $V$, let
$W_0\equiv0$ and $W_{t+1}=\mathcal T W_t$. Then $W_t$ is the
corresponding $t$-stage value, with rewards at times
$0,\ldots,t-1$, and
\begin{equation}
0\le V(\distr)-W_t(\distr)
\le \Discount^t\log_2|\States|
\qquad\text{for every }\distr\in\DistS.
\label{eq:bellman-tail}
\end{equation}
\end{restatable}

The proof combines standard contraction and finite-horizon arguments;
full details are given in Appendix~\ref{app:bellman-proof}.
These contraction and finite-horizon arguments use only continuity of the stage reward on $\DistS$, so they carry over to any continuous reward function of the distribution in place of $\Ent$.

The theorem immediately yields a value-iteration scheme.
Starting from $W_0\equiv0$, repeatedly apply the appropriate
Bellman operator. Once
$\Discount^t\log_2|\States|\le\varepsilon$,
the iterate $W_t$ approximates the corresponding value uniformly
within $\varepsilon$.
The remaining algorithmic difficulty is therefore the effective
evaluation of a Bellman step. In \cref{sec:hardness-main}, we turn
these recursions into finite computations and bound their space
requirements.

\section{Determinacy Across the Three Semantics}
\label{sec:determinacy-main}

The three semantics behave differently with respect to determinacy. Under
averaging, the continuation value is applied to a successor distribution
formed by the players' mixed choices, producing a nonlinear one-step
interaction. Under either sampled semantics, each pure profile induces a
scalar continuation payoff, so randomisation yields a zero-sum matrix
game.

\begin{restatable}
[Nondeterminacy under averaging]
{theorem}{avnondeterminacy}
\label{thm:nondetermined}
Stochastic games with distributional entropy objectives under averaged-distribution semantics are not determined.
\end{restatable}
The four-state game of \cref{fig:nondet-gadget} witnesses the result;
the full argument appears in Appendix~\ref{sec:determinacy}.
\begin{remark}[Strategy classes and mixed extensions]
\label{rem:strategy-classes}
Under averaged-distribution semantics each player chooses a mixed action at every state independently, and only the resulting distribution is observed. Sampled-profile strategies are mixtures over pure profiles, so they may correlate actions across states. Nondeterminacy is a statement about these averaged strategies. If $\MinPl$ may also randomise over its averaged strategies by a pre-game coin (a mixed extension), the payoff becomes bilinear in the two mixtures. In the game of \cref{fig:nondet-gadget}, a fair coin for $\MinPl$ over $x\in\{0,1\}$ gives expected payoff $(1-\gamma)+\gamma\cdot\frac12[h(y/2)+h((1-y)/2)]$, which is at most $(1-\gamma)+\gamma h(1/4)$ for every $y$, so the mixed extension is determined there, with value $V^-_{\mathsf{av}}$. This is consistent with \cref{thm:nondetermined}, which concerns averaged strategies: the coin is itself sampled randomness, so the mixed extension is no longer the averaged-distribution reading, in which the only randomness is the splitting of mass. This matches the determinacy of the sampled semantics (\cref{thm:sp-determined}): once randomisation is sampled, not averaged, the minimax argument applies. A natural attempt to restore determinacy by adding a dual variable for the entropy also fails; see \cref{rem:variational}.
\end{remark}

For sampled-profile and sampled-transition semantics, fixing a pure profile
reduces the one-step continuation payoff to a scalar. Thus the players'
mixed choices enter bilinearly, and von Neumann's minimax theorem applies
at each finite horizon.

\begin{restatable}[Sampled-profile and sampled-transition determinacy]{theorem}{sampleddeterminacy}
\label{thm:sp-determined}
\label{thm:st-determined}
For every $\Sem\in\{\mathsf{sp},\mathsf{st}\}$,
$V^{-}_{\Sem}=V^{+}_{\Sem}$.
Moreover, both players have optimal stationary randomised strategies, which remain optimal against arbitrary history-dependent randomised
strategies of the opponent.
\end{restatable}

\emph{Proof intuition (full proof in Appendix~\ref{sec:determinacy}).}
By \cref{thm:bellman}, the two sampled semantics are determined.
It remains to establish stationary optimal strategies. Using
$V_\Sem$ as the continuation value gives a finite matrix game at
every distribution. Measurable selection yields optimal mixed
choices as functions of the current distribution, and their
one-step guarantees telescope over time, proving optimality against
arbitrary history-dependent opponents. 

\begin{restatable}[Comparison of the three semantics]
{theorem}{semanticvaluedominance}
\label{thm:semantic-value-dominance}
The averaged values $V_{\mathsf{av}}^+$ and
$V_{\mathsf{av}}^-$ are concave and, for every
$\distr\in\DistS$,
$
V^+_{\mathsf{av}}(\distr)\ge
V^-_{\mathsf{av}}(\distr)\ge
\max\{V_{\mathsf{sp}}(\distr),V_{\mathsf{st}}(\distr)\}.
$
If $\States_{\max}=\emptyset$, then
$V_{\mathsf{av}}=V_{\mathsf{sp}}\ge V_{\mathsf{st}}$.
In general, $V_{\mathsf{sp}}$ and $V_{\mathsf{st}}$ are
incomparable: either strict inequality can occur.
\end{restatable}

The dominance follows from concavity and Jensen's inequality applied
to the finite-horizon Bellman recursions; the two sampled values are
separated by explicit examples. Full proofs appear in
Appendix~\ref{app:semantic-comparison}.

\section{Gap Value Upper Bounds}
\label{sec:hardness-main}

We show that the gap problem from \cref{prob:instance} lies in
$\PSPACE$ under all three semantics. Each proof truncates the
infinite horizon and evaluates the resulting finite recursion in
polynomial space. The main difference is how the one-step choices
and successor values are handled finitely.
Under averaged-distribution semantics, each player chooses a mixed action
from a continuous simplex, and the resulting entropy values may be
irrational. We therefore need both a finite discretisation of the action
space with controlled error and an efficient finite-precision evaluation
of entropy. The following lemma provides these two ingredients.

\begin{restatable}[Entropy tools]{lemma}{entropytools}
\label{ass:rational}
Let $n=|\States|$, and let $\mu,\nu\in\DistS$, with
$d_{\mathrm{TV}}(\mu,\nu)=\tfrac12\|\mu-\nu\|_1$.

\smallskip\noindent \textbf{Continuity.}
Let
$\omega_n(r)=r\log_2(n-1)+h(r)$,
where $h(r)=-r\log_2r-(1-r)\log_2(1-r)$ denotes binary entropy.
For $0{\le} r{\le}1-1/n$, $\omega_n$ is nondecreasing, and
$d_{\mathrm{TV}}(\mu,\nu)\le r$ implies
\[
|\Ent(\mu)-\Ent(\nu)|\le\omega_n(r).
\]
Moreover, $h(r)\le2\sqrt r$ for $0\le r\le1/2$.

\smallskip\noindent \textbf{Effective evaluation.}
Let $L{\defeq}\lceil\log_2 n\rceil$. 
For rational $\mu\in\DistS$ and integer
$p\ge1$, a deterministic algorithm computes a dyadic number
$\widehat H_p(\mu)\in[0,L]$, with denominator $2^p$, s.t.
\[
|\widehat H_p(\mu)-\Ent(\mu)|\le2^{-p},
\]
in time polynomial in $\operatorname{size}(\mu)+p$, where
$\operatorname{size}(\mu)$ is the total bit length of the coordinates of
$\mu$ written as reduced fractions.
\end{restatable}

Part~1 follows from an optimal coupling argument and the chain rule for
entropy. Part~2 uses a rapidly convergent series expansion for the
logarithm after rescaling each nonzero coordinate to a bounded range.
The full proof appears in Appendix \ref{sec:etr}.

The algorithms below and the hardness result use two properties specific to entropy, namely its modulus of continuity and effective evaluation (\cref{ass:rational}).

\begin{restatable}[Gap bound under averaged-distribution semantics]{theorem}{avgapupperbound}
\label{thm:etr-ub}
Fix rational $\Discount\in(0,1)$ and $\varepsilon_0>0$. Then, for both the upper and the lower value,
$\GapVal^{\diamond}_{\mathsf{av},\varepsilon_0}\in\PSPACE$ for $\diamond\in\{+,-\}$.
\end{restatable}

\begin{proof}[Proof intuition (full proof in Appendix \ref{sec:etr})]
Since entropy is bounded by $\log_2 n$, truncating the discounted objective
after polynomially many rounds incurs only a small tail error. The resulting
finite-horizon value satisfies a one-step recursion over the current
distribution.

The remaining difficulty is that each player chooses from a continuous
simplex. Using \cref{ass:rational}, we replace these mixed-action spaces by
sufficiently fine rational grids, while approximating entropy to polynomially
many bits of precision. The continuity bound controls the discretisation
error, and effective entropy evaluation controls the numerical error.

The resulting finite recursion is evaluated by an alternating Turing machine: one player's grid choice is guessed existentially, the other's universally, and the machine recurses on the successor distribution. 
The horizon and all required precisions are polynomially bounded, so the computation runs in alternating polynomial time and hence in $\PSPACE$. The total truncation, discretisation, and numerical error is kept below the promised gap.
\end{proof}

Under sampled-profile and sampled-transition semantics, no discretisation of
continuous mixed actions is needed. Instead, each one-step game is indexed
by pure profiles, whose number may be exponential. The following lemma shows
that near-optimal mixed strategies nevertheless admit polynomial-size sparse
supports.

\begin{restatable}[Sparse minimax certificates]{lemma}{sparseminimax}
\label{lem:sparse-minimax}
Let $M\in[0,U]^{R\times C}$ be a finite zero-sum game, with the
row player minimising.
Write $[R]=\{1,\dots,R\}$ and $[C]=\{1,\dots,C\}$ for the rows and columns, and $\operatorname{val}(M)=\min_{p}\max_{b\in[C]}\sum_ap_aM(a,b)=\max_q\min_{a\in[R]}\sum_bq_bM(a,b)$ over probability vectors $p,q$ on rows and columns. A distribution supported on columns $b_1,\dots,b_k$ is given by weights $q_1,\dots,q_k\ge0$ with $\sum_iq_i=1$, and likewise for rows with weights $p_i$.
\begin{enumerate}[(i)]
\item \textup{(Soundness.)}
If a distribution supported on columns $b_1,\ldots,b_k$ satisfies
$\sum_i q_iM(a,b_i)\ge v$ for every $a\in[R]$, then
$\operatorname{val}(M)\ge v$. Dually, if a distribution supported on
rows $a_1,\ldots,a_k$ satisfies
$\sum_i p_iM(a_i,b)\le v$ for every $b\in[C]$, then
$\operatorname{val}(M)\le v$.

\item \textup{(Existence.)}
If $\operatorname{val}(M)\ge v+\epsilon$, then for every
$k\ge\max\{1,\lceil2U^2\ln(2R)/\epsilon^2\rceil\}$ there exist columns
$b_1,\ldots,b_k$ such that
$\frac1k\sum_iM(a,b_i)\ge v$ for every $a\in[R]$.
Dually, if $\operatorname{val}(M)\le v-\epsilon$, then for every
$k\ge\max\{1,\lceil2U^2\ln(2C)/\epsilon^2\rceil\}$ there exist rows
$a_1,\ldots,a_k$ such that
$\frac1k\sum_iM(a_i,b)\le v$ for every $b\in[C]$.
Profiles may be repeated and every weight can be
chosen equal to $1/k$. The assertion holds for every integer $k$ at
least the displayed bound.
\end{enumerate}
\end{restatable}
Part~(ii) is due to Lipton and Young~\cite{lipton1994simple} and, independently, Alth\"ofer~\cite{althofer1994sparse}; we use it to sparsify one player's mixed strategies in each Bellman step.

Part~(i) is immediate from the definition of $\operatorname{val}(M)$.
For Part~(ii), sample $k$ pure strategies independently from an optimal
mixed strategy. Hoeffding's inequality ~\cite[Theorem 2]{hoeffding1963probability},
followed by a union bound over all responses, shows that some empirical
uniform mixture satisfies the required inequalities.

\begin{restatable}[Polynomial-space upper bound under sampled-profile semantics]{theorem}{spgapupperbound}
\label{thm:sp-quasi-pspace}
\label{thm:sp-pspace}
Fix rational $\Discount\in(0,1)$ and $\varepsilon_0>0$. Then
$\GapVal_{\mathsf{sp},\varepsilon_0}\in\PSPACE$.
\end{restatable}

\begin{proof}[Proof intuition (full proof in Appendix \ref{app:sp})]
As under averaged-distribution semantics, we truncate the horizon and evaluate
the resulting finite recursion. The difference is that each one-step game is
now a finite matrix game over pure profiles, but may have exponentially many
rows and columns.

By \cref{lem:sparse-minimax}, near-optimal mixed strategies can be replaced
by uniform mixtures over only polynomially many pure profiles. Since each
pure profile has a polynomial-size description, these sparse mixtures can be
guessed and checked in polynomial space. Evaluating the recursion depth-first
and recomputing subproblems as needed keeps only a polynomial-size recursion
stack. With the truncation, sparsification, and finite-precision errors chosen
below the promise gap, this yields a $\PSPACE$ algorithm.
\end{proof}

\begin{restatable}[Polynomial-space upper bound under sampled-transition semantics]{theorem}{stgapupperbound}
\label{thm:st-expspace}
\label{thm:st-pspace}
For fixed rational $\Discount\in(0,1)$ and $\varepsilon_0>0$,
$\GapVal_{\mathsf{st},\varepsilon_0}\in\PSPACE$.
\end{restatable}

The proof, given in Appendix~\ref{app:st}, follows
\cref{thm:sp-pspace} with one additional step. For a fixed pure-profile
pair, sampled-transition semantics may produce exponentially many successor
distributions, so each matrix entry is a weighted average over these
realisations rather than a single recursive value. Enumerating the
realisations one at a time and maintaining a running accumulator uses only
polynomial space. The remaining truncation, sparsification, and error
arguments are unchanged.
Together, \cref{thm:etr-ub,thm:sp-pspace,thm:st-pspace} place the gap-value
problem in $\PSPACE$ under all three semantics.

\medskip\noindent\textbf{Dependence on $\Discount$ and $\varepsilon_0$.} We treat $\Discount$ and $\varepsilon_0$ as fixed. Tracing the proofs, the horizon is $T=O(\log(L/e)/\log(1/\Discount))$, the grid denominator is polynomial in $n$, $A$, $1/e$ and $1/(1-\Discount)$, and the sample size is $k=O\bigl(n(\log n)^2\log(2A)/((1-\Discount)^2e^2)\bigr)$. The algorithms therefore run in space polynomial in the input length, $1/(1-\Discount)$ and $1/\varepsilon_0$; if $\Discount$ and $\varepsilon_0$ are given in binary, the bounds are exponential in their bit lengths.

\medskip\noindent\textbf{The complexity gap.} Theorems~\ref{thm:etr-ub}, \ref{thm:sp-pspace} and \ref{thm:st-pspace} give $\PSPACE$ upper bounds, while \cref{thm:set-cover-hard} gives $\mathrm{NP}$-hardness already for one player. Whether the two-player gap problems lie in a lower class is open. Our alternating machine uses $T=\Theta(\log n)$ rounds with polynomially long guesses, so it does not place the problem in a fixed level of the polynomial hierarchy.

\medskip\noindent\textbf{The One-Player Case.}
Suppose $\States_{\max}=\emptyset$. Then only \MinPl{} chooses actions, and
the one-step optimisation under every semantics is over \MinPl{}'s mixed
actions. The following lemma shows that pure profiles always suffice. 

\begin{restatable}[Pure strategies]{lemma}{puresuffices}
\label{lem:pure-suffices}
Suppose $\States_{\max}=\emptyset$. Define $W_0\equiv0$ and
$W_{t+1}=\mathcal T_\Sem W_t$, where $\mathcal T_\Sem$ is the one-step
Bellman operator under semantics $\Sem$. Then for every
$\Sem\in\{\mathsf{av},\mathsf{sp},\mathsf{st}\}$, the minimum defining
$W_{t+1}$ is attained by a pure profile. In particular,

$$
W_{t+1}(\mu)
=
(1-\Discount)\Ent(\mu)
+
\Discount\min_{a\in\mathcal P_{\min}}
C_t^\Sem(\mu,a),~\text{where}
$$
\[
C_t^{\mathsf{av}}(\mu,a)
=
C_t^{\mathsf{sp}}(\mu,a)
=
W_t(\Phi_a(\mu))
\text{ and }
C_t^{\mathsf{st}}(\mu,a)
=
\sum_{\xi\in\TransReal}
P_a(\xi)\,W_t(\Psi_\xi(\mu)).
\]

\end{restatable}

For $\mathsf{sp}$ and $\mathsf{st}$, the claim is immediate because the
one-step objective is linear in the mixed action and therefore attains its
minimum at a vertex of the simplex. For $\mathsf{av}$, induction shows that
each $W_t$ is concave; Jensen's inequality then implies that no mixed action
can improve on the best pure profile. The full proof appears in
Appendix~\ref{sec:onep}.

\begin{lemma}[One-player upper bounds]
\label{lem:onep-bounds}
For fixed rational $\Discount\in(0,1)$ and $\varepsilon_0>0$, if
$\States_{\max}=\emptyset$, then
$\GapVal_{\mathsf{av},\varepsilon_0}\in\NP$,
$\GapVal_{\mathsf{sp},\varepsilon_0}\in\NP$,
and 
$\GapVal_{\mathsf{st},\varepsilon_0}\in\PSPACE$.
\end{lemma}

\begin{proof}[Proof intuition (full proof in Appendix \ref{app:uponep})]
By \cref{lem:pure-suffices}, finite-horizon optimal play may be restricted
to pure profiles. Under $\mathsf{av}$ and $\mathsf{sp}$, a polynomial-length
sequence of pure profiles therefore serves as a certificate: the verifier
reconstructs the induced distributions, approximates their entropies, and
checks the threshold, giving $\NP$.

Under $\mathsf{st}$, fixing a pure profile still leaves stochastic transition
realisations, so such a deterministic certificate does not suffice. Pure
profiles nevertheless eliminate the sparsification step from the general
$\PSPACE$ algorithm, yielding the stated $\PSPACE$ upper bound.
\end{proof}

\section{Gap Value Lower Bounds}
\label{sec:lower-main}

We complement the upper bounds with a matching hardness result: even for
deterministic one-player games, approximating the gap value within a fixed
additive constant is $\NP$-hard.

\begin{restatable}[One-player hardness]{theorem}{onephardness}
\label{thm:set-cover-hard}
For every rational discount $\Discount\in(0,1)$ there exists a rational constant $\varepsilon_0>0$, depending only on $\Discount$, such that, for every $\Sem\in\{\mathsf{av},\mathsf{sp},\mathsf{st}\}$, $\GapVal_{\Sem,\varepsilon_0}$ is $\NP$-hard, already for deterministic one-player games. In the games constructed, pure memoryless strategies are optimal (\cref{lem:pure-suffices}), so the hardness does not rely on restricting the strategy class.
\end{restatable}

\begin{proof}[Proof intuition (full proof in Appendix~\ref{app:lower})]
The reduction starts from a gap version of Set Cover. It creates one state
for each universe element and one absorbing state for each candidate set.
The initial distribution is uniform over the element states, and the sole
player routes each element to a set containing it.

Because the game is deterministic and has only one player, all three
semantics coincide. After one round the process is absorbing, so the
objective differs from a constant only through the entropy of the induced
distribution over set states. Using fewer sets concentrates probability mass
and lowers entropy; using more sets spreads the mass and raises it.
A fixed gap in the minimum cover size is therefore converted into a fixed
additive gap in achievable entropy. The required Set-Cover gap follows from
Petrank's hardness result for bounded-occurrence three-dimensional
matching ~\cite[Theorem 4.4]{petrank1994gap}. Choosing the threshold between
the two resulting entropy ranges yields the claimed $\NP$-hardness for all
three semantics.
\end{proof}

\section{Exact-Threshold Complexity}
\label{sec:exact-main}

The upper-bound algorithms use the promise margin to absorb errors from
horizon truncation, entropy evaluation, and, where needed, discretisation or
sparsification. They therefore do not yield exact-threshold decision
procedures.
\begin{remark}[Exact upper bounds]
\label{rem:rexp}
One possible route is the structure
$\mathsf{FO}(\Real,+,\cdot,\exp,\leq)$, which can express $x\log x$ and is
conditionally decidable under Schanuel's conjecture~\cite{macintyre1996}.
For averaged-distribution and
sampled-profile semantics, however, we do not obtain a finite formula for the
infinite-horizon value.

Sampled-transition semantics admits such a finite representation. Under $\mathsf{st}$ every state sends all of its mass to one sampled successor, so each update is $\Psi_\xi$ for some map $\xi:\States\to\States$, that is, it pushes $\mu$ forward along $\xi$. After $t$ updates the distribution is $\mu_0$ pushed forward along the composition of $t$ such maps, which is again a map $\States\to\States$. There are at most $n^n$ such maps, so at most $n^n$ distributions are reachable from $\mu_0$, and one further update of a reachable distribution stays among them. The value is therefore described by a finite discounted matrix-game optimality system whose entropy constants are expressible in the real exponential field. For two players this gives a conditional route to exact decidability, but no unconditional upper bound.

With a single (minimising) player, pure profiles suffice (\cref{lem:pure-suffices}), so the at most $n^n$ reachable distributions form a finite discounted MDP in which a stationary pure policy is optimal. Hence $V(\mu_0)\le v$ iff some stationary pure policy $\pi$ has $V^\pi(\mu_0)\le v$.
Let $Q_\pi$ be the stochastic matrix on the reachable distributions under $\pi$, where $Q_\pi(\mu,\nu)$ is the probability that one update from $\mu$ with profile $\pi(\mu)$ leads to $\nu$; its entries are rational. Then $V^\pi=(1-\Discount)(I-\Discount Q_\pi)^{-1}\mathrm{Ent}$, where $\mathrm{Ent}$ is the vector of entropies of the reachable distributions, so each $V^\pi(\mu_0)$ is a rational linear combination of $\log_2 p$ for primes $p$;   the threshold $v$ is a rational multiple of $\log_2 2$, so $V^\pi(\mu_0)-v$ is such a combination too. It is zero iff all its prime coefficients vanish (unique factorisation); otherwise its sign is found by numerical approximation to increasing precision, which terminates because the value is nonzero. There are finitely many policies, so exact-threshold comparison is decidable unconditionally in this case; we do not bound the running time. The argument does not extend to two players, where the optimal strategies depend on the values and the entropy constants enter polynomial matrix-game equations.
\end{remark}

\begin{theorem}[Exact-value hardness]
\label{thm:set-cover-exact}
For every $\Sem\in\{\mathsf{av},\mathsf{sp},\mathsf{st}\}$ and $\diamond\in\{+,-\}$, $\ValDec^{\diamond}_{\Sem}$ is $\NP$-hard, already for deterministic single-player games. In the games constructed, pure memoryless strategies are optimal (\cref{lem:pure-suffices}).
\end{theorem}
\begin{proof}
The reduction of \cref{thm:set-cover-hard} produces a promise gap of width
$2\varepsilon_0$ around a midpoint $M$, which is generally irrational. Choose
a rational $v'$ with $|v'-M|<\varepsilon_0/2$, computable in polynomial time by the
logarithm series used for \cref{ass:rational}. Yes-instances have value at most $M-\varepsilon_0<v'$ and
no-instances at least $M+\varepsilon_0>v'$. Hence the exact comparison
$V^\diamond_\Sem(\distro)\le v'$ decides the same $\NP$-hard language; with a single player the upper and lower values coincide, so this holds for both $\diamond\in\{+,-\}$.
The construction is deterministic and single-player. By \cref{lem:pure-suffices}, pure profiles attain the same value as arbitrary history-dependent strategies, as shown at the start of Appendix~\ref{app:lower}; in the constructed games all mass is absorbed after one round, so a single pure profile suffices at every step. \cref{lem:set-cover-collapse} then computes the value of a pure profile, independently of the semantics.
\end{proof}

\section{Conclusion}
\label{sec:conclusion}
\label{sec:discussion}

We studied two-player zero-sum stochastic games whose stage reward is the Shannon entropy of the state distribution. Because entropy is nonlinear, the way randomisation enters the distribution matters: the three interpretations agree in expectation for affine rewards, but give genuinely different games here. Averaging before evaluating entropy can destroy determinacy, even on a four-state deterministic arena, whereas both sampled semantics are determined and admit optimal stationary randomised strategies. The values are only partly ordered: even the lower averaged-distribution value dominates both sampled values, while sampled-profile and sampled-transition values are incomparable in general. With a single entropy minimiser, averaged-distribution and sampled-profile semantics coincide and dominate sampled-transition semantics.

For fixed rational discount and positive gap, the gap-value problem lies in $\PSPACE$ under all three semantics, and in $\NP$ for a single entropy minimiser under averaged-distribution and sampled-profile semantics. All three semantics are $\NP$-hard already on deterministic one-player games, so the $\NP$ bounds are tight.

Several questions remain open: the exact complexity of the remaining $\PSPACE$ cases, and exact-threshold comparison, which we can decide unconditionally only for a single player under sampled-transition semantics (conditionally for two players under that semantics, and not at all for the others). Extending the theory beyond fixed-discount objectives, in particular to mean payoff, is another natural direction, since the geometric-tail arguments used here no longer apply. The fourth combination of the readings, averaging the players' mixed actions but sampling transitions, is also left open. More broadly, these questions ask how much of classical stochastic-game theory survives when the payoff is a nonlinear function of the evolving state distribution.

\bibliography{entropy-games}

\appendix
\clearpage
\section*{Guide to the appendix}
\label{app:guide}
Each row lists a section or subsection of the appendix, what it contains, and the part of the main text it supports.
\medskip

\footnotesize
\setlength{\tabcolsep}{3pt}
\begin{longtable}{@{}p{0.7cm}p{2.7cm}p{6.1cm}p{3.0cm}@{}}
\toprule
 & Title & Contents & Supports\\
\midrule
\endhead
\ref{sec:app-preliminaries} & Extended preliminaries & Notation for finite distributions, supports, expectations; Shannon entropy; norms and total variation. Proposition~\ref{prop:entropy-props}: bounds, strict concavity, continuity, binary entropy (with proofs). & Section~\ref{sec:model}; every later proof \\
\ref{app:sec:model-full} & The model in full & Arenas, strategies and the three semantics in detail. Contains the subsections B.1 to B.4 below. & Section~\ref{sec:model} \\
\ref{app:update-sem} & Distribution updates & The maps $\DistOp_{x,y}$, $\Phi_{a,b}$, $\Psi_\xi$ that update the current distribution under averaged, sampled-profile and sampled-transition semantics. & Section~\ref{sec:model} (semantics) \\
\ref{app:path-laws} & Canonical path laws & Kernels and canonical path laws for each semantics, and the conditional path-law identity used in the determinacy proofs. & Section~\ref{sec:model}; Appendix~\ref{sec:determinacy} \\
\ref{app:eg} & A running example & One round of a three-state arena under the three semantics: the same mean successor distribution, but expected entropies $1.4772>1.4056>1$ (\cref{lem:mean-profile-successor}, \cref{cor:entropy-ordering}). & Section~\ref{sec:model} (intuition) \\
\ref{app:relating} & Relating the three semantics & Lemmas relating the mean successor under averaging to sampled profiles and sampled transitions (\cref{lem:av-vs-sp,lem:sp-vs-st,lem:av-vs-st-direct}); one-step entropy ordering (\cref{cor:entropy-ordering}); compactness and continuity of the update maps (\cref{prop:compact}). & Section~\ref{sec:model}; Appendices~\ref{app:bellman-proof} and~\ref{app:semantic-comparison} \\
\ref{app:bellman-proof} & Bellman characterisations & Proof of \cref{thm:bellman}: continuity and attainment, contraction, finite-horizon identification (with Borel Markov policies), infinite-horizon identification and the error bound $\Discount^t\log_2|\States|$. Cited by Appendices~\ref{sec:determinacy}, \ref{sec:hardness} and~\ref{app:semantic-comparison} for the finite-horizon facts. & Theorem~\ref{thm:bellman} \\
\ref{sec:determinacy} & Determinacy & Contains D.1 and D.2 below. & Section~\ref{sec:determinacy-main} \\
\ref{app:det-av} & Averaged semantics: nondeterminacy & The four-state example with upper value $1$ and lower value $(1-\Discount)+\Discount h(1/4)$, and the proof that the averaged game is not determined. & Theorem~\ref{thm:nondetermined} \\
\ref{app:det-sampled} & Sampled semantics: determinacy & Matrix-game background; one-step identity for $V_\Sem$ from Appendix~\ref{app:bellman-proof}; Borel stationary saddle strategies; verification against history-dependent opponents. & Theorem~\ref{thm:sp-determined} (sp and st) \\
\ref{sec:hardness} & Gap value upper bounds & Contains E.1 to E.3 below. Common setting: fixed $\Discount,\varepsilon_0$, binary input. & Section~\ref{sec:hardness-main} \\
\ref{sec:etr} & Averaged semantics & Entropy tools (\cref{ass:rational}: continuity and effective evaluation of $\Ent$), horizon truncation, action grid, alternating machine; $\PSPACE$ upper bound for the upper value $V^+_{\mathsf{av}}$ and, symmetrically, for $V^-_{\mathsf{av}}$. & Lemma~\ref{ass:rational}, Theorem~\ref{thm:etr-ub} \\
\ref{app:sp} & Sampled-profile semantics & Contains the four parts below, proving Theorem~\ref{thm:sp-pspace}. & Theorem~\ref{thm:sp-pspace} \\
\ref{app:sec:sp-background} & \quad Background & $V_{\mathsf{sp}}$ as the unique fixed point of the Bellman operator $\mathcal T_{\mathsf{sp}}$ (from Appendix~\ref{sec:determinacy}), its range, and the matrix-game facts used by the algorithm. & Appendix~\ref{app:sp} \\
\ref{app:sec:sparse-lemma} & \quad Sparse certificate lemma & Sparse minimax strategies for zero-sum matrix games (Lipton-Young, Alth\"ofer), in the form used here (\cref{lem:sparse-minimax}). & Lemma~\ref{lem:sparse-minimax} \\
\ref{app:sec:compact-profile} & \quad Compact encoding of a pure profile & How a pure profile is stored in polynomial space. & Appendix~\ref{app:sec:sp-main} \\
\ref{app:sec:sp-main} & \quad Polynomial-space bound & Six steps: finite horizon and truncation error; sparse empirical replacement of the matrix value; finite-precision recursion; two-sided error invariant; correctness of the decision procedure; bit lengths and workspace. & Theorem~\ref{thm:sp-pspace} \\
\ref{app:st} & Sampled-transition semantics & Same scheme as sampled-profile, with the exact sum over transition realisations accumulated sequentially (error not multiplied by $n^n$); $\PSPACE$ bound. & Theorem~\ref{thm:st-pspace} \\
\ref{sec:onep} & The one-player case & Pure strategies suffice for one player (\cref{lem:pure-suffices}); averaged and sampled-profile values coincide (\cref{cor:av-sp-coincide}). & Lemma~\ref{lem:pure-suffices} \\
\ref{app:uponep} & Upper bounds for one player & $\NP$ upper bounds for the averaged and sampled-profile gap problems (\cref{cor:1p-av-np,cor:1p-sp-np}) and the sampled-transition bound (\cref{cor:1p-st-pspace}). & Lemma~\ref{lem:onep-bounds} \\
\ref{app:lower} & Lower bounds & Set-Cover gadget (deterministic one-player game), collapse to one-step entropy, entropy of a routing, bounded Set-Cover gap with perfect completeness, the hardness reduction and the exact-value version. & Theorem~\ref{thm:set-cover-hard} (Section~\ref{sec:lower-main}), Theorem~\ref{thm:set-cover-exact} (Section~\ref{sec:exact-main}) \\
\ref{app:semantic-comparison} & Comparing the semantics & Contains the two subsections below, plus remarks on deterministic transitions and pure strategies. & Section~\ref{sec:determinacy-main} (comparison) \\
\ref{app:cmp-openloop} & Open-loop comparison & Expected entropy ordering for prescribed open-loop policies (\cref{thm:open-loop-order}): averaged $\ge$ sampled-profile $\ge$ sampled-transition at every time. & Section~\ref{sec:determinacy-main} (comparison) \\
\ref{app:cmp-concavity} & Concavity and dominance & Mass-weighted concavity of the averaged values (\cref{lem:mass-weighted-concavity}); proof of \cref{thm:semantic-value-dominance}; Example~\ref{ex:sp-st-reversal} showing sp and st are incomparable. & Theorem~\ref{thm:semantic-value-dominance} \\
\bottomrule
\end{longtable}
\normalsize
\clearpage

\newpage
\section{Extended Preliminaries}
\label{sec:app-preliminaries}

\subsection{Finite distributions and entropy}
\label{app:prelim}
For a finite set $X$, let
\[
\Dist(X)
\defeq
\left\{
\mu:X\to[0,1]
\;\middle|\;
\sum_{x\in X}\mu(x)=1
\right\}
\]
denote the set of probability distributions over $X$. The \emph{support} of
$\mu\in\Dist(X)$ is
$\Supp(\mu)\defeq\{x\in X:\mu(x)>0\}$,
and the Dirac distribution concentrated at $x\in X$ is denoted by $\delta_x$.
For a function $g:X\to\mathbb R$, we write
\[
\mathbb E_{x\sim\mu}[g(x)]
\defeq
\sum_{x\in X}\mu(x)g(x).
\]
The same notation is used for vector-valued $g$, with expectation taken
coordinatewise.

\medskip\noindent\textbf{Shannon entropy.}
For a finite set $X$, the \emph{Shannon entropy} of a distribution
$\mu\in\Dist(X)$ is
\[
\Ent(\mu)
\defeq
-\sum_{x\in\Supp(\mu)}\mu(x)\log_2\mu(x),
\]
with the convention $0\log_2 0=0$.

\medskip\noindent\textbf{Norms and total variation distance.}
For a real-valued function $f$ on $\DistS$, write
$\|f\|_\infty\defeq\sup_{\distr\in\DistS}|f(\distr)|$ for its supremum
norm. For a vector $\mu\in\mathbb R^\States$, such as a signed difference
of two distributions, write
$\|\mu\|_1\defeq\sum_{s\in\States}|\mu(s)|$ for its $\ell_1$ norm. For
$\mu,\nu\in\DistS$, the \emph{total variation distance} is
$d_{\mathrm{TV}}(\mu,\nu)\defeq\tfrac12\|\mu-\nu\|_1$.

\begin{proposition}
\label{prop:entropy-props}
We will make repeated use of the following basic properties of Shannon entropy.
Let $X$ be a finite set with $|X|=n$, and let $\mu,\nu\in\Dist(X)$.
\begin{enumerate}[(i)]
  \item  \textbf{\emph{Bounds and extremizers.}}
  $0\leq\Ent(\mu)\leq\log_2 n$. Moreover, $\Ent(\mu)=0$ if and only if
  $\mu=\delta_x$ for some $x\in X$, and $\Ent(\mu)=\log_2 n$ if and only if
  $\mu$ is uniform on $X$.
  \item \textbf{\emph{Strict concavity.}}
  $\Ent$ is strictly concave: for every $\lambda\in[0,1]$,
  \[
  \Ent\bigl(\lambda\mu+(1-\lambda)\nu\bigr)
  \geq
  \lambda\Ent(\mu)+(1-\lambda)\Ent(\nu),
  \]
  with strict inequality whenever $\mu\neq\nu$ and $\lambda\in(0,1)$.
  \item \textbf{\emph{Continuity.}}
  $\Ent$ is continuous on the compact set $\Dist(X)$ and is therefore bounded
  and uniformly continuous.
\item \textbf{\emph{Binary entropy.}}
For a binary state space $X=\{0,1\}$, every distribution has the form
$(z,1-z)$ for some $z\in[0,1]$. We write its entropy as
$h(z)\defeq -z\log_2 z-(1-z)\log_2(1-z)$.
The function $h$ is symmetric about $1/2$ and restricts to a strictly
increasing bijection from $[0,\tfrac12]$ onto $[0,1]$.
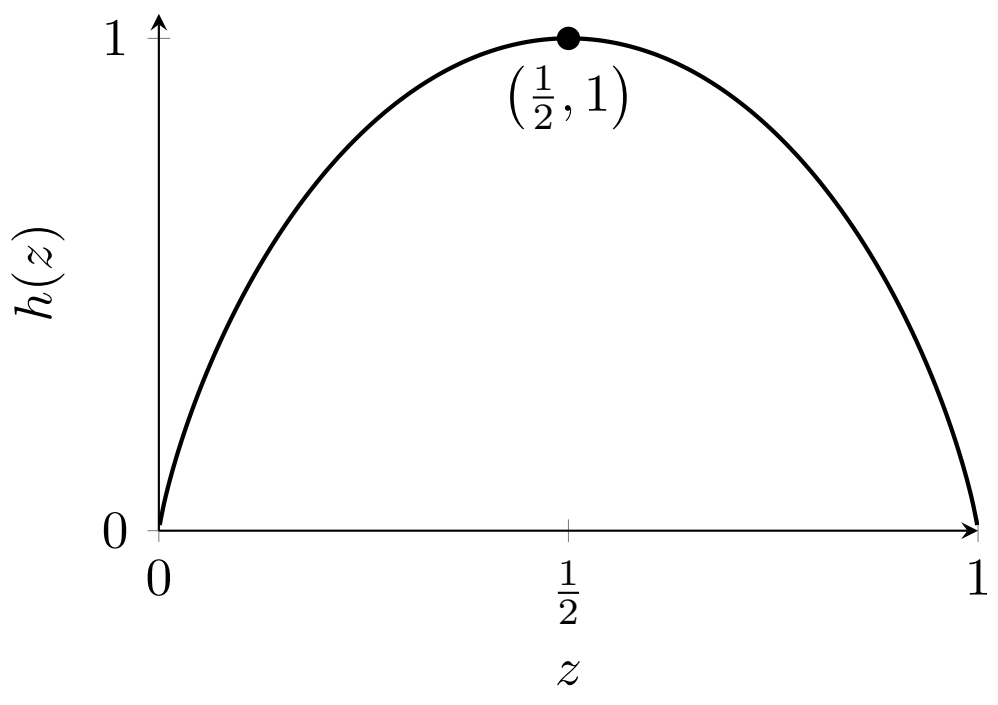
\begin{center}
\begin{minipage}{0.5\linewidth}
\centering
\begin{tikzpicture}
\begin{axis}[
    width=7cm,
    height=5cm,
    domain=0.001:0.999,
    samples=200,
    xlabel={$z$},
    ylabel={$h(z)$},
    xmin=0, xmax=1,
    ymin=0, ymax=1.05,
    xtick={0,0.5,1},
    xticklabels={$0$,$\tfrac12$,$1$},
    ytick={0,1},
    axis lines=left,
    smooth
]
\addplot[thick]
    {-x*ln(x)/ln(2) - (1-x)*ln(1-x)/ln(2)};
\addplot[only marks, mark=*] coordinates {(0.5,1)};
\node at (axis cs:0.5,0.88) {$\left(\tfrac12,1\right)$};
\end{axis}
\end{tikzpicture}
\end{minipage}
\end{center}
\end{enumerate}
\end{proposition}
\begin{proof}
These are standard facts about Shannon entropy. Write $f(p)\defeq-p\log_2p$
for $p\in(0,1]$ and $f(0)\defeq0$, so that $\Ent(\mu)=\sum_{x\in X}f(\mu(x))$;
$f$ is continuous on $[0,1]$ since $\lim_{p\to0^+}p\log_2p=0$.

\smallskip\noindent\textit{Part (i).}
Each term $f(\mu(x))$ is nonnegative, and $f(p)=0$ iff $p\in\{0,1\}$, so
$\Ent(\mu)=0$ iff $\mu=\delta_x$ for some $x$. For the upper bound let
$S\defeq\Supp(\mu)$, $m\defeq|S|\le n$, and let $Y\defeq1/\mu(x)$ on the event
$\{x\}$, $x\in S$. Then $\mathbb E_\mu[Y]=m$ and
$\mathbb E_\mu[\log_2Y]=\Ent(\mu)$, so Jensen's inequality for the strictly
concave $\log_2$ gives
$\Ent(\mu)\le\log_2\mathbb E_\mu[Y]=\log_2m\le\log_2n$. Equality in Jensen holds
iff $Y$ is $\mu$-a.s.\ constant, i.e.\ $\mu$ is uniform on $S$, and $m=n$ forces
$S=X$, so $\Ent(\mu)=\log_2n$ iff $\mu$ is uniform on $X$.

\smallskip\noindent\textit{Part (ii).}
On $(0,1)$, $f''(p)=-1/(p\ln2)<0$, so $f$ is strictly concave there. Since $f$ is continuous on $[0,1]$, it is concave on $[0,1]$. It is strictly concave on $[0,1]$: if $f$ met the chord over some segment $[p,q]$ at an interior point, concavity would force $f$ to be affine on $[p,q]$, which is impossible because $f$ is strictly concave on $(p,q)\subseteq(0,1)$. Applying this at $\mu(x),\nu(x)$ for each $x$ and summing gives concavity of $\Ent$. If $\mu\ne\nu$ and $\lambda\in(0,1)$, the term at some $x_0$ with $\mu(x_0)\ne\nu(x_0)$ is strict, hence so is the sum.

\smallskip\noindent\textit{Part (iii).}
$\Ent$ is a finite sum of the continuous maps $\mu\mapsto f(\mu(x))$, and
$\Dist(X)$ is closed and bounded in $\mathbb R^X$, hence compact; so $\Ent$ is
bounded and uniformly continuous (extreme value and Heine-Cantor theorems).

\smallskip\noindent\textit{Part (iv).}
$h'(z)=\log_2\frac{1-z}{z}>0$ on $(0,\tfrac12)$, so $h$ is strictly increasing
on $[0,\tfrac12]$; as $h(0)=0$ and $h(\tfrac12)=1$, continuity gives the
bijection.
\end{proof}

\section{Stochastic Games with Distributional Entropy Objectives} 
\label{app:sec:model-full}
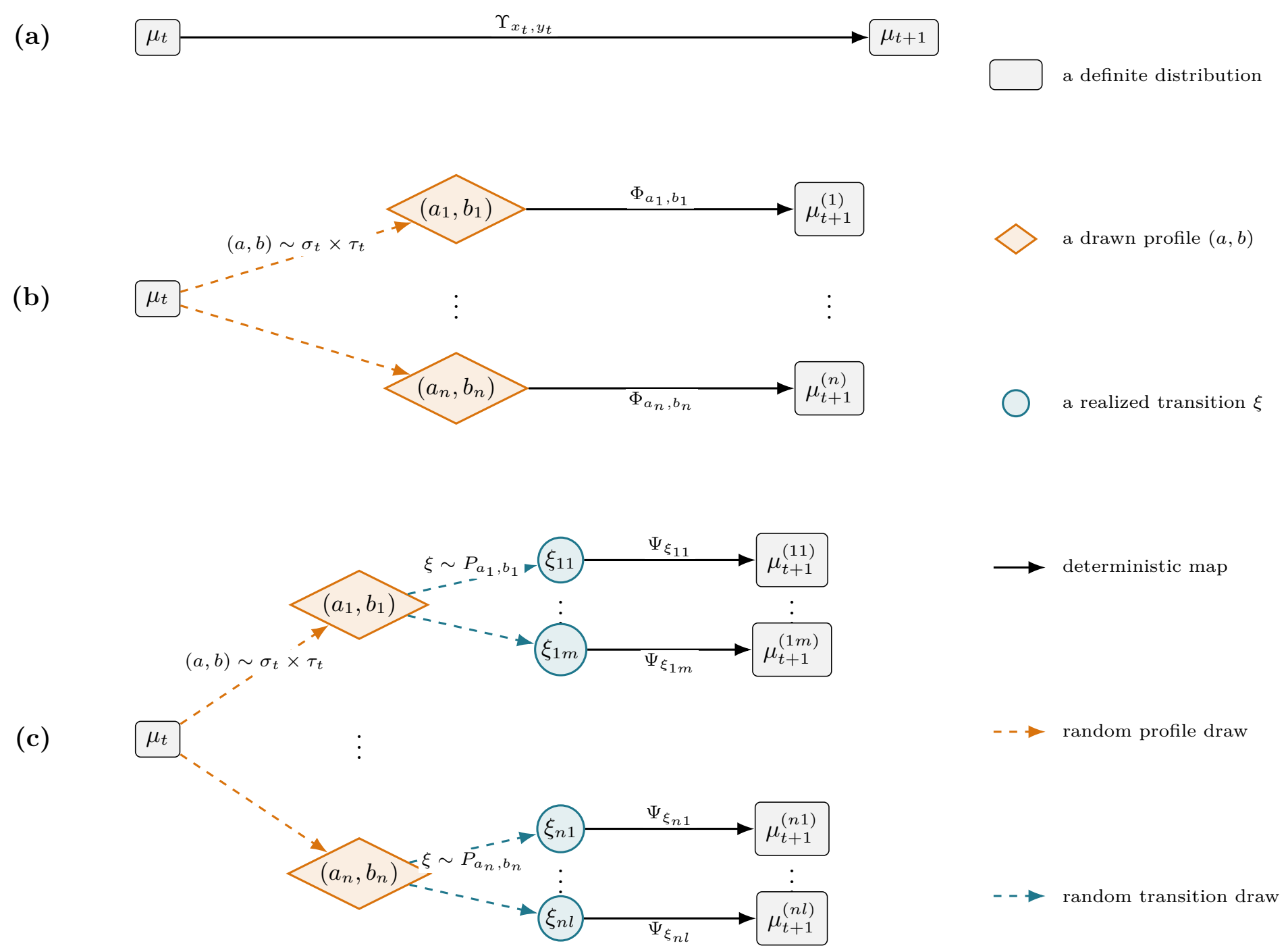
\begin{figure}[ht!]
\centering
\definecolor{profcol}{RGB}{217,115,13}   %
\definecolor{xicol}{RGB}{31,119,140}     %
\resizebox{0.9\textwidth}{!}{%
\begin{tikzpicture}[
  >=Latex,
  dist/.style={draw,rounded corners=2pt,inner sep=4pt,align=center,font=\small,fill=gray!10},
  choice/.style={draw=profcol,thick,diamond,aspect=2,inner sep=1pt,align=center,font=\small,fill=profcol!12},
  xi/.style={draw=xicol,thick,circle,inner sep=1pt,align=center,font=\small,fill=xicol!12},
  det/.style={-{Latex},thick,black},
  rndprof/.style={-{Latex},thick,dashed,draw=profcol},
  rndxi/.style={-{Latex},thick,dashed,draw=xicol},
  lab/.style={font=\scriptsize,fill=white,inner sep=1pt},
  dots/.style={font=\normalsize}
]
  \node[font=\bfseries,anchor=east] at (-1.3,10) {(a)};
  \node[dist] (a0) at (0,10) {$\distr_t$};
  \node[dist] (a1) at (10,10) {$\distr_{t+1}$};
  \draw[det] (a0) -- node[lab,above] {$\DistOp_{x_t,y_t}$} (a1);
  \node[font=\bfseries,anchor=east] at (-1.3,6.5) {(b)};
  \node[dist] (b0) at (0,6.5) {$\distr_t$};
  \node[choice] (bc1) at (4,7.7) {$(a_1,b_1)$};
  \node[dots] (bdots) at (4,6.5) {$\vdots$};
  \node[choice] (bcn) at (4,5.3) {$(a_n,b_n)$};
  \node[dist] (bo1) at (9,7.7) {$\distr_{t+1}^{(1)}$};
  \node[dots] (bodots) at (9,6.5) {$\vdots$};
  \node[dist] (bon) at (9,5.3) {$\distr_{t+1}^{(n)}$};
  \draw[rndprof] (b0) -- node[lab,above] {$(a,b)\sim\sigma_t\times\tau_t$} (bc1);
  \draw[rndprof] (b0) -- (bcn);
  \draw[det] (bc1) -- node[lab,above] {$\Phi_{a_1,b_1}$} (bo1);
  \draw[det] (bcn) -- node[lab,below] {$\Phi_{a_n,b_n}$} (bon);
  \node[font=\bfseries,anchor=east] at (-1.3,0.6) {(c)};
  \node[dist] (c0) at (0,0.6) {$\distr_t$};
  \node[choice] (cp1) at (2.7,2.4) {$(a_1,b_1)$};
  \node[dots] (cpdots) at (2.7,0.6) {$\vdots$};
  \node[choice] (cpn) at (2.7,-1.2) {$(a_n,b_n)$};
  \draw[rndprof] (c0) -- node[lab,above] {$(a,b)\sim\sigma_t\times\tau_t$} (cp1);
  \draw[rndprof] (c0) -- (cpn);
  \node[xi] (cx11) at (5.4,3.0) {$\xi_{11}$};
  \node[dots] (cx1dots) at (5.4,2.4) {$\vdots$};
  \node[xi] (cx1m) at (5.4,1.8) {$\xi_{1m}$};
  \draw[rndxi] (cp1) -- node[lab,above] {$\xi\sim P_{a_1,b_1}$} (cx11);
  \draw[rndxi] (cp1) -- (cx1m);
  \node[xi] (cxn1) at (5.4,-0.6) {$\xi_{n1}$};
  \node[dots] (cxndots) at (5.4,-1.2) {$\vdots$};
  \node[xi] (cxnl) at (5.4,-1.8) {$\xi_{nl}$};
  \draw[rndxi] (cpn) -- node[lab,below] {$\xi\sim P_{a_n,b_n}$} (cxn1);
  \draw[rndxi] (cpn) -- (cxnl);
  \node[dist] (co11) at (8.5,3.0) {$\distr_{t+1}^{(11)}$};
  \node[dots] (codots1) at (8.5,2.4) {$\vdots$};
  \node[dist] (co1m) at (8.5,1.8) {$\distr_{t+1}^{(1m)}$};
  \node[dist] (con1) at (8.5,-0.6) {$\distr_{t+1}^{(n1)}$};
  \node[dots] (codotsn) at (8.5,-1.2) {$\vdots$};
  \node[dist] (conl) at (8.5,-1.8) {$\distr_{t+1}^{(nl)}$};
  \draw[det] (cx11) -- node[lab,above] {$\Psi_{\xi_{11}}$} (co11);
  \draw[det] (cx1m) -- node[lab,below] {$\Psi_{\xi_{1m}}$} (co1m);
  \draw[det] (cxn1) -- node[lab,above] {$\Psi_{\xi_{n1}}$} (con1);
  \draw[det] (cxnl) -- node[lab,below] {$\Psi_{\xi_{nl}}$} (conl);
  \node[dist,minimum width=0.7cm,minimum height=0.4cm] (lg1) at (11.5,9.5) {};
  \node[font=\scriptsize,anchor=west] at (12.0,9.5) {a definite distribution};
  \node[choice,minimum width=0.55cm,minimum height=0.4cm] (lg2) at (11.5,7.3) {};
  \node[font=\scriptsize,anchor=west] at (12.0,7.3) {a drawn profile $(a,b)$};
  \node[xi,minimum size=0.35cm] (lg3) at (11.5,5.1) {};
  \node[font=\scriptsize,anchor=west] at (12.0,5.1) {a realized transition $\xi$};
  \draw[det] (11.2,2.9) -- (11.9,2.9);
  \node[font=\scriptsize,anchor=west] at (12.0,2.9) {deterministic map};
  \draw[rndprof] (11.2,0.7) -- (11.9,0.7);
  \node[font=\scriptsize,anchor=west] at (12.0,0.7) {random profile draw};
  \draw[rndxi] (11.2,-1.5) -- (11.9,-1.5);
  \node[font=\scriptsize,anchor=west] at (12.0,-1.5) {random transition draw};
\end{tikzpicture}%
}
\caption{\small The three semantics as pipelines from $\distr_t$ to $\distr_{t+1}$. Orange marks the profile draw $(a,b)\sim\sigma_t\times\tau_t$, teal the transition draw $\xi\sim P_{a,b}$, black the deterministic maps. (a)~Averaged: no random draw. (b)~Sampled profile: one draw, then $\Phi_{a_i,b_i}$. (c)~Sampled transition: a profile draw and then a transition draw, then $\Psi_{\xi_{ij}}$; the numbers $m,l$ of realizable $\xi$ may differ between profiles. Only representative branches are drawn, with $\vdots$ for the rest.}
\label{fig:three-semantics-branching}
\end{figure}

First, we give an illustration that describes the three semantics in \cref{fig:three-semantics-branching}.

\subsection{Distribution Updates across the 3 semantics}
\label{app:update-sem}
Four maps describe how one round of play updates the current
distribution, depending on how much randomness has already been
resolved at the point the update is taken. $\Phi_{a,b}$
(\eqref{eq:phi-def}) gives the successor distribution once a pure
profile $(a,b)$ is fixed and the arena's own transition probabilities
are averaged in directly. $\DistOp_{x,y}$ (\eqref{eq:distribution-update})
does the same starting from a statewise-mixed profile $(x,y)$, averaging
over the players' randomised choices as well. $P_{a,b}$
(\eqref{eq:transition-realization-prob}) is instead the probability law
of a fully realised transition outcome $\xi$, given a fixed pure profile
$(a,b)$; and $\Psi_\xi$ (\eqref{eq:psi-def}) gives the successor
distribution once that particular realisation $\xi$ has actually
occurred.

\begin{equation}
\label{eq:phi-def}
\Phi_{a,b}(\distr)(s')
\defeq
\sum_{s\in\States}
\distr(s)\,
\Trans\bigl(s,\rho_{a,b}(s),s'\bigr).
\end{equation}

\begin{equation}
\label{eq:distribution-update}
\DistOp_{x,y}(\distr)(s')
\defeq
\sum_{s\in\States}
\distr(s)
\sum_{\alpha\in\Acts(s)}
\pi_{x,y}(s)(\alpha)\,
\Trans(s,\alpha,s'),
\end{equation}

\begin{equation}
\label{eq:transition-realization-prob}
P_{a,b}(\xi)
\defeq
\prod_{s\in\States}
\Trans\bigl(s,\rho_{a,b}(s),\xi(s)\bigr).
\end{equation}

\begin{equation}
\label{eq:psi-def}
\Psi_\xi(\distr)(s')
\defeq
\sum_{\substack{s\in\States\\ \xi(s)=s'}}
\distr(s).
\end{equation}

A history is $h_t=(\distr_0,\ldots,\distr_t)\in\DistS^{t+1}$. Let $\Hist_t$ denote the set of histories.

 A strategy under averaged-distribution semantics (denoted $\mathsf{av}$)
is a sequence of Borel-measurable maps
\[
\stratminat{\mathsf{av}}{t}:\Hist_t\to\mathcal X,
\qquad
\stratmaxat{\mathsf{av}}{t}:\Hist_t\to\mathcal Y.
\]

Under the average distribution semantics, let 
$x_t=\stratminat{\mathsf{av}}{t}(h_t)$ and
$y_t=\stratmaxat{\mathsf{av}}{t}(h_t)$, where $h_t$ is a history. The next distribution is the
deterministic update

\begin{equation}
\label{eq:av-update}
\distr_{t+1}
=
\DistOp_{x_t,y_t}(\distr_t).
\end{equation}
For $\Sem\in\{\mathsf{sp},\mathsf{st}\}$, where
$\mathsf{sp},\mathsf{st}$ respectively denote the sampled-profile and the
sampled-transition semantics, a strategy under semantics
$\Sem$ is a sequence of Borel-measurable maps
\[
\stratminat{\Sem}{t}:\Hist_t\to\Dist(\mathcal P_{\min}),
\qquad
\stratmaxat{\Sem}{t}:\Hist_t\to\Dist(\mathcal P_{\max}).
\]

Under sampled-profile semantics, at history $h_t$, put
$\sigma_t=\stratminat{\mathsf{sp}}{t}(h_t)$ and
$\tau_t=\stratmaxat{\mathsf{sp}}{t}(h_t)$. Conditional on $h_t$,
independently sample $a_t\sim\sigma_t$ and $b_t\sim\tau_t$, that is,
$a_t\in\mathcal P_{\min}$ and $b_t\in\mathcal P_{\max}$ are random variables
with $\Pr[a_t=a]=\sigma_t(a)$ and $\Pr[b_t=b]=\tau_t(b)$ for every $a,b$, and
$\Pr[a_t=a,b_t=b]=\sigma_t(a)\tau_t(b)$, and set

\begin{equation}
\label{eq:sp-update}
\distr_{t+1}
=
\Phi_{a_t,b_t}(\distr_t).
\end{equation}

Under sampled-transition semantics, at history $h_t$, put
$\sigma_t=\stratminat{\mathsf{st}}{t}(h_t)$ and
$\tau_t=\stratmaxat{\mathsf{st}}{t}(h_t)$. Conditional on $h_t$, first draw
$a_t\sim\sigma_t$ and $b_t\sim\tau_t$ independently, then draw
$\xi_t\sim P_{a_t,b_t}$ conditionally on $(a_t,b_t)$, and set

\begin{equation}
\label{eq:st-update}
\distr_{t+1}
=
\Psi_{\xi_t}(\distr_t).
\end{equation}

\subsection{Canonical path laws}
\label{app:path-laws}
The kernels below instantiate the background from
\cref{sec:preliminaries}. 
All three semantics can  be placed on the same measurable path space. Let
\[
\PathSpace\defeq\DistS^{\mathbb N_0},
\qquad
\PathSigma\defeq\mathcal B(\DistS)^{\otimes\mathbb N_0},
\]
where $\PathSigma$ is the \emph{product $\sigma$-algebra} on $\PathSpace$:
the smallest $\sigma$-algebra on $\DistS^{\mathbb N_0}$ making every
coordinate map $\omega\mapsto\omega_t$ measurable, equivalently the
$\sigma$-algebra generated by the cylinder sets
$\{\omega\in\PathSpace:\omega_0\in C_0,\ldots,\omega_n\in C_n\}$ for Borel
sets $C_0,\ldots,C_n\subseteq\DistS$ and $n\geq0$  i.e.\ events that depend
on only finitely many coordinates of the path. 
Let $\RandDistr_t:\PathSpace\to\DistS$ be the coordinate map
$\RandDistr_t(\omega)=\omega_t$. 
Write $\mathcal F_t$ for the canonical filtration generated by the
coordinate process $\RandDistr_0,\ldots,\RandDistr_t$ alone, namely the
smallest $\sigma$-algebra with respect to which each of
$\RandDistr_0,\ldots,\RandDistr_t$ is measurable (this is the standard
probabilistic notion of a $\sigma$-algebra, unrelated to the
strategy-sampling distribution $\sigma$ used elsewhere in the paper).
This makes $(\RandDistr_t)_{t\geq0}$ adapted: at each time $t$, whether
$\RandDistr_t$ has landed in any given Borel set is already decided by
the information in $\mathcal F_t$, i.e.\ by the path so far up to and
including round $t$; nothing about later rounds is needed to know it,
and nothing about it depends on information from the future.

 For a Borel set $C\subseteq\DistS$, write $\mathbf 1_C(z)=1$ if
$z\in C$ and $\mathbf 1_C(z)=0$ otherwise, and define the one-step kernels
induced by a strategy pair as follows:
\begin{align}
Q^{\mathsf{av}}_t(C\mid h_t)
&\defeq
\mathbf 1_C\!\left(
\DistOp_{\stratminat{\mathsf{av}}{t}(h_t), \stratmaxat{\mathsf{av}}{t}(h_t)}(\distr_t)
\right),
\label{eq:path-kernel-av}
\\
Q^{\mathsf{sp}}_t(C\mid h_t)
&\defeq
\sum_{a\in\mathcal P_{\min}}
\sum_{b\in\mathcal P_{\max}}
\sigma_t(a)\tau_t(b)
\mathbf 1_C\!\left(\Phi_{a,b}(\distr_t)\right),
\label{eq:path-kernel-sp}
\\
Q^{\mathsf{st}}_t(C\mid h_t)
&\defeq
\sum_{a\in\mathcal P_{\min}}
\sum_{b\in\mathcal P_{\max}}
\sigma_t(a)\tau_t(b)
\sum_{\xi\in\TransReal}P_{a,b}(\xi)
\mathbf 1_C\!\left(\Psi_\xi(\distr_t)\right).
\label{eq:path-kernel-st}
\end{align}
In words: \eqref{eq:path-kernel-av} says that, since $\mathsf{av}$-semantics has no
randomness, the probability of $C$ is either $1$ or $0$, according to whether
the single deterministic successor $\DistOp_{x_t,y_t}(\distr_t)$ happens to
lie in $C$, the Dirac mass at that one point.
\eqref{eq:path-kernel-sp} says that the probability of $C$ is obtained by
adding up $\sigma_t(a)\tau_t(b)$, the probability of drawing the pure profile
$(a,b)$, over exactly those profiles whose resulting successor
$\Phi_{a,b}(\distr_t)$ happens to land in $C$.
\eqref{eq:path-kernel-st} repeats this one layer deeper: for every profile
$(a,b)$ (weighted by $\sigma_t(a)\tau_t(b)$) and every realized transition
$\xi$ (weighted, conditional on that profile, by $P_{a,b}(\xi)$), add up the
combined weight over exactly those pairs whose resulting successor
$\Psi_\xi(\distr_t)$ lands in $C$.
In \eqref{eq:path-kernel-sp} and \eqref{eq:path-kernel-st}, $\sigma_t$ and
$\tau_t$ are the distributions prescribed at $h_t$ by the strategies for the
indicated semantics.
These are Borel probability kernels: the successor maps
are continuous, the strategies are Borel measurable, and all displayed sums
are finite. This is the precise sense in
which the first semantics has a deterministic distribution trajectory and the
two sampled semantics produce stochastic processes.
For each semantics $\Sem\in\{\mathsf{av},\mathsf{sp},\mathsf{st}\}$,
initial distribution $\distr\in\DistS$, and strategy pair
$\strpair\in\StratMin^{\Sem}\times\StratMax^{\Sem}$, the
Ionescu-Tulcea theorem therefore gives a unique probability measure
\[
\PathProb{\distr}{\Sem}{\stratmin}{\stratmax}
\quad\text{on }(\PathSpace,\PathSigma)
\]
such that $\RandDistr_0=\distr$ almost surely and, for every $t\geq0$,
\begin{equation}
\label{eq:path-law-conditional}
\PathProb{\distr}{\Sem}{\stratmin}{\stratmax}
\bigl(\RandDistr_{t+1}\in C\mid\mathcal F_t\bigr)
=
Q_t^{\Sem}(C\mid\RandDistr_0,\ldots,\RandDistr_t)
\end{equation}
We write $\PathExpect{\distr}{\Sem}{\stratmin}{\stratmax}$ for expectation
with respect to this measure: for a $\PathSigma$-measurable function
$Y:\PathSpace\to\mathbb R$ (bounded, or nonnegative so the integral is
well-defined in $[0,\infty]$),
\[
\PathExpect{\distr}{\Sem}{\stratmin}{\stratmax}[Y]
\defeq
\int_{\PathSpace}Y(\omega)\,\PathProb{\distr}{\Sem}{\stratmin}{\stratmax}(d\omega).
\]
Under averaged-distribution semantics every kernel is
Dirac, so this path measure is concentrated on the unique trajectory generated
by \eqref{eq:av-update}. Under the two sampled semantics,
$(\RandDistr_t)_{t\geq0}$ is generally a non-deterministic stochastic process,
with laws distinguished precisely by \eqref{eq:path-kernel-sp} and
\eqref{eq:path-kernel-st}.
\subsection{A running example}
\label{app:eg}
We illustrate the three semantics on a three-state arena. State $A$ belongs to $\MinPl$ and has actions $\alpha,\beta$ leading deterministically to $B$ and to $C$ respectively. State $B$ belongs to $\MaxPl$: under $\alpha$ it moves to $A$ with probability $\frac34$ and stays at $B$ with probability $\frac14$, under $\beta$ it moves to $A$ with probability $\frac14$ and stays with probability $\frac34$. State $C$ is absorbing. Start from $\distr_0=(0,\frac12,\frac12)$ (coordinates $A,B,C$), so $\Ent(\distr_0)=1$, and let $\MaxPl$ choose $\alpha$ at $B$ with probability $y=\frac14$ (the actions at $A$ and $C$ do not affect round~1).

\smallskip\noindent\emph{Averaged distribution.} The mixed action at $B$ gives the transition row $\frac14(\frac34,\frac14)+\frac34(\frac14,\frac34)=(\frac38,\frac58)$ on $(A,B)$, so $\distr_1=(\frac3{16},\frac5{16},\frac12)$ with $\Ent(\distr_1)\approx1.4772$.

\smallskip\noindent\emph{Sampled profile.} The pure action at $B$ is sampled. With probability $\frac14$ it is $\alpha$ and $\distr_1=(\frac38,\frac18,\frac12)$, with probability $\frac34$ it is $\beta$ and $\distr_1=(\frac18,\frac38,\frac12)$. Both have entropy $\approx1.4056$, so $\Ent(\distr_1)\approx1.4056$ almost surely.

\smallskip\noindent\emph{Sampled transition.} The successor of $B$ is sampled as well: $\xi(B)=A$ with probability $\frac14\cdot\frac34+\frac34\cdot\frac14=\frac38$, and $\xi(B)=B$ otherwise. Then $\distr_1=(\frac12,0,\frac12)$ or $(0,\frac12,\frac12)$, and $\Ent(\distr_1)=1$ almost surely.

\smallskip
In all three cases the mean of $\distr_1$ is $(\frac3{16},\frac5{16},\frac12)$, as \cref{lem:mean-profile-successor} predicts, but the expected entropies are ordered
\[
1.4772>1.4056>1.0000,
\]
which is the one-step inequality of \cref{cor:entropy-ordering}. The update $\Phi_{a,b}$ keeps the row of $B$ as fractional mass, while $\Psi_\xi$ realises one successor and produces only a two-point distribution. That corollary compares the three updates from a common current distribution and does not by itself assert an ordering along whole trajectories.

\subsection{Relating the three semantics}
\label{app:relating}

Every $x\in\mathcal X$ and
$y\in\mathcal Y$ induce product laws on pure profiles $a,b$:
\begin{equation}
\label{eq:induced-profile-laws}
\widehat x(a)
\defeq
\prod_{s\in\States_{\min}}x(s)(a(s)),
\qquad
\widehat y(b)
\defeq
\prod_{s\in\States_{\max}}y(s)(b(s)).
\end{equation}
Here and below an empty product equals $1$: this is the standard
convention that a product with no factors equals the multiplicative
identity, exactly as an empty sum equals $0$. It is needed precisely
when a player controls no states at all, i.e.\ $\States_{\min}=\emptyset$
or $\States_{\max}=\emptyset$: then $\mathcal P_{\min}$ (respectively
$\mathcal P_{\max}$) contains only the single, vacuous profile assigning
no actions, and the product in \eqref{eq:induced-profile-laws} ranges
over this empty index set. The convention gives $\widehat x(a)=1$ for
that unique profile, correctly making $\widehat x$ the point mass
required for it to be a valid element of $\Dist(\mathcal P_{\min})$.

To see that \eqref{eq:induced-profile-laws} defines an element of
$\Dist(\mathcal P_{\min})$ in general: each factor $x(s)(a(s))$ is
non-negative, so $\widehat x(a)\ge 0$ for every profile $a$; and since
$\mathcal P_{\min}=\prod_{s\in\States_{\min}}A(s)$ is exactly the
Cartesian product of the per-state action sets, summing over all
profiles distributes over the product (using that $\States_{\min}$ is
finite):
\[
\sum_{a\in\mathcal P_{\min}}\widehat x(a)
=\sum_{a\in\mathcal P_{\min}}\prod_{s\in\States_{\min}}x(s)(a(s))
=\prod_{s\in\States_{\min}}\sum_{\alpha\in A(s)}x(s)(\alpha)
=\prod_{s\in\States_{\min}}1
=1.
\]
So $\widehat x$ has non-negative entries summing to $1$, i.e.\ is a
genuine element of $\Dist(\mathcal P_{\min})$; the argument for
$\widehat y$ is identical.

This construction produces only \emph{product} distributions: by
\eqref{eq:induced-profile-laws} the choice prescribed at each state is
independent of the choice at every other state. A general
$\mu\in\Dist(\mathcal P_{\min})$ need not factor this way. For instance,
with two states $s_1,s_2\in\States_{\min}$ each offering actions
$\{0,1\}$, let $\mu$ place mass $\tfrac12$ on the profile $(0,0)$ and
mass $\tfrac12$ on $(1,1)$. This $\mu$ has the same single-state
marginals as the product $\widehat x$ with $x(s_1)=x(s_2)$ the fair
coin on $\{0,1\}$, yet $\mu(0,1)=0$ while $\widehat x(0,1)=\tfrac14$:
under $\mu$ the two states are perfectly correlated, under $\widehat x$
they are independent. So $x\mapsto\widehat x$ picks out exactly the
product-form (uncorrelated) distributions inside the strictly larger
space $\Dist(\mathcal P_{\min})$, which also contains distributions
like $\mu$ that correlate the actions at different states.

Thus
$\widehat x\in\Dist(\mathcal P_{\min})$ and
$\widehat y\in\Dist(\mathcal P_{\max})$.
These are product distributions
across controlled states; by contrast, an arbitrary element of
$\Dist(\mathcal P_i)$ may correlate the actions prescribed at different
states.

For the remainder of this subsection, let $\mathsf A$ and $\mathsf B$ be
random variables taking values in $\mathcal P_{\min}$ and $\mathcal P_{\max}$
respectively, with joint law $\widehat x\times\widehat y$ (i.e.\ $\mathsf A$
has law $\widehat x$, $\mathsf B$ has law $\widehat y$, and $\mathsf A,\mathsf B$
are independent), and let $\boldsymbol\xi$ be a random variable taking values
in $\TransReal$ such that, conditional on $(\mathsf A,\mathsf B)=(a,b)$,
$\boldsymbol\xi$ has law $P_{a,b}$. Indeed, by
the chain rule of probability,
\[
\Pr[\mathsf A=a,\mathsf B=b,\boldsymbol\xi=\xi]
=\Pr[\mathsf A=a,\mathsf B=b]\cdot\Pr[\boldsymbol\xi=\xi\mid \mathsf A=a,\mathsf B=b].
\]
The first factor is $\widehat x(a)\widehat y(b)$, since $(\mathsf A,\mathsf B)\sim\widehat x\times\widehat y$; the second factor is $P_{a,b}(\xi)$, by definition of ``$\boldsymbol\xi$ has law $P_{a,b}$ conditional on $(\mathsf A,\mathsf B)=(a,b)$.''  Hence the joint law is
\begin{equation}
\label{eq:joint-one-step-law}
\Pr[\mathsf A=a,\mathsf B=b,\boldsymbol\xi=\xi]
=
\widehat x(a)\widehat y(b)P_{a,b}(\xi).
\end{equation}

All expectations below are finite sums under this law.
\begin{lemma}[Mean successor under product profile sampling]
\label{lem:mean-profile-successor}
For every $\distr\in\DistS$, $x\in\mathcal X$, and $y\in\mathcal Y$,
\[
\DistOp_{x,y}(\distr)
=
\mathbb E\bigl[\Phi_{\mathsf A,\mathsf B}(\distr)\bigr].
\]
The equality is coordinatewise in $\mathbb R^{\States}$.
\end{lemma}
\begin{proof}
Under $\widehat x$, the marginal law of $\mathsf A(s)$ is $x(s)$. Indeed, summing the product measure $\widehat x$ over every profile $a$ agreeing with $\alpha$ at $s$,
\[
\Pr[\mathsf A(s)=\alpha]
=\sum_{\substack{a\in\mathcal P_{\min}\\ a(s)=\alpha}}\widehat x(a)
=\sum_{\substack{a\in\mathcal P_{\min}\\ a(s)=\alpha}}\ x(s)(a(s))\prod_{r\in\States_{\min}\setminus\{s\}}x(r)(a(r)).
\]
On this restricted sum $a(s)$ is pinned to $\alpha$, so the factor $x(s)(a(s))$ is the constant $x(s)(\alpha)$ and pulls out; what remains is a sum over every unconstrained choice of $a(r)$ at the other Min-states $r\neq s$, which factors coordinatewise exactly as in the proof of \eqref{eq:transition-realization-prob}:
\[
\Pr[\mathsf A(s)=\alpha]
=x(s)(\alpha)\prod_{r\in\States_{\min}\setminus\{s\}}\Big(\sum_{\beta\in\Acts(r)}x(r)(\beta)\Big)
=x(s)(\alpha),
\]
since each inner sum is $1$ because $x(r)\in\Dist(\Acts(r))$. The analogous marginal identity holds for $\mathsf B(s)$ and $y(s)$.
Crucially, $\rho_{\mathsf A,\mathsf B}(s)$, meaning the deterministic
function $\rho_{a,b}(s)$ evaluated at the random profile pair
$(a,b)=(\mathsf A,\mathsf B)$, depends only on the single coordinate of
the profile at $s$: by definition $\rho_{a,b}(s)=a(s)$ if
$s\in\States_{\min}$ and $\rho_{a,b}(s)=b(s)$ if $s\in\States_{\max}$, so
substituting $(a,b)=(\mathsf A,\mathsf B)$ gives
$\rho_{\mathsf A,\mathsf B}(s)=\mathsf A(s)$ if $s\in\States_{\min}$ and
$\rho_{\mathsf A,\mathsf B}(s)=\mathsf B(s)$ if $s\in\States_{\max}$;
hence $\Trans(s,\rho_{\mathsf A,\mathsf B}(s),s')$ is a function of
$\mathsf A(s)$ alone (respectively $\mathsf B(s)$ alone), not of the
full profile. Its expectation is therefore determined entirely by the
marginal law just computed.

 Explicitly, if $s\in\States_{\min}$,
\[
\mathbb E\bigl[\Trans(s,\rho_{\mathsf A,\mathsf B}(s),s')\bigr]
=\mathbb E\bigl[\Trans(s,\mathsf A(s),s')\bigr]
=\sum_{\alpha\in\Acts(s)}\Pr[\mathsf A(s)=\alpha]\,\Trans(s,\alpha,s')
=\sum_{\alpha\in\Acts(s)}x(s)(\alpha)\Trans(s,\alpha,s'),
\]
and symmetrically with $\mathsf B(s)$ and $y(s)$ if $s\in\States_{\max}$. Since $\pi_{x,y}(s)=x(s)$ for $s\in\States_{\min}$ and $\pi_{x,y}(s)=y(s)$ for $s\in\States_{\max}$, both cases read uniformly as
\[
\mathbb E\bigl[\Trans(s,\rho_{\mathsf A,\mathsf B}(s),s')\bigr]
=\sum_{\alpha\in\Acts(s)}\pi_{x,y}(s)(\alpha)\Trans(s,\alpha,s').
\]
 Hence,
using finiteness to exchange sums and expectations,
\begin{align*}
\mathbb E\bigl[\Phi_{\mathsf A,\mathsf B}(\distr)(s')\bigr]
&=
\sum_{s\in\States}\distr(s)
\mathbb E\bigl[
\Trans(s,\rho_{\mathsf A,\mathsf B}(s),s')
\bigr]
\\
&=
\sum_{s\in\States}\distr(s)
\sum_{\alpha\in\Acts(s)}
\pi_{x,y}(s)(\alpha)\Trans(s,\alpha,s')
\\
&=
\DistOp_{x,y}(\distr)(s'),
\end{align*}
where the last equality is \eqref{eq:distribution-update}. This holds for
every coordinate $s'$.
\end{proof}
\begin{lemma}[Averaged distribution versus sampled profile]
\label{lem:av-vs-sp}
For every $\distr\in\DistS$, $x\in\mathcal X$, and $y\in\mathcal Y$,
\[
\Ent\bigl(\DistOp_{x,y}(\distr)\bigr)
\geq
\mathbb E\bigl[\Ent(\Phi_{\mathsf A,\mathsf B}(\distr))\bigr].
\]
\end{lemma}
\begin{proof}
By \cref{lem:mean-profile-successor}, $\DistOp_{x,y}(\distr)=\mathbb E[\Phi_{\mathsf A,\mathsf B}(\distr)]$, so the left-hand side is $\Ent$ evaluated at the mean of the $\DistS$-valued random variable $\Phi_{\mathsf A,\mathsf B}(\distr)$.
Since $\mathcal P_{\min}$ and $\mathcal P_{\max}$ are finite and $\mathsf A,\mathsf B$ are independent with laws $\widehat x,\widehat y$, the random variable $\Phi_{\mathsf A,\mathsf B}(\distr)$ takes only the finitely many values $\Phi_{a,b}(\distr)$, $(a,b)\in\mathcal P_{\min}\times\mathcal P_{\max}$, with $\Pr[\mathsf A=a,\mathsf B=b]=\widehat x(a)\widehat y(b)$. Hence its mean is the explicit finite convex combination
\[
\DistOp_{x,y}(\distr)
=\mathbb E[\Phi_{\mathsf A,\mathsf B}(\distr)]
=\sum_{(a,b)\in\mathcal P_{\min}\times\mathcal P_{\max}}\widehat x(a)\widehat y(b)\,\Phi_{a,b}(\distr),
\]
a weighted average, with nonnegative weights summing to $1$, of the points $\Phi_{a,b}(\distr)\in\DistS$.
By \cref{prop:entropy-props}(ii), $\Ent$ is concave on $\DistS$; concavity for a two-point combination extends by induction to any finite convex combination, giving
\[
\Ent\Big(\sum_i p_i z_i\Big)\ \ge\ \sum_i p_i\,\Ent(z_i)
\]
for any weights $p_i\ge0$ with $\sum_i p_i=1$ and points $z_i\in\DistS$. Applying this with $p_{(a,b)}=\widehat x(a)\widehat y(b)$ and $z_{(a,b)}=\Phi_{a,b}(\distr)$ yields
\[
\Ent\bigl(\DistOp_{x,y}(\distr)\bigr)
=\Ent\Big(\sum_{(a,b)}\widehat x(a)\widehat y(b)\,\Phi_{a,b}(\distr)\Big)
\ge
\sum_{(a,b)}\widehat x(a)\widehat y(b)\,\Ent\bigl(\Phi_{a,b}(\distr)\bigr)
=\mathbb E\bigl[\Ent(\Phi_{\mathsf A,\mathsf B}(\distr))\bigr],
\]
the last equality again being the definition of expectation of the finite-valued real random variable $\Ent(\Phi_{\mathsf A,\mathsf B}(\distr))$. This is the claimed inequality.
\end{proof}

\begin{lemma}[Connecting sampled profile and sampled transition]
For every fixed pure-profile pair $(a,b)$, 
\begin{equation}
\label{eq:psi-phi-expectation}
\mathbb E_{\xi\sim P_{a,b}}
\bigl[\Psi_\xi(\distr)\bigr]
=
\Phi_{a,b}(\distr).
\end{equation}
\end{lemma}
\begin{proof}
Indeed, by definition the expectation of the finitely-valued $\DistS$-valued
random variable $\Psi_\xi(\distr)$ is the probability-weighted sum over its
possible values,
\[
\mathbb E_{\xi\sim P_{a,b}}\bigl[\Psi_\xi(\distr)\bigr]
=\sum_{\xi\in\TransReal}P_{a,b}(\xi)\,\Psi_\xi(\distr),
\]
a vector equation in $\mathbb R^{\States}$. Taking the $s'$ coordinate of both
sides and substituting \eqref{eq:psi-def},
\[
\mathbb E_{\xi\sim P_{a,b}}\bigl[\Psi_\xi(\distr)\bigr](s')
=\sum_{\xi\in\TransReal}P_{a,b}(\xi)\sum_{s:\,\xi(s)=s'}\distr(s)
=\sum_{s\in\States}\distr(s)\sum_{\substack{\xi\in\TransReal\\ \xi(s)=s'}}P_{a,b}(\xi)
=\sum_{s\in\States}\distr(s)\,\Pr_{\xi\sim P_{a,b}}[\xi(s)=s'],
\]
where the sums over $\xi$ and $s$ were exchanged (both finite), and the
inner sum was recognized as $\Pr_{\xi\sim P_{a,b}}[\xi(s)=s']$ by definition
of probability.
To evaluate $\Pr_{\xi\sim P_{a,b}}[\xi(s)=s']$, marginalize the product measure $P_{a,b}$ over every coordinate but $s$:
\[
\Pr_{\xi\sim P_{a,b}}[\xi(s)=s']
=\sum_{\substack{\xi\in\TransReal\\ \xi(s)=s'}}P_{a,b}(\xi)
=\sum_{\substack{\xi\in\TransReal\\ \xi(s)=s'}}\ \Trans(s,\rho_{a,b}(s),\xi(s))\prod_{s''\neq s}\Trans(s'',\rho_{a,b}(s''),\xi(s'')).
\]
On this restricted sum $\xi(s)$ is pinned to $s'$, so the factor $\Trans(s,\rho_{a,b}(s),\xi(s))$ is the constant $\Trans(s,\rho_{a,b}(s),s')$ and pulls out; what remains is a sum over every unconstrained choice of $\xi(s'')$ at the other states $s''\neq s$, which again factors as in the proof of \eqref{eq:transition-realization-prob}:
\[
\Pr_{\xi\sim P_{a,b}}[\xi(s)=s']
=\Trans(s,\rho_{a,b}(s),s')\cdot
\prod_{s''\neq s}\Big(\sum_{s'''\in\States}\Trans(s'',\rho_{a,b}(s''),s''')\Big)
=\Trans(s,\rho_{a,b}(s),s')\cdot 1,
\]
since each inner sum equals $1$ because $\Trans(s'',\rho_{a,b}(s''),\cdot)\in\Dist(\States)$. Hence $\xi(s)$ has marginal law $\Trans(s,\rho_{a,b}(s),\cdot)$, i.e.\ $\Pr_{\xi\sim P_{a,b}}[\xi(s)=s']=\Trans(s,\rho_{a,b}(s),s')$, and so the $s'$ coordinate of the left-hand side is
\[
\sum_{s\in\States}\distr(s)
\Pr_{\xi\sim P_{a,b}}[\xi(s)=s']
=
\sum_{s\in\States}\distr(s)
\Trans(s,\rho_{a,b}(s),s'),
\]
which is exactly the $s'$ coordinate of $\Phi_{a,b}(\distr)$ by \eqref{eq:phi-def}.
Thus sampled-profile semantics uses the conditional mean of the
sampled-transition successor, whereas sampled-transition semantics retains the
individual realized successor before the nonlinear continuation value is
evaluated. See also \cref{fig:three-semantics-branching}(c).
\end{proof}

\begin{lemma}[Sampled profile versus sampled transition]
\label{lem:sp-vs-st}
For every $\distr\in\DistS$ and fixed
$(a,b)\in\mathcal P_{\min}\times\mathcal P_{\max}$,
\begin{equation}
\label{eq:sp-st-fixed-profile}
\Ent(\Phi_{a,b}(\distr))
\geq
\mathbb E_{\xi\sim P_{a,b}}
\bigl[\Ent(\Psi_\xi(\distr))\bigr].
\end{equation}
Consequently,
\[
\mathbb E\bigl[\Ent(\Phi_{\mathsf A,\mathsf B}(\distr))\bigr]
\geq
\mathbb E\bigl[\Ent(\Psi_{\boldsymbol\xi}(\distr))\bigr].
\]
\end{lemma}
\begin{proof}
Fix $(a,b)$. By \eqref{eq:psi-phi-expectation}, $\Phi_{a,b}(\distr)=\mathbb E_{\xi\sim P_{a,b}}[\Psi_\xi(\distr)]$, so the left-hand side of \eqref{eq:sp-st-fixed-profile} is $\Ent$ evaluated at the mean of the $\DistS$-valued random variable $\Psi_\xi(\distr)$, $\xi\sim P_{a,b}$.
Since $\TransReal$ is finite, this mean is the explicit finite convex combination
\[
\Phi_{a,b}(\distr)
=\mathbb E_{\xi\sim P_{a,b}}\bigl[\Psi_\xi(\distr)\bigr]
=\sum_{\xi\in\TransReal}P_{a,b}(\xi)\,\Psi_\xi(\distr),
\]
a weighted average, with nonnegative weights summing to $1$, of the points $\Psi_\xi(\distr)\in\DistS$. By \cref{prop:entropy-props}(ii) and the same finite-concavity inequality used in \cref{lem:av-vs-sp},
\[
\Ent\bigl(\Phi_{a,b}(\distr)\bigr)
=\Ent\Big(\sum_{\xi\in\TransReal}P_{a,b}(\xi)\,\Psi_\xi(\distr)\Big)
\ge
\sum_{\xi\in\TransReal}P_{a,b}(\xi)\,\Ent\bigl(\Psi_\xi(\distr)\bigr)
=\mathbb E_{\xi\sim P_{a,b}}\bigl[\Ent(\Psi_\xi(\distr))\bigr],
\]
which is \eqref{eq:sp-st-fixed-profile}.
For the consequence, multiply \eqref{eq:sp-st-fixed-profile} by $\widehat x(a)\widehat y(b)\ge0$ and sum over every $(a,b)\in\mathcal P_{\min}\times\mathcal P_{\max}$:
\[
\sum_{(a,b)}\widehat x(a)\widehat y(b)\,\Ent\bigl(\Phi_{a,b}(\distr)\bigr)
\ \ge\
\sum_{(a,b)}\widehat x(a)\widehat y(b)\sum_{\xi\in\TransReal}P_{a,b}(\xi)\,\Ent\bigl(\Psi_\xi(\distr)\bigr).
\]
The left-hand side is exactly $\mathbb E\bigl[\Ent(\Phi_{\mathsf A,\mathsf B}(\distr))\bigr]$, by the same definition of expectation of a finite-valued real random variable used in \cref{lem:av-vs-sp}.
For the right-hand side, distribute the outer sum and swap the (finite) order of summation over $(a,b)$ and $\xi$:
\[
\sum_{(a,b)}\sum_{\xi\in\TransReal}\bigl[\widehat x(a)\widehat y(b)\,P_{a,b}(\xi)\bigr]\Ent\bigl(\Psi_\xi(\distr)\bigr)
=
\sum_{\xi\in\TransReal}\Big(\sum_{(a,b)}\widehat x(a)\widehat y(b)\,P_{a,b}(\xi)\Big)\Ent\bigl(\Psi_\xi(\distr)\bigr).
\]
By the joint law \eqref{eq:joint-one-step-law}, $\Pr[\mathsf A=a,\mathsf B=b,\boldsymbol\xi=\xi]=\widehat x(a)\widehat y(b)\,P_{a,b}(\xi)$ (first the profile is drawn, then the transition is drawn conditionally on it), so summing this joint probability over all $(a,b)$ for fixed $\xi$ gives exactly the marginal law of the fully randomized transition $\boldsymbol\xi$:
\[
\sum_{(a,b)}\widehat x(a)\widehat y(b)\,P_{a,b}(\xi)=\Pr[\boldsymbol\xi=\xi].
\]
Substituting, the right-hand side becomes
\[
\sum_{\xi\in\TransReal}\Pr[\boldsymbol\xi=\xi]\,\Ent\bigl(\Psi_\xi(\distr)\bigr)
=\mathbb E\bigl[\Ent(\Psi_{\boldsymbol\xi}(\distr))\bigr],
\]
again by definition of expectation. Combining the two displays gives
\[
\mathbb E\bigl[\Ent(\Phi_{\mathsf A,\mathsf B}(\distr))\bigr]
\ge
\mathbb E\bigl[\Ent(\Psi_{\boldsymbol\xi}(\distr))\bigr],
\]
as claimed. 
Equivalently, this whole computation is an instance of the tower property
(law of total expectation), applied to the real-valued random variable
$Z\defeq\Ent(\Psi_{\boldsymbol\xi}(\distr))$:
\[
\mathbb E[Z]=\mathbb E\bigl[\mathbb E[Z\mid \mathsf A,\mathsf B]\bigr].
\]
Here the inner conditional expectation, evaluated at $(\mathsf A,\mathsf B)=(a,b)$, is exactly $\mathbb E_{\xi\sim P_{a,b}}[\Ent(\Psi_\xi(\distr))]$, since $\boldsymbol\xi$ has law $P_{a,b}$ conditional on that event; averaging this conditional expectation against the law $\widehat x\times\widehat y$ of $(\mathsf A,\mathsf B)$ recovers the unconditional expectation $\mathbb E[Z]=\mathbb E[\Ent(\Psi_{\boldsymbol\xi}(\distr))]$, which is exactly the computation carried out above by multiplying by $\widehat x(a)\widehat y(b)$, summing, and recognizing the result as the marginal law of $\boldsymbol\xi$.
\end{proof}
\begin{lemma}[Averaged distribution as the mean realized successor]
\label{lem:av-vs-st-direct}
For every $\distr\in\DistS$, $x\in\mathcal X$, and $y\in\mathcal Y$,
\begin{equation}
\label{eq:av-realized-mean}
\DistOp_{x,y}(\distr)
=
\mathbb E\bigl[\Psi_{\boldsymbol\xi}(\distr)\bigr].
\end{equation}
Consequently,
\[
\Ent\bigl(\DistOp_{x,y}(\distr)\bigr)
\geq
\mathbb E\bigl[\Ent(\Psi_{\boldsymbol\xi}(\distr))\bigr].
\]
\end{lemma}
\begin{proof}
We first establish \eqref{eq:av-realized-mean} by applying the tower property to the $\DistS$-valued random variable $\Psi_{\boldsymbol\xi}(\distr)$, coordinatewise: for each $s'\in\States$, $\Psi_{\boldsymbol\xi}(\distr)(s')$ is a real-valued random variable, so the ordinary (scalar) tower property gives
\[
\mathbb E\bigl[\Psi_{\boldsymbol\xi}(\distr)(s')\bigr]
=
\mathbb E\Bigl[\,\mathbb E\bigl[\Psi_{\boldsymbol\xi}(\distr)(s')\mid\mathsf A,\mathsf B\bigr]\Bigr],
\]
and since this holds for every coordinate $s'$, the corresponding vector identity holds:
\[
\mathbb E\bigl[\Psi_{\boldsymbol\xi}(\distr)\bigr]
=
\mathbb E\!\left[
\mathbb E\bigl[\Psi_{\boldsymbol\xi}(\distr)
\mid\mathsf A,\mathsf B\bigr]
\right].
\]
Evaluate the inner conditional expectation at $(\mathsf A,\mathsf B)=(a,b)$: since $\boldsymbol\xi$ has law $P_{a,b}$ conditional on this event,
\[
\mathbb E\bigl[\Psi_{\boldsymbol\xi}(\distr)\mid\mathsf A=a,\mathsf B=b\bigr]
=
\mathbb E_{\xi\sim P_{a,b}}\bigl[\Psi_\xi(\distr)\bigr]
=
\Phi_{a,b}(\distr),
\]
the last equality being \eqref{eq:psi-phi-expectation}. Thus the inner conditional expectation, as a function of $(\mathsf A,\mathsf B)$, is exactly the random variable $\Phi_{\mathsf A,\mathsf B}(\distr)$, and so
\[
\mathbb E\bigl[\Psi_{\boldsymbol\xi}(\distr)\bigr]
=
\mathbb E\bigl[\Phi_{\mathsf A,\mathsf B}(\distr)\bigr]
=
\DistOp_{x,y}(\distr),
\]
the last equality being \cref{lem:mean-profile-successor}. This proves \eqref{eq:av-realized-mean}.
For the entropy inequality, note $\TransReal$ is finite, so $\Psi_{\boldsymbol\xi}(\distr)$ takes only finitely many values, and its mean is the explicit finite convex combination
\[
\mathbb E\bigl[\Psi_{\boldsymbol\xi}(\distr)\bigr]
=
\sum_{\xi\in\TransReal}\Pr[\boldsymbol\xi=\xi]\,\Psi_\xi(\distr),
\]
a weighted average, with nonnegative weights summing to $1$, of the points $\Psi_\xi(\distr)\in\DistS$. By \cref{prop:entropy-props}(ii) and the same finite-concavity inequality used in \cref{lem:av-vs-sp,lem:sp-vs-st},
\[
\Ent\Bigl(\mathbb E\bigl[\Psi_{\boldsymbol\xi}(\distr)\bigr]\Bigr)
=
\Ent\Bigl(\sum_{\xi\in\TransReal}\Pr[\boldsymbol\xi=\xi]\,\Psi_\xi(\distr)\Bigr)
\ge
\sum_{\xi\in\TransReal}\Pr[\boldsymbol\xi=\xi]\,\Ent\bigl(\Psi_\xi(\distr)\bigr)
=
\mathbb E\bigl[\Ent(\Psi_{\boldsymbol\xi}(\distr))\bigr].
\]
Combining this with \eqref{eq:av-realized-mean} gives
\[
\Ent\bigl(\DistOp_{x,y}(\distr)\bigr)
=
\Ent\Bigl(\mathbb E\bigl[\Psi_{\boldsymbol\xi}(\distr)\bigr]\Bigr)
\ge
\mathbb E\bigl[\Ent(\Psi_{\boldsymbol\xi}(\distr))\bigr],
\]
as claimed.
\end{proof}
\begin{corollary}[One-step entropy ordering]
\label{cor:entropy-ordering}
For every $\distr\in\DistS$, $x\in\mathcal X$, and $y\in\mathcal Y$, under
the joint law \eqref{eq:joint-one-step-law},
\[
\Ent\bigl(\DistOp_{x,y}(\distr)\bigr)
\geq
\mathbb E\bigl[\Ent(\Phi_{\mathsf A,\mathsf B}(\distr))\bigr]
\geq
\mathbb E\bigl[\Ent(\Psi_{\boldsymbol\xi}(\distr))\bigr].
\]
\end{corollary}
\begin{proof}
Combine \cref{lem:av-vs-sp,lem:sp-vs-st}. The direct comparison of the first
and third terms is also \cref{lem:av-vs-st-direct}.
\end{proof}
\begin{proposition}[Compactness and continuity of the distribution dynamics]
\label{prop:compact}
The simplex $\DistS$ is compact and convex. Moreover:
\begin{enumerate}[(i)]
\item
for every statewise mixed-action pair $(x,y)$,
$\distr\mapsto\DistOp_{x,y}(\distr)$ is affine and continuous, and
$(\distr,x,y)\mapsto\DistOp_{x,y}(\distr)$ is jointly continuous;
\item
for every pure-profile pair $(a,b)$,
$\Phi_{a,b}:\DistS\to\DistS$ is affine and continuous;
\item
for every transition realization $\xi$,
$\Psi_\xi:\DistS\to\DistS$ is affine and continuous.
\end{enumerate}
\end{proposition}
\begin{proof}
$\DistS=\{\distr\in\mathbb R^{\States}:\distr(s)\geq0\text{ for all }s\in\States,\ \sum_{s\in\States}\distr(s)=1\}$ is the intersection of finitely many closed half-spaces (one for each nonnegativity constraint $\distr(s)\geq0$) with a single hyperplane (the constraint $\sum_s\distr(s)=1$); an intersection of convex sets is convex, so $\DistS$ is convex. It is also closed (an intersection of closed sets) and bounded (each coordinate lies in $[0,1]$), hence compact in the finite-dimensional space $\mathbb R^{\States}$ by Heine--Borel.
(i) Fix $(x,y)$. By \eqref{eq:distribution-update}, the $s'$ coordinate of $\DistOp_{x,y}(\distr)$ is
\[
\DistOp_{x,y}(\distr)(s')=\sum_{s\in\States}\distr(s)\underbrace{\sum_{\alpha\in\Acts(s)}\pi_{x,y}(s)(\alpha)\Trans(s,\alpha,s')}_{\text{constant in }\distr,\text{ call it }c_s(x,y,s')}.
\]
For fixed $(x,y)$, each coefficient $c_s(x,y,s')$ does not depend on $\distr$, so $\DistOp_{x,y}(\distr)(s')=\sum_s c_s(x,y,s')\,\distr(s)$ is a linear (hence affine) function of $\distr$; this holds for every coordinate $s'$, so $\distr\mapsto\DistOp_{x,y}(\distr)$ is linear, and every linear map between finite-dimensional real vector spaces is continuous. For joint continuity in $(\distr,x,y)$: expanding $c_s(x,y,s')=\sum_\alpha\pi_{x,y}(s)(\alpha)\Trans(s,\alpha,s')$, each term $\distr(s)\,\pi_{x,y}(s)(\alpha)\,\Trans(s,\alpha,s')$ is a product of the coordinate $\distr(s)$, the coordinate $\pi_{x,y}(s)(\alpha)$ (equal to $x(s)(\alpha)$ or $y(s)(\alpha)$, a coordinate of $x$ or $y$), and the constant $\Trans(s,\alpha,s')$; a finite sum of such products is a polynomial (degree $2$) in the finitely many real coordinates of $(\distr,x,y)$, and every polynomial in finitely many real variables is continuous, since addition and multiplication are continuous on $\mathbb R$. Hence each coordinate of $(\distr,x,y)\mapsto\DistOp_{x,y}(\distr)$ is jointly continuous, and therefore so is the map into $\DistS\subset\mathbb R^{\States}$.\\
(ii) Fix $(a,b)$. By \eqref{eq:phi-def}, the $s'$ coordinate of $\Phi_{a,b}(\distr)$ is
\[
\Phi_{a,b}(\distr)(s')=\sum_{s\in\States}\distr(s)\,\Trans(s,\rho_{a,b}(s),s'),
\]
where the coefficient $\Trans(s,\rho_{a,b}(s),s')$ is a fixed constant (it depends only on the fixed $(a,b)$ and the arena, not on $\distr$). So, as in (i), each coordinate is a linear function of $\distr$, hence $\Phi_{a,b}$ is linear (affine) and, being linear on a finite-dimensional space, continuous.\\
(iii) Fix $\xi$. By \eqref{eq:psi-def}, the $s'$ coordinate of $\Psi_\xi(\distr)$ is
\[
\Psi_\xi(\distr)(s')=\sum_{s:\,\xi(s)=s'}\distr(s)=\sum_{s\in\States}\mathbf 1_{\{\xi(s)=s'\}}\,\distr(s),
\]
using the indicator notation of \eqref{eq:path-kernel-av}, here applied
to the condition $\xi(s)=s'$ rather than to a named set: $\mathbf
1_{\{\xi(s)=s'\}}=1$ if $\xi(s)=s'$ and $0$ otherwise. Each coefficient
$\mathbf 1_{\{\xi(s)=s'\}}\in\{0,1\}$ is fixed once $\xi$ is fixed and
does not depend on $\distr$. Again each coordinate is a linear function
of $\distr$, so $\Psi_\xi$ is linear (affine) and continuous by the same
finite-dimensional argument.
\end{proof}

\section{Bellman characterisations}
\label{app:bellman-proof}

\bellmancharacterisations*

\begin{proof}
Write $K\defeq\DistS$ and $L\defeq\log_2|\States|$.
Since $K$ is compact, $C(K)$ equipped with the supremum norm is a
Banach space. We establish continuity, contraction, the
finite-horizon recursions, and finally the infinite-horizon
identification.

\medskip

\noindent\textbf{Continuity and attainment.}
Fix $f\in C(K)$. Since $\DistS,\mathcal X,\mathcal Y$ are each compact
(\cref{prop:compact} for $\DistS$; $\mathcal X,\mathcal Y$ are finite
products of simplices) and $(\distr,x,y)\longmapsto f(\DistOp_{x,y}(\distr))$
is continuous (\cref{prop:compact} again, composed with $f$), this map
is uniformly continuous on the compact set $K\times\mathcal X\times\mathcal Y$.

Fix $\eta>0$. Uniform continuity gives $\rho>0$ such that
$d(\distr,\distr')<\rho$ implies
$|f(\DistOp_{x,y}(\distr))-f(\DistOp_{x,y}(\distr'))|<\eta$ for every
$x\in\mathcal X,y\in\mathcal Y$ simultaneously. For such $\distr,\distr'$,
writing $g(x,y)\defeq f(\DistOp_{x,y}(\distr))$ and
$g'(x,y)\defeq f(\DistOp_{x,y}(\distr'))$, $g(x,y)\le g'(x,y)+\eta$ for
every $x,y$; taking $\max_y$ of both sides and then $\min_x$ gives
$\min_x\max_yg(x,y)\le\min_x\max_yg'(x,y)+\eta$, and the symmetric
argument with $g,g'$ exchanged gives the reverse inequality, so
\[
\Bigl|\min_{x\in\mathcal X}\max_{y\in\mathcal Y}f(\DistOp_{x,y}(\distr))
-\min_{x\in\mathcal X}\max_{y\in\mathcal Y}f(\DistOp_{x,y}(\distr'))\Bigr|
\le\eta,
\]
and the identical argument with $\min$ and $\max$ exchanged bounds the
other ordered extremum by $\eta$ as well. Since $\eta$ was arbitrary,
both $\distr\mapsto\min_{x}\max_{y}f(\DistOp_{x,y}(\distr))$ and
$\distr\mapsto\max_{y}\min_{x}f(\DistOp_{x,y}(\distr))$ are
(uniformly) continuous. The same bound, applied with $\distr$ fixed and
only $x$ (or only $y$) varying, shows
$x\mapsto\max_{y\in\mathcal Y}f(\DistOp_{x,y}(\distr))$ is continuous
on the compact set $\mathcal X$ for fixed $\distr$, so by the extreme
value theorem its minimum over $\mathcal X$ is attained; the dual
argument attains the maximum, then the minimum, for the other ordered
extremum. Since $\Ent$ is continuous, both averaged operators preserve
$C(K)$.

For either sampled semantics, every entry of
$M_f^\Sem(\distr)$ is continuous in $\distr$, since
$\Phi_{a,b}$ and $\Psi_\xi$ are continuous and the sum over
$\xi\in\TransReal$ is finite. For finite matrices $M,N$ of the
same dimensions,
\begin{equation}
\bigl|\operatorname{val}(M)-\operatorname{val}(N)\bigr|
\le
\max_{a,b}|M_{a,b}-N_{a,b}|.
\label{eq:bellman-proof-matrix-stability}
\end{equation}
Indeed, if the right-hand side is $\delta$, every mixed-strategy
payoff changes by at most $\delta$, and taking the minimax value
preserves this bound. Hence $\operatorname{val}$ is continuous in
the matrix entries, and $\mathcal T_\Sem$ also preserves $C(K)$.

We also require measurable optimal choices. 
The measurable maximum
theorem~\cite[Theorem~18.19]{aliprantis2006infinite} gives Borel selectors for the outer extrema of the averaged
operators. For the sampled operators, applying the same theorem to
\[
(\distr,\sigma)\longmapsto
\max_{\tau\in\Dist(\mathcal P_{\max})}
\sigma^{\mathsf T}M_f^\Sem(\distr)\tau
\]
and
\[
(\distr,\tau)\longmapsto
\min_{\sigma\in\Dist(\mathcal P_{\min})}
\sigma^{\mathsf T}M_f^\Sem(\distr)\tau
\]
gives Borel selectors of optimal mixed rows and columns.

\medskip
\noindent\textbf{Contraction.}
Let $f,g\in C(K)$ and put $\delta=\|f-g\|_\infty$.
For every $\distr,x,y$,
\[
\bigl|
f(\DistOp_{x,y}(\distr))
-g(\DistOp_{x,y}(\distr))
\bigr|\le\delta.
\]
Minimisation and maximisation preserve a uniform difference bound,
and the current entropy terms cancel. Hence
\[
\|\mathcal T_{\mathsf{av}}^\pm f
-\mathcal T_{\mathsf{av}}^\pm g\|_\infty
\le\Discount\delta.
\]

For sampled-profile semantics, corresponding matrix entries differ
by at most $\delta$. Under sampled-transition semantics,
\[
\begin{aligned}
\bigl|M_f^{\mathsf{st}}(\distr)_{a,b}
-M_g^{\mathsf{st}}(\distr)_{a,b}\bigr|
&\le
\sum_{\xi\in\TransReal}P_{a,b}(\xi)
\bigl|f(\Psi_\xi(\distr))-g(\Psi_\xi(\distr))\bigr|\\
&\le\delta.
\end{aligned}
\]
Together with \eqref{eq:bellman-proof-matrix-stability}, this gives
\[
\|\mathcal T_\Sem f-\mathcal T_\Sem g\|_\infty
\le\Discount\delta.
\]
Thus every operator is a $\Discount$-contraction. Since
$0<\Discount<1$, Banach's fixed-point theorem gives each operator a
unique fixed point, and its iterates from zero converge uniformly
to that fixed point.

\medskip
\noindent\textbf{Finite-horizon identification.}
For $t\ge0$, let
\[
\mathsf{Pay}_{\Discount,t}
\defeq
(1-\Discount)\sum_{j=0}^{t-1}
\Discount^j\Ent(\distr_j),
\]
with the empty sum equal to zero, and let
$V_{\Sem,t}^+$ and $V_{\Sem,t}^-$ be its upper and lower values
over the same history-dependent strategy spaces as in the
infinite-horizon game.

For averaged-distribution semantics, backward induction gives
\begin{align}
V_{\mathsf{av},0}^+
&=V_{\mathsf{av},0}^-=0,\\
V_{\mathsf{av},t+1}^+(\distr)
&=
(1-\Discount)\Ent(\distr)
+\Discount\min_x\max_y
V_{\mathsf{av},t}^+(\DistOp_{x,y}(\distr)),
\label{eq:bellman-proof-av-plus}\\
V_{\mathsf{av},t+1}^-(\distr)
&=
(1-\Discount)\Ent(\distr)
+\Discount\max_y\min_x
V_{\mathsf{av},t}^-(\DistOp_{x,y}(\distr)).
\label{eq:bellman-proof-av-minus}
\end{align}
We recall the argument for completeness. For the upper value,
a Borel minimiser of the right-hand side of
\eqref{eq:bellman-proof-av-plus}, followed after the successor
distribution by an optimal $t$-stage policy, gives the ``$\le$''
inequality. Conversely, fix any $(t+1)$-stage strategy of
\MinPl{} and let $x$ be its first action. Against any first-round
response $y$, with
$\nu=\DistOp_{x,y}(\distr)$, its continuation has worst-case
$t$-stage payoff at least $V_{\mathsf{av},t}^+(\nu)$.
\MaxPl{} can therefore choose a continuation arbitrarily close to
this worst case. Maximising over $y$ and then taking the infimum
over \MinPl{} gives the reverse inequality.

The lower recursion is dual. A Borel maximiser of the right-hand
side of \eqref{eq:bellman-proof-av-minus}, followed by an optimal
$t$-stage policy, gives the ``$\ge$'' inequality. Conversely,
fix a $(t+1)$-stage strategy of \MaxPl{} and let $y$ be its first
action. For each response $x$, with
$\nu=\DistOp_{x,y}(\distr)$, \MinPl{} can choose a continuation
whose payoff against the fixed continuation of \MaxPl{} is
arbitrarily close from above to
$V_{\mathsf{av},t}^-(\nu)$, since
\[
\inf_{\pi_{\min}}
\mathsf{Obj}_{\Discount,t}
(\pi_{\min},\pi_{\max};\nu)
\le V_{\mathsf{av},t}^-(\nu)
\]
for every fixed $\pi_{\max}$.
Minimising over $x$ and then taking the supremum over
\MaxPl{} gives \eqref{eq:bellman-proof-av-minus}.
The measurable selectors above supply the required Markov policies.
Thus
\[
V_{\mathsf{av},t}^+
=(\mathcal T_{\mathsf{av}}^+)^t0,
\qquad
V_{\mathsf{av},t}^-
=(\mathcal T_{\mathsf{av}}^-)^t0.
\]

Now fix $\Sem\in\{\mathsf{sp},\mathsf{st}\}$.
We prove inductively that
\[
V_{\Sem,t}^-=V_{\Sem,t}^+=W_t,
\qquad
W_0=0,\qquad
W_{t+1}=\mathcal T_\Sem W_t.
\]
For $t+1$ stages, committing in the first round to a pure-profile
pair $(a,b)$ gives continuation value
\[
W_t(\Phi_{a,b}(\distr))
\]
under sampled-profile semantics, and
\[
\sum_{\xi\in\TransReal}
P_{a,b}(\xi)W_t(\Psi_\xi(\distr))
\]
under sampled-transition semantics. These are precisely the entries
of $M_{W_t}^\Sem(\distr)$.

The first-round interaction is therefore the finite zero-sum matrix
game $M_{W_t}^\Sem(\distr)$. Finite minimax gives a common value
and optimal mixed rows and columns; the selectors above make these
choices Borel in $\distr$. After the successor distribution is
observed, each player follows its inductively supplied optimal
$t$-stage policy. Hence, against arbitrary history-dependent
opponents,
\[
V_{\Sem,t+1}^+(\distr)
\le(\mathcal T_\Sem W_t)(\distr)
\le V_{\Sem,t+1}^-(\distr).
\]
Together with the minimax inequality
$V_{\Sem,t+1}^-\le V_{\Sem,t+1}^+$, this proves equality and
completes the induction.

\medskip
\noindent\textbf{Infinite-horizon identification and error bound.}
Since $0\le\Ent(\distr_j)\le L$, every path satisfies
\[
0\le
\mathsf{Pay}_{\Discount}
-\mathsf{Pay}_{\Discount,t}
\le
(1-\Discount)\sum_{j=t}^{\infty}\Discount^jL
=\Discount^tL.
\]
The bound is uniform over initial distributions and strategy pairs.
Taking expectations and then either ordered optimisation gives
\[
0\le
V_\Sem^\pm(\distr)-V_{\Sem,t}^\pm(\distr)
\le\Discount^tL.
\]

For averaged semantics, the finite-horizon identities therefore
show that the iterates of $\mathcal T_{\mathsf{av}}^\pm$ converge
uniformly to $V_{\mathsf{av}}^\pm$. Hence these values are the
respective unique fixed points.

For sampled semantics, both finite-horizon values equal $W_t$, so
\[
W_t(\distr)
\le V_\Sem^-(\distr)
\le V_\Sem^+(\distr)
\le W_t(\distr)+\Discount^tL.
\]
Letting $t\to\infty$ proves $V_\Sem^-=V_\Sem^+$ and identifies the
common value with the unique fixed point of $\mathcal T_\Sem$.
The same inequalities give
\[
0\le V(\distr)-W_t(\distr)
\le\Discount^t\log_2|\States|,
\]
as claimed.
\end{proof}

\section{Determinacy Across the Three Semantics (with details)}
\label{sec:determinacy}
\subsection{Averaged-distribution semantics}
\label{app:det-av}
\avnondeterminacy*

For $p\in[0,1]$, write $(1-p,p)$ for the distribution assigning mass $1-p$ to
one outcome and mass $p$ to the other, and define the \emph{binary entropy
function} (proposition \ref{prop:entropy-props} (iv))
\[
h(p)\ \defeq\ \Ent(1-p,p)\ =\ -p\log_2 p-(1-p)\log_2(1-p),
\]
\begin{proof}
Consider the four-state game of \cref{fig:nondet-gadget}. Write
$e=s_{\min}$ and $a=s_{\max}$. Initially
$\distro(e)=\distro(a)=\tfrac12$, while $q_0$ and $q_1$ have no mass. At both
$e$ and $a$, action $i\in\{0,1\}$ moves deterministically to $q_i$; the two
$q$-states have only an absorbing self-loop. Let $x$ be the probability with
which \MinPl chooses action $1$ at $e$, and let $y$ be the corresponding
probability for \MaxPl at $a$.
Only $e$ and $a$ contribute to the first update. Consequently,
\[
\distr_1(q_1)
=\distro(e)x+\distro(a)y
=\frac{x+y}{2},
\qquad
\distr_1(q_0)=1-\frac{x+y}{2}.
\]
The two strategic probabilities occur in separate summands: no product $xy$
is formed. Since $q_0$ and $q_1$ are absorbing,
$\distr_t=\distr_1$ for all $t\geq1$. Moreover,
$\Ent(\distro)=1$, because $\distro$ is uniform on two states, and
$\Ent(\distr_1)=h((x+y)/2)$ since $\distr_1$ is supported on $\{q_0,q_1\}$
with mass $(x+y)/2$ on $q_1$. The entire discounted payoff is therefore
\[
\mathsf{Obj}^{\mathsf{av}}_{\Discount}(x,y)
=(1-\Discount)\Ent(\distro)+(1-\Discount)h\!\left(\frac{x+y}2\right)\sum_{t=1}^\infty\Discount^t
=(1-\Discount)+\Discount\,h\!\left(\frac{x+y}{2}\right).
\]

For calculating $V^+_{\mathsf{av}}(\distro)$, fix any $x\in[0,1]$. The response $y=1-x$ makes the
successor distribution uniform on $\{q_0,q_1\}$. Since $h\leq1$ everywhere,
\[
\sup_{y\in[0,1]}h\!\left(\frac{x+y}{2}\right)=1
\quad\text{for every }x,
\]
hence
\[
\sup_{y\in[0,1]}\mathsf{Obj}^{\mathsf{av}}_{\Discount}(x,y)
=(1-\Discount)+\Discount\cdot1=1
\quad\text{for every }x,
\]
and therefore $$V^+_{\mathsf{av}}(\distro)=\inf_{x\in[0,1]}1=1$$.

For calculating $V^-_{\mathsf{av}}(\distro)$ value, fix $y$. As $x$ ranges over $[0,1]$, the argument of
$h$ ranges over $[y/2,(1+y)/2]$. The map $x\mapsto h((x+y)/2)$ is concave
(a concave function composed with an affine map of $x$), and a concave
function on a compact interval lies on or above the chord joining its
endpoint values, so every interior point dominates the smaller of the two
endpoints; hence its minimum over $[0,1]$ occurs at $x=0$ or $x=1$, giving
\[
\inf_{x\in[0,1]}h\!\left(\frac{x+y}{2}\right) 
=\min\!\left\{h\!\left(\frac y2\right),
h\!\left(\frac{1+y}{2}\right)\right\}
=\min\!\left\{h\!\left(\frac y2\right),
h\!\left(\frac{1-y}{2}\right)\right\}
\]
after using symmetry $h(1-p)=h(p)$ to rewrite $h((1+y)/2)=h((1-y)/2)$.
Both $y/2$ and $(1-y)/2$ lie in $[0,\tfrac12]$, where $h$ is increasing, so
the smaller argument gives the smaller value:
\[
\inf_{x\in[0,1]}h\!\left(\frac{x+y}{2}\right)
=h\!\left(\frac{\min\{y,1-y\}}2\right).
\]
The quantity $\min\{y,1-y\}$ equals $y$ (increasing in $y$) for $y\le\tfrac12$
and $1-y$ (decreasing in $y$) for $y\ge\tfrac12$, so it is maximised at
$y=\tfrac12$, where it equals $\tfrac12$. Since $h$ is strictly increasing on
$[0,\tfrac12]\supseteq[0,\tfrac14]$ and $\min\{y,1-y\}/2$ ranges over
$[0,\tfrac14]$, composing with $h$ preserves this maximiser: the expression
above is maximised uniquely at $y=\tfrac12$, with value $h(1/4)$. Hence
\[
V^-_{\mathsf{av}}(\distro)
=\sup_{y\in[0,1]}\Bigl[(1-\Discount)+\Discount\inf_{x\in[0,1]}h\bigl(\tfrac{x+y}2\bigr)\Bigr]
=(1-\Discount)+\Discount\,h\!\left(\frac14\right).
\]
Finally,
\[
h\!\left(\frac14\right)
=2-\frac34\log_2 3
\approx0.8113,
\]
so
\[
V^+_{\mathsf{av}}(\distro)-V^-_{\mathsf{av}}(\distro)
= 1- \left((1-\gamma) +\gamma h\!\left(\frac14\right)\right)
\]
\[
V^+_{\mathsf{av}}(\distro)-V^-_{\mathsf{av}}(\distro)
= \Discount\!\left(1-h\!\left(\frac14\right)\right)
\approx0.1887\Discount>0.
\]
Because $q_0,q_1$ absorb all mass after round $0$, only the round-$0$
actions at $e$ and $a$ affect $\Ent(\distr_t)$ for any $t\ge1$; every Borel
strategy in $\StratMin^{\mathsf{av}}$ or $\StratMax^{\mathsf{av}}$ therefore
induces the same value of $\mathsf{Obj}^{\mathsf{av}}_\Discount$ as some
constant pair $(x,y)\in[0,1]^2$ specifying that single round-$0$ mixed
action. Consequently $V^+_{\mathsf{av}}(\distro)=\inf_\sigma\sup_\tau
\mathsf{Obj}^{\mathsf{av}}_\Discount(\sigma,\tau;\distro)$, the infimum
and supremum taken over the full Borel strategy spaces, equals
$\inf_{x\in[0,1]}\sup_{y\in[0,1]}\mathsf{Obj}^{\mathsf{av}}_\Discount(x,y)$,
the quantity computed above; likewise $V^-_{\mathsf{av}}(\distro)$
equals $\sup_{y\in[0,1]}\inf_{x\in[0,1]}\mathsf{Obj}^{\mathsf{av}}_\Discount(x,y)$.
Each of these is an equality between a full-strategy-space value and the
constant-pair value of the \emph{same} polarity ($V^+$ with $V^+$,
$V^-$ with $V^-$); it is not a claim that $V^+_{\mathsf{av}}=V^-_{\mathsf{av}}$,
which the computation above already shows is false. There is no history
to react to, so history-dependent strategies cannot close either gap.

\end{proof}
\begin{remark}[A variational reformulation]
\label{rem:variational}
Since $\Ent(\mu)=\min_q\sum_s\mu(s)\log_2\frac1{q(s)}$, with the minimum at $q=\mu$, one may try to replace distributions by states and let an extra minimiser choice $q$ play the dual variable. This does not give an equivalent game on $\States$. If $q$ is fixed together with $\MinPl$'s action, so that it cannot depend on $\MaxPl$'s choice in the same round, then in the game of \cref{fig:nondet-gadget} the upper value is $1$ and the lower value is $(1-\Discount)+\Discount h(1/4)$, so the values still differ. If $q$ may depend on the distribution it evaluates, which is determined by both players' choices in the round, the game is the one we study, on distributions.
\end{remark}

\subsection{Sampled-profile and sampled-transition semantics}
\label{app:det-sampled}
The two sampled semantics have different one-step dynamics, but they share the
feature that makes determinacy possible: once the current pure profiles have
been fixed, all remaining one-step randomness produces a single scalar
expected continuation payoff. This scalar can be placed in a finite payoff
matrix, with one row for each pure profile of \MinPl and one column for each
pure profile of \MaxPl.
To make this precise, fix
$\Sem\in\{\mathsf{sp},\mathsf{st}\}$ and let
$K_{\Sem}(\cdot\mid\distr,a,b)$ denote the law of the next distribution when
the current distribution is $\distr$ and the pure-profile pair $(a,b)$ has
been selected. 
\[
K_{\mathsf{sp}}(\cdot\mid\distr,a,b)
=
\delta_{\Phi_{a,b}(\distr)}.
\]
\[
K_{\mathsf{st}}(\cdot\mid\distr,a,b)
=
\sum_{\xi\in\TransReal}
P_{a,b}(\xi)\,
\delta_{\Psi_\xi(\distr)},
\]
We briefly describe the matrix-game background used below. A finite zero-sum
matrix game is a one-round game specified by a real matrix
$M=(M_{a,b})_{a\in\mathcal P_{\min},b\in\mathcal P_{\max}}$. The row player
\MinPl chooses a row and seeks to minimise the entry, while the column player
\MaxPl chooses a column and seeks to maximise it. The players may randomise:
\MinPl chooses
$\sigma\in\Dist(\mathcal P_{\min})$ and \MaxPl chooses
$\tau\in\Dist(\mathcal P_{\max})$, independently. Their expected payoff is
\[
G_M(\sigma,\tau)
\defeq
\sum_{a,b}\sigma(a)\tau(b)M_{a,b}.
\]
This expression is bilinear: it is linear in $\sigma$ when $\tau$ is fixed
and linear in $\tau$ when $\sigma$ is fixed.
The finite minimax theorem, also called von Neumann's minimax theorem
\cite{vonneumann1928theorie}, states that every finite zero-sum matrix game
has a common mixed-strategy value:
\begin{equation}
\label{eq:matrix-minimax-value}
\min_{\sigma\in\Dist(\mathcal P_{\min})}
\max_{\tau\in\Dist(\mathcal P_{\max})}
G_M(\sigma,\tau)
=
\max_{\tau\in\Dist(\mathcal P_{\max})}
\min_{\sigma\in\Dist(\mathcal P_{\min})}
G_M(\sigma,\tau).
\end{equation}

We denote their common
number by $\operatorname{val}(M)$. Since the mixed-strategy simplices are
compact and $G_M$ is continuous, the minimum and maximum are attained. Thus
there are optimal mixed strategies $\sigma^*$ and $\tau^*$ satisfying the
saddle inequalities
\begin{equation}
\label{eq:matrix-saddle-inequalities}
G_M(\sigma^*,\tau)
\leq
\operatorname{val}(M)
\leq
G_M(\sigma,\tau^*)
\quad
\text{for every }\sigma,\tau.
\end{equation}

This theorem applies naturally to the sampled semantics. 
If $F:\DistS\to
\mathbb R$ is a continuation payoff, then the entry associated with $(a,b)$
is the scalar
\[
\int_{\DistS}F(\nu)\,K_{\Sem}(d\nu\mid\distr,a,b).
\]
Randomising over the current pure profiles averages these already-evaluated
scalar entries and therefore produces exactly the bilinear payoff $G_M$.

\sampleddeterminacy*

\begin{proof}
Let $\mathcal E=(\States,\States_{\min},\States_{\max},\Acts,\Trans,\distro)$
be an entropy game, and let $\Sem\in\{\mathsf{sp},\mathsf{st}\}$. Put
$L\defeq\log_2|\States|$.

\cref{thm:sp-determined} makes two claims: (1) $V^-_{\Sem}=V^+_{\Sem}$, and (2) both players have optimal
stationary strategies that stay optimal against arbitrary history-dependent
opponents. The proof has three steps. Step 1 collects what the proof of
\cref{thm:bellman} (\cref{app:bellman-proof}) already establishes, which gives
claim (1) and a one-step identity for the true value $V_{\Sem}$. Step 2 builds
Borel measurable stationary saddle strategies from that identity, and Step 3
shows that they are optimal against the whole history-dependent strategy
spaces, which is claim (2).

\begin{enumerate}
\item [Step 1:] \emph{determinacy and the one-step identity.}
For the kernel $K_{\Sem}$ above and a continuous $f:\DistS\to\mathbb R$, the
entry $\int_{\DistS}f(\nu)\,K_{\Sem}(d\nu\mid\distr,a,b)$ equals
$f(\Phi_{a,b}(\distr))$ for $\Sem=\mathsf{sp}$ and
$\sum_{\xi\in\TransReal}P_{a,b}(\xi)f(\Psi_\xi(\distr))$ for $\Sem=\mathsf{st}$.
These are exactly the entries of the matrix $M_f^{\Sem}(\distr)$ of
\cref{thm:bellman}, so $\mathcal T_{\Sem}f=(1-\Discount)\Ent
+\Discount\operatorname{val}(M^{\Sem}_f)$. The proof of \cref{thm:bellman}
(\cref{app:bellman-proof}) shows, for $L\defeq\log_2|\States|$, that the
$n$-stage upper and lower values coincide with $W^{\Sem}_n=(\mathcal
T_{\Sem})^n0$, which is continuous, with Borel Markov optimal strategies at
every horizon, and that
\begin{equation}
\label{eq:sampled-value-limit}
\|V_{\Sem}-W^{\Sem}_n\|_\infty\leq\Discount^nL,
\qquad
V^-_{\Sem}=V^+_{\Sem}=:V_{\Sem},
\end{equation}
where $\|\cdot\|_\infty$ is as in \cref{app:prelim}. This proves claim (1).
It also shows that $V_{\Sem}$ is continuous (as a uniform limit of continuous
functions), that $0\leq V_{\Sem}\leq L$, and that $V_{\Sem}$ is the unique
fixed point of $\mathcal T_{\Sem}$. Define the limiting continuation matrix
\begin{equation}
\label{eq:sampled-value-matrix}
M^{\Sem}(\distr)_{a,b}
\defeq
\int_{\DistS}V_{\Sem}(\nu)\,
K_{\Sem}(d\nu\mid\distr,a,b),
\end{equation}
that is, $M^{\Sem}=M^{\Sem}_{V_{\Sem}}$, with entries
$V_{\mathsf{sp}}(\Phi_{a,b}(\distr))$ under sampled-profile semantics and
$\sum_{\xi\in\TransReal}P_{a,b}(\xi)V_{\mathsf{st}}(\Psi_\xi(\distr))$ under
sampled-transition semantics. The fixed-point equation
$V_{\Sem}=\mathcal T_{\Sem}V_{\Sem}$ then reads
\begin{equation}
\label{eq:sampled-one-step-value}
V_{\Sem}(\distr)
=
(1-\Discount)\Ent(\distr)
+\Discount\operatorname{val}\bigl(M^{\Sem}(\distr)\bigr).
\end{equation}

\item [Step 2: ] \emph{measurable stationary saddle strategies.}
For each $\distr$, let $\Sigma^*_{\min}(\distr)$ and
$\Sigma^*_{\max}(\distr)$ be the sets of optimal mixed rows and columns in
$M^{\Sem}(\distr)$. The entries of this matrix depend continuously on
$\distr$: this follows from continuity of $V_{\Sem}$ and of the successor
maps, as in the proof of \cref{thm:bellman}. Hence the optimal-strategy sets are nonempty and
compact by the finite minimax theorem and vary measurably with $\distr$. The
same measurable maximum theorem used in \cref{app:bellman-proof} therefore provides Borel
selectors
\[
\sigma^*(\distr)\in\Sigma^*_{\min}(\distr),
\qquad
\tau^*(\distr)\in\Sigma^*_{\max}(\distr).
\]
Because these selectors depend only on the current distribution, using them at
every round defines stationary randomised strategies of the type specified in
\cref{sec:model}. Measurability is what ensures that the corresponding
one-step kernels and canonical path laws remain well defined.
Optimality in the one-step matrix game and
\eqref{eq:sampled-one-step-value} imply,
for every $\distr$ and every mixed profile $\tau$ of \MaxPl,
\begin{align}
V_{\Sem}(\distr)
\geq{}(1-\Discount)\Ent(\distr)
+\Discount\sum_{a,b}\sigma^*(\distr)(a)\tau(b)
M^{\Sem}(\distr)_{a,b},
\label{eq:sampled-min-one-step}
\end{align}
and, for every mixed profile $\sigma$ of \MinPl,
\begin{align}
V_{\Sem}(\distr)
\leq{}(1-\Discount)\Ent(\distr)
+\Discount\sum_{a,b}\sigma(a)\tau^*(\distr)(b)
M^{\Sem}(\distr)_{a,b}.
\label{eq:sampled-max-one-step}
\end{align}
Indeed, the first inequality is obtained from
$G_{M^{\Sem}(\distr)}(\sigma^*(\distr),\tau)
\leq\operatorname{val}(M^{\Sem}(\distr))$, and the second from the other
saddle inequality in \eqref{eq:matrix-saddle-inequalities}.

\item [Step 3: ] \emph{verification against history-dependent opponents.}
Let the minimiser use $\sigma^*$ against an arbitrary history-dependent
randomised strategy of the maximiser. Conditional on $\mathcal F_t$, the
observed history $(\RandDistr_0,\ldots,\RandDistr_t)$ and therefore the current
distribution $\RandDistr_t$ are fixed, while the opponent's current
prescription is some $\mathcal F_t$-measurable mixed profile $\tau_t$. By the
path-law identity \eqref{eq:path-law-conditional}, conditional on
$\mathcal F_t$ the law of $\RandDistr_{t+1}$ is the mixture
$\sum_{a,b}\sigma^*(\RandDistr_t)(a)\tau_t(b)\,
K_{\Sem}(\cdot\mid\RandDistr_t,a,b)$ of one-step kernels, weighted by the
current prescriptions $\sigma^*(\RandDistr_t)$ and $\tau_t$. Taking the
conditional expectation of $V_{\Sem}$ against this law, and swapping the
finite sum with the integral, gives
\begin{align*}
&\mathbb E\!\left[
V_{\Sem}(\RandDistr_{t+1})\mid\mathcal F_t
\right]
\\
&\qquad=
\sum_{a,b}
\sigma^*(\RandDistr_t)(a)\tau_t(b)
\int_{\DistS}V_{\Sem}(\nu)\,K_{\Sem}(d\nu\mid\RandDistr_t,a,b).
\end{align*}
Each inner integral is exactly $M^{\Sem}(\RandDistr_t)_{a,b}$ by its
defining equation \eqref{eq:sampled-value-matrix}, so
\begin{align*}
&\mathbb E\!\left[
V_{\Sem}(\RandDistr_{t+1})\mid\mathcal F_t
\right]
\\
&\qquad=
\sum_{a,b}
\sigma^*(\RandDistr_t)(a)\tau_t(b)
M^{\Sem}(\RandDistr_t)_{a,b}.
\end{align*}

Substituting this equality into \eqref{eq:sampled-min-one-step} gives, almost
surely,
\[
V_{\Sem}(\RandDistr_t)
\geq
(1-\Discount)\Ent(\RandDistr_t)
+\Discount\,
\mathbb E\!\left[V_{\Sem}(\RandDistr_{t+1})\mid\mathcal F_t\right].
\]
Multiply by $\Discount^t$, take expectations, and use the tower property
$\mathbb E[\mathbb E[X\mid\mathcal F_t]]=\mathbb E[X]$. Summing for
$t=0,\ldots,T-1$ cancels every intermediate continuation term: the term
$\Discount^t\mathbb E[V_{\Sem}(\RandDistr_t)]$ on the left at round $t$
cancels the corresponding term produced on the right at round $t-1$. Only
the initial and terminal terms remain, giving
\[
(1-\Discount)\sum_{t=0}^{T-1}\Discount^t
\mathbb E[\Ent(\RandDistr_t)]
\leq
V_{\Sem}(\distro)
-\Discount^T\mathbb E[V_{\Sem}(\RandDistr_T)]
\leq V_{\Sem}(\distro).
\]
Recall from \cref{sec:preliminaries} that  
\begin{equation}
 \label{eq:objective-sum-of-expectations}
 \mathsf{Obj}^{\Sem}_{\Discount}(\stratmin,\stratmax;\distr)=(1-\Discount)\sum_{t=0}^{\infty}\Discount^t\,\PathExpect{\distr}{\Sem}{\stratmin}{\stratmax}\bigl[\Ent(\RandDistr_t)\bigr].
 \end{equation}

Letting $T\to\infty$ and using
\cref{eq:objective-sum-of-expectations} yields
\[
\mathsf{Obj}^{\Sem}_{\Discount}(\sigma^*,\stratmax;\distro)
\leq V_{\Sem}(\distro)
\quad\text{for every }\stratmax.
\]
For completeness, let the maximiser use $\tau^*$ against an arbitrary
history-dependent minimising strategy. Conditional on $\mathcal F_t$, let
$\sigma_t$ be the minimiser's current mixed profile. Then
\[
\mathbb E\!\left[
V_{\Sem}(\RandDistr_{t+1})\mid\mathcal F_t
\right]
=
\sum_{a,b}\sigma_t(a)\tau^*(\RandDistr_t)(b)
M^{\Sem}(\RandDistr_t)_{a,b}.
\]
Substitution into \eqref{eq:sampled-max-one-step}, followed by the same
discounting, expectation, and telescoping calculation, yields
\[
(1-\Discount)\sum_{t=0}^{T-1}\Discount^t
\mathbb E[\Ent(\RandDistr_t)]
\geq
V_{\Sem}(\distro)
-\Discount^T\mathbb E[V_{\Sem}(\RandDistr_T)].
\]
Since $0\leq V_{\Sem}\leq L$,
\[
0\leq
\Discount^T\mathbb E[V_{\Sem}(\RandDistr_T)]
\leq\Discount^T L
\longrightarrow0.
\]
Thus \cref{eq:objective-sum-of-expectations} gives
\[
\mathsf{Obj}^{\Sem}_{\Discount}(\stratmin,\tau^*;\distro)
\geq V_{\Sem}(\distro)
\quad\text{for every }\stratmin.
\]
Thus $\sigma^*$ guarantees at most $V_{\Sem}$ and $\tau^*$ guarantees at
least $V_{\Sem}$. They are therefore optimal against the full
history-dependent strategy spaces. The argument used only the common kernel
description: substituting $K_{\mathsf{sp}}$ proves the $\mathsf{sp}$ case of
\cref{thm:sp-determined}, and substituting $K_{\mathsf{st}}$ proves the
$\mathsf{st}$ case.
\end{enumerate}
\end{proof}

\section{Gap Value Upper Bounds}
\label{sec:hardness}
We study the gap problem from \cref{prob:instance}, grouping the results by
semantics. Throughout this section, the discount $\Discount\in(0,1)$ and
gap $\varepsilon_0>0$ are fixed rational constants. The input consists of
the explicitly listed arena, rational transition entries
$\Trans(s,a,s')$, rational initial distribution, and rational threshold
$v$, all encoded in binary. Write $\ell$ for the total input length,
$n=|\States|$, $A=\max_s|\Acts(s)|$, and $e=\min\{\varepsilon_0,1\}$.
For $n\ge2$ put $L=\lceil\log_2n\rceil$; one-state games have value zero.

\subsection{Averaged-distribution semantics}
\label{sec:etr}
The upper and lower values can differ under this semantics, so we explicitly
study $\GapVal^{+}_{\mathsf{av},\varepsilon_0}$, without assuming determinacy.

Entropy is a nonlinear, transcendental stage reward: $-x\log_2x$ has no
closed rational form except at $x\in\{0,1\}$. Lemma~\ref{ass:rational}
packages two tools needed to tame it, each used later for a different
purpose. Part~1 (continuity) bounds how much entropy can change between
nearby distributions. Part~2 (effective evaluation) computes $\Ent(\mu)$
to any desired number of bits in time polynomial in that number;  this is
what lets \cref{thm:etr-ub} below evaluate entropy exactly enough to
decide the gap problem, without ever needing its closed, transcendental
form.
\entropytools*

\begin{proof}
\noindent \textbf{Part 1 : \emph{Continuity.}} Couple $X\sim\mu$ and $Y\sim\nu$ on a common
probability space so as to agree as often as possible, as follows.
Since $\min(\mu(s),\nu(s))=\tfrac12\bigl(\mu(s)+\nu(s)-|\mu(s)-\nu(s)|\bigr)$
for each $s$, summing over $s\in\States$ gives
\[
\sum_{s\in\States}\min(\mu(s),\nu(s))
=\tfrac12(1+1)-\tfrac12\sum_{s\in\States}|\mu(s)-\nu(s)|
=1-d_{\mathrm{TV}}(\mu,\nu),
\]
using $\sum_s\mu(s)=\sum_s\nu(s)=1$. With probability
$1-d_{\mathrm{TV}}(\mu,\nu)$, draw a single state $s$ with probability
$\min(\mu(s),\nu(s))/(1-d_{\mathrm{TV}}(\mu,\nu))$ and set $X=Y=s$;
otherwise, with the complementary probability $d_{\mathrm{TV}}(\mu,\nu)$,
draw $X$ and $Y$ independently from the normalized leftover distributions
$(\mu(s)-\min(\mu(s),\nu(s)))/d_{\mathrm{TV}}(\mu,\nu)$ and
$(\nu(s)-\min(\mu(s),\nu(s)))/d_{\mathrm{TV}}(\mu,\nu)$ respectively.
These two leftover distributions have disjoint supports, wherever
$\mu(s)>\nu(s)$ the second is $0$ at $s$, and vice versa, so $X\ne Y$
almost surely in this second branch. Under this construction $X$ has
marginal law $\mu$ and $Y$ has marginal law $\nu$ (each coordinate
recovers $\mu(s)$, respectively $\nu(s)$, as the weighted sum of its
shared and leftover parts), and $X=Y$ occurs exactly in the first
branch, so $q=\Pr[X\ne Y]=d_{\mathrm{TV}}(\mu,\nu)$.

Let $E=\mathbf1_{\{X\ne Y\}}$ be the
indicator of this coupling's disagreement event, a single Bernoulli($q$)
random variable. Extending $\Ent(\cdot)$ to random
variables in the usual way, so that $\Ent(X\mid Y)$ denotes conditional
entropy: since $E$ is a deterministic function of $(X,Y)$,
\[
\Ent(X,E\mid Y)=\Ent(X\mid Y)+\Ent(E\mid X,Y)=\Ent(X\mid Y),
\]
while the chain rule expands the same quantity as
$\Ent(X,E\mid Y)=\Ent(E\mid Y)+\Ent(X\mid Y,E)$. Equating the two and using
$\Ent(E\mid Y)\le\Ent(E)$ (conditioning cannot increase entropy) gives
\[
\Ent(X\mid Y)\le\Ent(E)+\Ent(X\mid Y,E).
\]
Split the remaining term on the value of $E$: if $E=0$ then $X=Y$, so $X$ is
determined once $Y$ and $E=0$ are known, contributing $0$; if $E=1$ then
$X$ is confined to the $n-1$ states other than $Y$'s value, so it
contributes at most $\log_2(n-1)$ (the max-entropy bound on a support of
size $n-1$). Averaging over $E$ with weights $1-q,q$ gives
$\Ent(X\mid Y,E)\le q\log_2(n-1)$, and since $E$ is Bernoulli($q$),
$\Ent(E)=h(q)$. Together,
\[
\Ent(X\mid Y)\le h(q)+q\log_2(n-1)=\omega_n(q).
\]
Thus $\Ent(X)-\Ent(Y)\le\Ent(X\mid Y)\le\omega_n(q)$; interchanging $X,Y$
proves the absolute-value bound. On $(0,1-1/n)$,
$\omega_n'(r)=\log_2((n-1)(1-r)/r)\ge0$, so $q\le r$ gives the stated
modulus. The elementary binary-entropy bound
$h(r)\le2\sqrt{r(1-r)}\le2\sqrt r$ is a standard bound; for instance, $h(r)\le r\log_2(e/r)$ (as $(1-r)\ln\frac1{1-r}\le r$) and
$r\log_2(e/r)\le2\sqrt r$ on $(0,1]$.

Computing $\Ent(\mu)$ exactly is not an option: for most rational $\mu$ the
value involves $\log_2$ of a rational number, which is irrational. Any
algorithm that later needs to compare $\Ent(\mu)$ against a threshold, or
feed it into a larger computation, can only work with rationals, so what is
actually needed is a way to compute $\Ent(\mu)$ to any desired number of
bits, with an explicit error bound, in time that stays polynomial as that
number of bits grows. The natural tool is a series expansion for $\log_2$,
but the standard one converges too slowly (or not at all) unless its
argument is small, which is what step~2 below arranges.

\noindent \textbf{Part 2 : \emph{Effective dyadic evaluation.}} On input $(\mu,p)$, compute
$\widehat H_p(\mu)$ as follows.

\begin{enumerate}
\item \emph{Zero coordinates.} For each $s$ with $\mu(s)=0$, contribute $0$
(by convention $0\log_20=0$).
\item \emph{Per-coordinate logarithm.} For each $s$ with $q=\mu(s)>0$, write
$q=2^{-j}u$ with $1\le u<2$, where $j$'s size is bounded by the encoding of
$q$. Compute
\[
\ln u=2\sum_{i=0}^{m}\frac{z^{2i+1}}{2i+1}+R_m,
\qquad z=\frac{u-1}{u+1}\in[0,1/3],
\qquad
|R_m|\le\frac{2z^{2m+3}}{(2m+3)(1-z^2)},
\]
with $m=O(p+\operatorname{size}(q)+\log n)$ terms; the same expansion at
$u=2$ gives $\ln2$, bounded away from zero. Since
$\log_2q=-j+\ln u/\ln2$, this yields $-q\log_2q$ to within $2^{-p-1}/n$,
using only polynomially many rational operations on rationals of
polynomial bit length.
\item \emph{Summation.} Add the (at most $n$) per-coordinate
approximations; the total evaluation error is at most
$n\cdot2^{-p-1}/n=2^{-p-1}$.
\item \emph{Rounding.} Round the sum to the nearest multiple of $2^{-p}$,
adding at most $2^{-p-1}$ more error.
\item \emph{Clipping.} Clip the result to $[0,L]$; since
$\Ent(\mu)\in[0,L]$, this cannot increase the distance to $\Ent(\mu)$.
\end{enumerate}
Steps~3-4 give total error at most $2^{-p-1}+2^{-p-1}=2^{-p}$, and step~5
does not increase it, so the output satisfies the claimed bound. Every step
uses only polynomially many rational operations on rationals of polynomial
bit length, so the whole procedure runs in time polynomial in
$\operatorname{size}(\mu)+p$. This proves the claimed algorithm.
\end{proof}

\avgapupperbound*

\begin{proof}
Let $\ell$ denote the total input encoding length (the arena, the rational
transition probabilities, the initial distribution, and the threshold $v$,
all written in binary), and recall $n=|\States|$, $A=\max_s|\Acts(s)|$, and
$e=\min\{\varepsilon_0,1\}$. Put
$L=\lceil\log_2n\rceil$, the maximum possible entropy on $n$ states. If
$n=1$, entropy is identically zero on the single state, so
$V^+_{\mathsf{av}}(\distro)=0$ regardless of either player's strategy;
comparing $0$ against $v$ decides the promise directly, with no further
construction needed. Henceforth assume $n\ge2$.
The proof has two parts. Steps~1-2 reduce the infinite-horizon value to a
finite, memoryless recursion on the current distribution. Step~3 then gives a
$\PSPACE$ algorithm deciding the promise gap directly, by
evaluating that recursion on a finite grid of actions via an alternating
Turing machine.

\medskip\noindent\textbf{Step 1: uniform horizon truncation.}
The value $V^+_{\mathsf{av}}$ is defined by summing over infinitely many
rounds, but any algorithm can only refer to finitely many of them. So
before anything else, the infinite horizon must be cut down to some
finite $T$, with a guarantee that the discarded tail is too small to
affect the promise-gap decision. For every strategy pair the omitted tail
satisfies
\[
0\le(1-\Discount)\sum_{t=T}^{\infty}\Discount^t\Ent(\distr_t)
\le\Discount^T\log_2n\le\Discount^T L.
\]

Choose the least positive integer $T$ with $\Discount^T L\le e/4$.
It can be found by exact rational comparisons, and
$T=O(1+\log(n/e))$. Both $\Discount$ and $e=\min\{\varepsilon_0,1\}$ are
themselves fixed constants, not part of the varying instance; the
$1/e$ dependence inside the logarithm is displayed explicitly to show
how $T$ scales for other fixed choices of the gap $\varepsilon_0$.
There are $T$ rewards, at times $0,\ldots,T-1$, but only $T-1$ relevant
action pairs $(x_t,y_t)$ (one per transition $\mu_t\to\mu_{t+1}$), since
the last rewarded distribution $\mu_{T-1}$ needs no outgoing action.

\medskip\noindent\textbf{Step 2: finite-horizon values and public histories.}
For $r\ge0$ let $U_r(\mu)$ be the upper value, over strategies whose action at
time $t$ may depend on the whole observed distribution history
$(\mu_0,\ldots,\mu_t)$, of the truncated payoff
$(1-\Discount)\sum_{t=0}^{r-1}\Discount^t\Ent(\mu_t)$ started at $\mu$. This is
the quantity $V^{+}_{\mathsf{av},r}$ of the proof of \cref{thm:bellman}
(\cref{app:bellman-proof}), where the following facts are proved, with the same
history-dependent strategy spaces. Each $U_r$ is continuous, the minimiser has a
Borel Markov policy attaining it from every distribution, and
\begin{equation}
\label{eq:finite-upper-recursion}
U_0=0,\qquad
U_{r+1}(\mu)=(1-\Discount)\Ent(\mu)
+\Discount\min_{x\in\mathcal X}\max_{y\in\mathcal Y}
U_r(\DistOp_{x,y}(\mu)),
\end{equation}
that is, $U_r=(\mathcal T^+_{\mathsf{av}})^r0$. The same proof bounds the
truncation error uniformly in the strategy pair: the omitted tail of the payoff
is at most $\Discount^rL'$ with $L'=\log_2|\States|\le L$. Taking either ordered
optimisation of this pointwise bound, and using $\Discount^TL\le e/4$ from
Step~1, gives
\begin{equation}
\label{eq:finite-infinite-error}
0\le V^+_{\mathsf{av}}(\mu)-U_T(\mu)\le e/4.
\end{equation}
Step~3 below uses only \eqref{eq:finite-upper-recursion}, continuity of $U_r$, and \eqref{eq:finite-infinite-error}.

\medskip\noindent\textbf{Step 3: a finite-grid PSPACE algorithm.}
Step 2 characterised $U_T(\mu)$ via a recursion over $x\in\mathcal
X,y\in\mathcal Y$, continuous, infinite sets of mixed actions. A Turing
machine cannot search over a continuum, so before this recursion can
become an algorithm, the action spaces must be replaced by a finite grid,
and the entropy reward must be evaluated to finite precision rather than
exactly. Both replacements introduce error, which has to be tracked and
kept within budget alongside Step 1's truncation error. The construction
below does this in three stages: first, bound how much rounding an action
to the nearest grid point can perturb the resulting distribution and, in
turn, the value $U_r$; second, choose the grid fine enough that this
perturbation costs at most $e/8$ of the total budget; third, build an
alternating Turing machine that evaluates the grid-restricted recursion
exactly, guessing the minimiser's action existentially and the
maximiser's universally at each round, and confirm the whole computation
fits in polynomial space via $\AP=\PSPACE$.

\emph{Rounding actions to a grid.} This step produces the basic tool
everything else relies on: a way to replace an arbitrary mixed action by
one from a finite grid, with an explicit, quantified bound on how much
the resulting distribution can change as a result.
Let $\mathcal X_Q,\mathcal Y_Q$ be the
products of statewise action simplices with coordinates in
$\{0,1/Q,\ldots,1\}$: grids of mixed actions with denominator $Q$. Any
mixed action $x\in\mathcal X$ can be rounded to a grid point $x^Q$ as
follows: at each state $s\in\States_{\min}$ with $k=|\Acts(s)|\le A$
actions, let $q_a=Q\times x(s)(a)$ for each $a\in\Acts(s)$, so $\sum_aq_a=Q$.
Set $x^Q(s)(a)=\lfloor q_a\rfloor/Q$ initially; since the fractional parts
$q_a-\lfloor q_a\rfloor$ sum to an integer $R\in[0,k-1]$, add $1/Q$ to the
$R$ coordinates with the largest fractional parts, restoring the total to
$1$. Each single coordinate changes by less than $1/Q$ under this
procedure (rounding down or up moves $q_a/Q$ by less than $1/Q$), so
summing over at most $A$ coordinates,
\[
\|x(s)-x^Q(s)\|_1=\sum_{a\in\Acts(s)}|x(s)(a)-x^Q(s)(a)|\le\frac{|\Acts(s)|}{Q}\le\frac AQ.
\]
To convert this into a bound on the successor distribution, write out the
difference coordinatewise using
\[
\DistOp_{x,y}(\mu)(s')=\sum_{s\in\States_{\min}}\mu(s)\sum_ax(s)(a)\Trans(s,a,s')
+\sum_{s\in\States_{\max}}\mu(s)\sum_ay(s)(a)\Trans(s,a,s')
\]
(only the Min-controlled part depends on $x$, so the Max-controlled part
cancels):
\[
\DistOp_{x,y}(\mu)(s')-\DistOp_{x^Q,y}(\mu)(s')
=\sum_{s\in\States_{\min}}\mu(s)\sum_a\bigl[x(s)(a)-x^Q(s)(a)\bigr]\Trans(s,a,s').
\]
Summing absolute values over $s'$ and using $\sum_{s'}\Trans(s,a,s')=1$,
\[
\|\DistOp_{x,y}(\mu)-\DistOp_{x^Q,y}(\mu)\|_1
\le\sum_{s\in\States_{\min}}\mu(s)\sum_a|x(s)(a)-x^Q(s)(a)|
=\sum_{s\in\States_{\min}}\mu(s)\|x(s)-x^Q(s)\|_1.
\]
Dividing by $2$ and using the coordinate bound above, with
$\sum_{s\in\States_{\min}}\mu(s)\le1$,
\begin{equation}
\label{eq:action-rounding-transition}
d_{\mathrm{TV}}(\DistOp_{x,y}(\mu),\DistOp_{x^Q,y}(\mu))
\le\tfrac12\sum_{s\in\States_{\min}}\mu(s)\|x(s)-x^Q(s)\|_1
\le A/(2Q).
\end{equation}
The same holds for rounding $y$.
 
 \emph{From rounding error to value error.} The previous step bounded how
much rounding perturbs a distribution; this step upgrades that into a
bound on how much the value $U_r$ itself can shift as a result, and uses
it to compare the exact one-step value against its grid-restricted
counterpart.
 
 For fixed $x,y$, the successor
map is nonexpansive in total variation: writing
$\DistOp_{x,y}(\mu)-\DistOp_{x,y}(\nu)=(\mu-\nu)K$ for the stochastic
matrix $K(s,s')=\sum_a\pi_{x,y}(s)(a)\Trans(s,a,s')$ whose rows sum to
$1$,
\[
\|\DistOp_{x,y}(\mu)-\DistOp_{x,y}(\nu)\|_1
=\sum_{s'}\Bigl|\sum_s(\mu(s)-\nu(s))K(s,s')\Bigr|
\le\sum_s|\mu(s)-\nu(s)|\sum_{s'}K(s,s')
=\|\mu-\nu\|_1,
\]
so $d_{\mathrm{TV}}(\DistOp_{x,y}(\mu),\DistOp_{x,y}(\nu))\le
d_{\mathrm{TV}}(\mu,\nu)$: applying the same actions to two distributions
cannot push their successors further apart than the originals were.
We prove, for $0\le r_0\le1-1/n$ and every $r\ge0$,
\begin{equation}
\label{eq:finite-value-modulus}
d_{\mathrm{TV}}(\mu,\nu)\le r_0
\ \Longrightarrow\
|U_r(\mu)-U_r(\nu)|\le\omega_n(r_0).
\end{equation}
For $r=0$, $U_0\equiv0$, so $|U_0(\mu)-U_0(\nu)|=0\le\omega_n(r_0)$, since
$\omega_n\ge0$; this is the base case. Assume \eqref{eq:finite-value-modulus}
for $r$ and suppose $d_{\mathrm{TV}}(\mu,\nu)\le r_0$. By Part~1 of
\cref{ass:rational}, the entropy term contributes
$(1-\Discount)|\Ent(\mu)-\Ent(\nu)|\le(1-\Discount)\omega_n(r_0)$. For the
continuation term, nonexpansiveness gives
$d_{\mathrm{TV}}(\DistOp_{x,y}(\mu),\DistOp_{x,y}(\nu))\le r_0$ for
\emph{every} fixed $(x,y)$, so the inductive hypothesis applies to that
pair of successors:
$|U_r(\DistOp_{x,y}(\mu))-U_r(\DistOp_{x,y}(\nu))|\le\omega_n(r_0)$ for
every $x,y$. Adding a constant bound to a function does not change which
point maximizes or minimizes it, only shifts the extremal value by that
constant, so taking $\max_y$ and then $\min_x$ of both sides (and the
symmetric argument swapping $\mu,\nu$) gives
\[
\Bigl|\min_x\max_yU_r(\DistOp_{x,y}(\mu))-\min_x\max_yU_r(\DistOp_{x,y}(\nu))\Bigr|
\le\omega_n(r_0).
\]
Combining the two terms of \eqref{eq:finite-upper-recursion},
\[
|U_{r+1}(\mu)-U_{r+1}(\nu)|
\le(1-\Discount)\omega_n(r_0)+\Discount\,\omega_n(r_0)
=\omega_n(r_0),
\]
completing the induction.
This is what lets grid-rounding error be converted into value error. Fix
$\mu$ and write $F(x,y)=U_r(\DistOp_{x,y}(\mu))$ for the one-step
continuation value as a function of the two players' actions. Let
$M=\min_x\max_yF$ be the exact value, $M_X=\min_{x\in\mathcal X_Q}\max_yF$
the value once the minimiser alone is restricted to the grid, and
$M_Q=\min_{x\in\mathcal X_Q}\max_{y\in\mathcal Y_Q}F$ the value once both
players are restricted. Since $\mathcal X_Q\subseteq\mathcal X$, minimising
over the smaller set $\mathcal X_Q$ cannot decrease the minimum, so
$M\le M_X$. For the other direction, let $x^*$ achieve $M=\max_yF(x^*,y)$
and round it to $x^{*Q}\in\mathcal X_Q$; by
\eqref{eq:action-rounding-transition} with $\eta=A/(2Q)\le1-1/n$ and
\eqref{eq:finite-value-modulus}, $F(x^{*Q},y)\le F(x^*,y)+\omega_n(\eta)$
for every $y$, so $\max_yF(x^{*Q},y)\le M+\omega_n(\eta)$; since
$x^{*Q}\in\mathcal X_Q$, $M_X\le\max_yF(x^{*Q},y)\le M+\omega_n(\eta)$.
Together,
\[
M\le M_X\le M+\omega_n(\eta).
\]
The same argument, applied to rounding the maximiser's action instead
(restricting $\mathcal Y$ to $\mathcal Y_Q$ can only decrease a maximum,
giving $M_Q\le M_X$; rounding the maximiser's best response at the
minimiser's grid-optimal $x$ gives the other direction), gives
\[
M_Q\le M_X\le M_Q+\omega_n(\eta).
\]
By the triangle inequality,
$|M-M_Q|\le|M-M_X|+|M_X-M_Q|\le\omega_n(\eta)+\omega_n(\eta)=2\omega_n(\eta)$:
restricting both players to the grid changes the one-step value by at
most $2\omega_n(\eta)$.

 \emph{Choosing the grid.} The previous step bounded the value error
introduced by the grid in a single round; this step accumulates that
error across all $T$ rounds of the recursion and picks concrete
constants so the total stays within the $e/8$ budget.
 Define $U_r^Q$ by the same recursion, with both
action sets restricted to the grid throughout. Since each round's
discrepancy from the exact value is at most $2\omega_n(\eta)$, and the
recursion discounts the previous round's error by $\Discount$, the total
discrepancy after $r$ rounds, $D_r=\|U_r-U_r^Q\|_\infty$, satisfies
$D_{r+1}\le\Discount D_r+2\Discount\omega_n(\eta)$ and $D_0=0$. Unrolling
this recursion, $D_r\le2\Discount\omega_n(\eta)(1+\Discount+\cdots+
\Discount^{r-1})=2\Discount\omega_n(\eta)\cdot\frac{1-\Discount^r}{1-\Discount}
\le\frac{2\Discount}{1-\Discount}\omega_n(\eta)$. Choose the rational
constants and integer
\[
c=\frac{e(1-\Discount)}{16\Discount},\qquad
r_* =\min\{1/2,c/(2n),c^2/16\},\qquad
Q=\left\lceil\frac{A}{2r_*}\right\rceil.
\]
This choice is engineered so that $\eta=A/(2Q)\le r_*\le1-1/n$ (the
modulus bound applies), and, since $\omega_n$ is nondecreasing on this
range and $\omega_n(r)\le nr+2\sqrt r$ (using $\log_2(n-1)\le n$ and
$h(r)\le2\sqrt r$ from \cref{ass:rational}),
$\omega_n(\eta)\le\omega_n(r_*)\le nr_*+2\sqrt{r_*}\le\frac c2+\frac
c2=c$, using $r_*\le c/(2n)$ for the first term and $r_*\le c^2/16$ for
the second. Substituting $\omega_n(\eta)\le c$ into the unrolled bound
above,
\begin{equation}
\label{eq:grid-value-error}
\|U_T-U_T^Q\|_\infty
\le\frac{2\Discount}{1-\Discount}\omega_n(\eta)
\le\frac{2\Discount}{1-\Discount}\cdot\frac{e(1-\Discount)}{16\Discount}
=e/8.
\end{equation}
Since $\Discount,e$ are fixed, $c$ is a fixed positive constant not
depending on $n$ or $A$. If $c/(2n)$ is the smallest of the three terms
in $r_*=\min\{1/2,c/(2n),c^2/16\}$, then $r_*=c/(2n)$; otherwise $r_*$
equals one of the two constants $1/2,c^2/16$, which forces
$n\le c/(2r_*)=O(1)$ for fixed $c$. In either case $r_*=\Omega(1/n)$ for
fixed $\Discount,e$, so $Q=\lceil A/(2r_*)\rceil=O(nA)$: discretising the
action spaces to a grid of denominator $Q$ costs at most $e/8$ of the
total error budget.

 \emph{The alternating machine.} With the action spaces now finite and the
rounding error accounted for, this step builds the actual algorithm: a
machine that evaluates the grid-restricted recursion round by round,
guessing each player's action, and checks its acceptance condition
decides the promise gap correctly.
With grid actions in hand, the recursion can finally be evaluated
directly. Choose the least integer $p\ge1$ with $2^{-p}\le e/8$, so
$p=O(\log(1/e))$ bits suffice, and entropy can be evaluated to error
$2^{-p}$ via $\widehat H_p$ (Part~2 of \cref{ass:rational}).

        \smallskip 
\emph{Why $\min$ becomes $\exists$ and $\max$ becomes $\forall$.} For any
threshold $v$ and any function $g$ on a nonempty finite set,
$\min_xg(x)\le v$ iff some $x$ has $g(x)\le v$, i.e.\
$\min_xg(x)\le v\iff\exists x:g(x)\le v$; dually, $\max_yh(y)\le v$ iff
\emph{every} $y$ has $h(y)\le v$, i.e.\
$\max_yh(y)\le v\iff\forall y:h(y)\le v$. These hold against the same
fixed $v$ on both sides, so they compose under nesting: for
$g(x)=\max_yf(x,y)$, $\min_x\max_yf(x,y)\le v\iff\exists x\,[\,g(x)\le
v\,]\iff\exists x\,[\,\max_yf(x,y)\le v\,]\iff\exists x\forall
y:f(x,y)\le v$. This composition argument used only that both sides
compare against the same fixed threshold $v$, never the specific form
of the payoff, so it applies unchanged with any function of $\mu_t$ in
place of $f$.

Let $z$ denote the quantity obtained from the same recursion as
$U_r^Q$ (grid-restricted actions, as defined above) but with $\Ent$
replaced throughout by the dyadic $\widehat H_p$ from
\cref{ass:rational}. Iterating the substitution above down all $T-1$
rounds gives
\[
z\le v
\quad\Longleftrightarrow\quad
\exists x_0\forall y_0\cdots\exists x_{T-2}\forall y_{T-2}:\
(1-\Discount)\sum_{t=0}^{T-1}\Discount^t\widehat H_p(\mu_t)\le v,
\]
with $\mu_0=\distro$ and $\mu_{t+1}=\DistOp_{x_t,y_t}(\mu_t)$: the
entire nested $\min$-$\max$ recursion, evaluated once, is equivalent to
a single alternating quantifier prefix over the one payoff inequality
at the end. Since an alternating machine can evaluate and compare
$\widehat H_p$ exactly, this is exactly the shape such a machine
decides natively: an existential state accepts iff some child does
(matching $\exists$/$\min$), a universal state accepts iff every child
does (matching $\forall$/$\max$).

\emph{What happens at each round.} The machine carries the current
distribution $\mu_t$, the round index $t$, and the running weighted sum
$(1-\Discount)\sum_{t'=0}^{t}\Discount^{t'}\widehat H_p(\mu_{t'})$ on its
tape. At round $t=0,\ldots,T-2$ it: (1) computes the current reward
$\widehat H_p(\mu_t)$ and adds $\Discount^t\widehat H_p(\mu_t)$, weighted
by $(1-\Discount)$, to the running sum; (2) enters an existential state
and guesses the numerators of $x_t\in\mathcal X_Q$, one per
$(s,\alpha)$ with $s\in\States_{\min}$, $\alpha\in\Acts(s)$, each using
$\lceil\log_2(Q+1)\rceil$ bits; (3) validates them: each numerator must
lie in $[0,Q]$, and at each state $s\in\States_{\min}$ the numerators
over $\alpha\in\Acts(s)$ must sum to exactly $Q$ (encoding a valid grid
probability vector); an invalid guess makes this branch reject
immediately, since a malformed $x_t$ is not a legitimate witness for the
$\exists$ and must not be able to manufacture acceptance; (4) on a valid
guess, enters a universal state and guesses the numerators of
$y_t\in\mathcal Y_Q$ the same way, this time over $\States_{\max}$; (5)
validates them identically, except an invalid guess now makes this branch
\emph{accept} immediately, since a malformed $y_t$ is not a legitimate
move for the $\forall$ and must not be able to manufacture rejection; (6)
on a valid guess, computes the exact rational successor
$\mu_{t+1}=\DistOp_{x_t,y_t}(\mu_t)$ and recurses to round $t+1$. At $\mu_{T-1}$ there is no outgoing action, matching Step~1's observation
that $\mu_{T-1}$ needs no outgoing action; the machine adds the final term $(1-\Discount)\Discount^{T-1}\widehat H_p(\mu_{T-1})$ to the running sum, which is then complete over all $T$ reward terms, and the branch accepts exactly when
\[
(1-\Discount)\sum_{t=0}^{T-1}\Discount^t\widehat H_p(\mu_t)\le v.
\]

Recall that $z$ is the value of this numerically evaluated grid game, so the machine accepts exactly when $z\le v$, by the equivalence displayed above. By construction $z$ differs from $U_T^Q(\distro)$
by at most $2^{-p}\le e/8$, the entropy precision fixed by the choice of
$p$ above. By \eqref{eq:grid-value-error}, $U_T^Q(\distro)$ differs from
$U_T(\distro)$ by at most $e/8$. By \eqref{eq:finite-infinite-error},
$U_T(\distro)$ differs from $V^+_{\mathsf{av}}(\distro)$ by at most
$e/4$. Hence
\[
|z-V^+_{\mathsf{av}}(\distro)|\le |z-U_T^Q(\distro)| + |U_T^Q(\distro)-U_T(\distro)| + |U_T(\distro)-V^+_{\mathsf{av}}(\distro)| \le e/8+e/8+e/4=e/2<\varepsilon_0.
\]
If $V^+_{\mathsf{av}}(\distro)\le v-\varepsilon_0$, then
$z\le V^+_{\mathsf{av}}(\distro)+e/2\le v-\varepsilon_0+e/2<v$, so the
machine accepts. If $V^+_{\mathsf{av}}(\distro)\ge v+\varepsilon_0$,
then $z\ge V^+_{\mathsf{av}}(\distro)-e/2\ge v+\varepsilon_0-e/2>v$, so
the machine rejects. This proves promise correctness: the machine
accepts every yes-instance and rejects every no-instance.

\emph{Space bound.} It remains to check the machine runs in polynomial
time along every branch, not just that its numbers are individually
accurate; this is what $\AP=\PSPACE$ actually requires.

Let $D$ be the product of the denominators of all input transition
probabilities, and $D_0$ the product of the denominators of the initial
distribution's coordinates. Each individual denominator is a substring
of the $\ell$-bit input encoding, so the sum of their bit lengths is at
most $\ell$; since the bit length of a product of positive integers is
at most the sum of the bit lengths of the factors, both $D$ and $D_0$
have $O(\ell)$ bits.

Fix a branch, determined by the guessed grid actions $(x_t,y_t)$ at
each round $t$. Each numerator of $x_t,y_t$ has denominator $Q$ by
construction, and each $\Trans(s,\alpha,s')$ has denominator dividing
$D$; by \eqref{eq:distribution-update},
$\DistOp_{x_t,y_t}(\mu)(s')=\sum_{s}\mu(s)\sum_\alpha\pi_{x_t,y_t}(s)(\alpha)\Trans(s,\alpha,s')$
is a sum of products of a denominator-$Q$ term and a denominator-$D$
term, so the transition matrix for this fixed pair has common
denominator dividing $QD$. 

This allows an induction on $t$: the base
case $\mu_0=\distro$ has common denominator dividing $D_0$. For the
inductive step, if every coordinate of $\mu_t$ has the form
$m_s/(D_0(QD)^t)$ for an integer $m_s$, and
$\sum_\alpha\pi_{x_t,y_t}(s)(\alpha)\Trans(s,\alpha,s')=q_{s,s'}/(QD)$
for an integer $q_{s,s'}$, then
\[
\mu_{t+1}(s')=\sum_{s\in\States}\frac{m_s}{D_0(QD)^t}\cdot\frac{q_{s,s'}}{QD}
=\frac{\sum_sm_sq_{s,s'}}{D_0(QD)^{t+1}},
\]
an integer numerator over $D_0(QD)^{t+1}$. So
$\mu_t \text{ has common denominator dividing } D_0(QD)^t$
and each of its $n$ coordinates uses $O(\ell+t(\ell+\log Q))$ bits.
Exact updates are therefore additions and multiplications of integers
of this bit length, which elementary arithmetic performs in time
polynomial in $\ell+T(\ell+\log Q)$, itself polynomial in $\ell$ since
$T,Q$ are polynomial in $\ell$.

The discounted dyadic sum also has polynomial bit length: if $g$ is the
fixed denominator of $\Discount$, then $(1-\Discount)$ and $\Discount^t$
each have denominator dividing $g^{t+1}$, and $\widehat H_p(\mu_t)$ has
denominator $2^p$ by construction, so each summand
$(1-\Discount)\Discount^t\widehat H_p(\mu_t)$ has denominator dividing
$g^{t+1}2^p$, hence dividing $g^T2^p$ for every $t\le T-1$; the sum of
the $T$ summands therefore has denominator dividing $g^T2^p$, a
polynomial-bit rational since $T,p$ are polynomial in $\ell$.

There are only $T-1$ action pairs along any branch. At each, validating
the guessed numerators (checking they lie in $[0,Q]$ and sum to $Q$ at
every controlled state) is a comparison of integers already bounded
above, and computing $\widehat H_p(\mu_t)$ takes time polynomial in
$\operatorname{size}(\mu_t)+p$ by \cref{ass:rational}, itself polynomial
in $\ell$ since $\mu_t$'s bit length and $p$ both are. Summing these
$T-1$ polynomial-time steps gives a polynomial total time along the
branch. This is an alternating polynomial-time computation, hence is in
$\PSPACE$ by $\AP=\PSPACE$. Equivalently, since the branch depth $T-1$
and every configuration's bit length are both polynomial in $\ell$, a
deterministic depth-first traversal of its polynomial-depth binary
computation tree uses a polynomial stack of polynomial-size
configurations.

\medskip\noindent\textbf{The lower value.} For $V^-_{\mathsf{av}}$ the same proof applies with the order of the extrema swapped. Let $U^-_r=(\mathcal T^-_{\mathsf{av}})^r0$, so that $U^-_{r+1}(\mu)=(1-\Discount)\Ent(\mu)+\Discount\max_{y}\min_{x}U^-_r(\DistOp_{x,y}(\mu))$. By the finite-horizon identification and error bound in the proof of \cref{thm:bellman}, $0\le V^-_{\mathsf{av}}-U^-_T\le\Discount^TL\le e/4$. The modulus \eqref{eq:finite-value-modulus} holds for $U^-_r$ with the same proof, since adding a uniform bound to the argument shifts $\max_y\min_x$ by at most that bound. For the grid, put $F(x,y)=U^-_r(\DistOp_{x,y}(\mu))$ and $M^-=\max_y\min_xF$. Restricting $y$ to $\mathcal Y_Q$ gives $M^-_Y\le M^-$, and rounding a maximiser $y^*$ to $y^{*Q}$ changes $\min_xF(x,\cdot)$ by at most $\omega_n(\eta)$, by \eqref{eq:action-rounding-transition} applied for every $x$ and by \eqref{eq:finite-value-modulus}; so $M^-\le M^-_Y+\omega_n(\eta)$. Restricting the inner minimum to $\mathcal X_Q$ can only increase it, and for every $y$ rounding a best response raises it by at most $\omega_n(\eta)$. Hence the value changes by at most $2\omega_n(\eta)$ per round, and \eqref{eq:grid-value-error} holds for $U^-_T$ with the same $c$ and $Q$. Since $\max_yg(y)\le v\iff\forall y\,[g(y)\le v]$ and $\min_xh(x)\le v\iff\exists x\,[h(x)\le v]$, the machine uses the prefix $\forall y_0\exists x_0\cdots\forall y_{T-2}\exists x_{T-2}$: at each round it guesses $y_t$ universally and then $x_t$ existentially. A malformed universal guess accepts and a malformed existential guess rejects. Correctness ($|z-V^-_{\mathsf{av}}(\distro)|\le e/2<\varepsilon_0$) and the polynomial bounds are unchanged.
\end{proof}
\subsection{Sampled-profile semantics}
\label{app:sp}
By \cref{thm:sp-determined}, the sampled-profile upper and lower values
coincide, so the polarity may be omitted from
$\GapVal_{\mathsf{sp},\varepsilon_0}$ and $V_{\mathsf{sp}}$. 

\spgapupperbound*

Under sampled profile semantics, a sampled profile mixture 
has many successor distributions, and its value is the average of their 
 continuation values, not the value of their mean, so it cannot be
evaluated the way it is under averaged-distribution semantics. We retain
these different successors but evaluate them one at a time.

Deciding $\GapVal_{\mathsf{sp},\varepsilon_0}$ comes down to computing
$V_{\mathsf{sp}}(\distro)$ to enough accuracy to compare it with $v$. In
Appendix \ref{app:sec:sp-background}, we show that $V_{\mathsf{sp}}$ satisfies a recursion in
which each step solves a matrix game $\operatorname{val}(M)$ over the pure
profile sets $\mathcal P_{\min},\mathcal P_{\max}$, and these sets can
have up to $A^n$ elements: writing down even a single mixed strategy over
them, or a single profile as an explicit vector, would already cost
exponential space. We describe below the steps that help overcome this :
\begin{enumerate}[(i)]
\item Appendix \ref{app:sec:sparse-lemma} removes the need for a full mixed
strategy ranging over all of $\mathcal P_{\min}$ or $\mathcal P_{\max}$:
it shows a near-optimal choice can always be taken to be a
\emph{uniform} mixture of only polynomially many pure profiles, so
solving $\operatorname{val}(M)$ approximately becomes a search over
polynomially many candidates instead of an exponential simplex.

\item Appendix \ref{app:sec:compact-profile} gives each pure profile an
$O(n\log(2A))$-bit description, so listing the polynomially many
profiles from (i) still costs only polynomially many bits, not
exponentially many.
\item Appendix \ref{app:sec:sp-main} puts the two together: it evaluates the
finite-horizon recursion for $V_{\mathsf{sp}}$ depth first, at each step
searching only over the polynomially many candidate mixtures from (i),
each written down using the compact profile encoding from (ii). It
proves a uniform bound on the rounding and truncation error this
introduces, and it bounds the space of the whole computation by showing
that only the current path of recursive calls needs to be held in
memory at once, not the entire recursion tree. The result is a
deterministic polynomial-space procedure that decides
$\GapVal_{\mathsf{sp},\varepsilon_0}$.
\end{enumerate}

\subsubsection{Background}
\label{app:sec:sp-background}
Recall that $n=|\States|$ and $A=\max_{s\in\States}|\Acts(s)|$, and fix a
rational discount $\Discount\in(0,1)$.

Under sampled-profile semantics, \cref{thm:sp-determined} shows that the
upper and lower values coincide; write $V_{\mathsf{sp}}:\DistS\to[0,\log_2n]$
for this common value (Step~2 of that proof). Its range follows because
$V_{\mathsf{sp}}$ is a discounted convex combination of entropy values, each
lying in $[0,\log_2n]$ by \cref{prop:entropy-props}(i). Step~1 of the same
proof derives the one-step identity \eqref{eq:sampled-one-step-value} for
$V_{\mathsf{sp}}$, which is exactly the statement that $V_{\mathsf{sp}}$ is a
fixed point of the operator $\mathcal T_{\mathsf{sp}}$ defined below by that
same formula; since $\mathcal T_{\mathsf{sp}}$ is a $\Discount$-contraction,
$V_{\mathsf{sp}}$ is in fact its unique fixed point, its \emph{Bellman
operator}:
\begin{equation}
(\mathcal T_{\mathsf{sp}}V)(\mu)
=(1-\Discount)\Ent(\mu)
+\Discount\cdot\operatorname{val}\bigl(M_V(\mu)\bigr),
\qquad
M_V(\mu)_{a,b}\defeq V(\Phi_{a,b}(\mu)).
\label{eq:Tsp-def}
\end{equation}

where, for a matrix $M\in\Real^{\mathcal P_{\min}\times\mathcal P_{\max}}$,
$\operatorname{val}(M)\defeq
\max_{\tau\in\Dist(\mathcal P_{\max})}\min_{\sigma\in\Dist(\mathcal P_{\min})}
\sigma^{\mathsf T}M\tau$
is the value of the associated zero-sum matrix game (\MinPl minimising,
\MaxPl maximising), well defined by von Neumann's minimax theorem
\cite{vonneumann1928theorie,sion1958general}  since $\mathcal P_{\min},\mathcal
P_{\max}$ are finite. Here $\sigma$ and $\tau$ play the same role as the
sampling strategies $\sigma_t,\tau_t$ from the definition of sampled-profile
semantics (see Appendix \ref{app:update-sem}).
$\sigma^{\mathsf T}M\tau=\sum_{a,b}\sigma(a)\tau(b)\,M(a,b)
=\mathbb E_{a\sim\sigma,\,b\sim\tau}[M(a,b)]$ is exactly the expected
continuation value under the same sampling distribution $\sigma\times\tau$
from which the round's profile $(a_t,b_t)$ is drawn and then executed
(\cref{eq:sp-update}); $\operatorname{val}(M)$ asks what
expected value each player can guarantee by choosing that sampling
distribution optimally, \emph{before} the profile is actually sampled.

This $\mathcal T_{\mathsf{sp}}$ specialises, at
$\Sem=\mathsf{sp}$, the general one-step identity
\eqref{eq:sampled-one-step-value} from Step~1 of \cref{thm:sp-determined}'s
proof,
$$
V_{\Sem}(\distr)=(1-\Discount)\Ent(\distr)+\Discount\operatorname{val}
\bigl(M^{\Sem}(\distr)\bigr),
$$
where $M^{\mathsf{sp}}(\distr)_{a,b}=V_{\mathsf{sp}}(\Phi_{a,b}(\distr))$ is the
sampled-profile case of \eqref{eq:sampled-value-matrix} there. This is the
same formula, evaluated at the same function, as $M_V(\distr)_{a,b}\defeq
V(\Phi_{a,b}(\distr))$ above with $V=V_{\mathsf{sp}}$ substituted, so
\begin{equation}
\label{eq:M-identification}
M_{V_{\mathsf{sp}}}(\distr)=M^{\mathsf{sp}}(\distr)\qquad\text{for every }\distr\in\DistS.
\end{equation}
Substituting $V=V_{\mathsf{sp}}$ into the definition of $\mathcal T_{\mathsf{sp}}$
and applying \eqref{eq:M-identification} gives, for every $\distr$,
$$
(\mathcal T_{\mathsf{sp}}V_{\mathsf{sp}})(\distr)
=(1-\Discount)\Ent(\distr)+\Discount\operatorname{val}\bigl(M_{V_{\mathsf{sp}}}(\distr)\bigr)
\overset{\eqref{eq:M-identification}}{=}
(1-\Discount)\Ent(\distr)+\Discount\operatorname{val}\bigl(M^{\mathsf{sp}}(\distr)\bigr),
$$
and the right-hand side here is exactly the right-hand side of
\eqref{eq:sampled-one-step-value} at $\Sem=\mathsf{sp}$, hence equals
$V_{\mathsf{sp}}(\distr)$. So $(\mathcal T_{\mathsf{sp}}V_{\mathsf{sp}})(\distr)
=V_{\mathsf{sp}}(\distr)$ for every $\distr\in\DistS$, i.e.\
$V_{\mathsf{sp}}=\mathcal T_{\mathsf{sp}}V_{\mathsf{sp}}$: $V_{\mathsf{sp}}$
is a fixed point of $\mathcal T_{\mathsf{sp}}$; since $\mathcal T_{\mathsf{sp}}$
is a $\Discount$-contraction ($\Discount<1$) in sup norm on the complete space
of bounded functions $\DistS\to\Real$, this fixed point is automatically
unique.

Recall $\operatorname{val}(M)=\max_\tau\min_\sigma\sigma^{\mathsf T}M\tau$
from its definition above; the minimax theorem invoked there
allows opposite order of computation as well,
$\operatorname{val}(M)=\min_\sigma\max_\tau\sigma^{\mathsf T}M\tau$. In each
order, the \emph{inner} optimisation is over a variable that enters the
objective only linearly, so it is attained at a vertex of its simplex, i.e.\
at a pure strategy: fixing $\tau$ in the first order,
$\sigma^{\mathsf T}M\tau=\sum_{a}\sigma(a)(M\tau)_a$ is linear in $\sigma$, so
$\min_\sigma\sigma^{\mathsf T}M\tau=\min_{a}(M\tau)_a$; fixing $\sigma$ in the
second order, $\sigma^{\mathsf T}M\tau=\sum_{b}(\sigma^{\mathsf T}M)_b\,\tau(b)$
is linear in $\tau$, so $\max_\tau\sigma^{\mathsf T}M\tau=\max_{b}(\sigma^{\mathsf
T}M)_b$. This yields the two equivalent \emph{one-sided} forms of
$\operatorname{val}$: only one player's strategy remains
mixed, the other having collapsed to a pure-strategy optimum
\begin{equation}
\label{eq:val-two-forms}
\operatorname{val}(M)
=\max_{\tau\in\Dist(\mathcal P_{\max})}\ \min_{a\in\mathcal P_{\min}}(M\tau)_a
=\min_{\sigma\in\Dist(\mathcal P_{\min})}\ \max_{b\in\mathcal P_{\max}}(\sigma^{\mathsf T}M)_b .
\end{equation}

Having recalled this background, we now turn to the problem of interest, 
$\GapVal_{\mathsf{sp},\varepsilon_0}$. 

\subsubsection{A Sparse Certificate Lemma for Zero-Sum Matrix Games}
\label{app:sec:sparse-lemma}
Direct evaluation of $\operatorname{val}(M)$ via \eqref{eq:val-two-forms}
requires quantifying over a full mixed strategy in $\Dist(\mathcal P_{\min})$
or $\Dist(\mathcal P_{\max})$: simplices whose vertex count can be as large
as $A^n$ (Appendix \ref{app:sec:sp-background}). So representing even a single such
mixed strategy would already cost exponentially many variables, incompatible
with the space bound targeted here. \Cref{lem:sparse-minimax}(ii) removes
this obstacle: for any target accuracy $\varepsilon>0$, a mixed strategy
that is optimal to within $\varepsilon$ can always be taken \emph{sparse},
a uniform mixture of only $k=\mathrm{poly}(n,\log A,1/\varepsilon)$ pure
strategies, which is what lets Appendix \ref{app:sec:sp-main} below quantify over
a near-optimal strategy compactly.

Only one player's
strategy is ever sparsified in a single application of \Cref{lem:sparse-minimax} : as the
one-sided forms \eqref{eq:val-two-forms} show, whichever player's
strategy is treated as the \emph{outer} optimization there is the one
that may need full, possibly exponential, support, while the other
player's best response, the \emph{inner} optimization, is attained
exactly at a single pure strategy with no approximation at all.
Appendix \ref{app:sec:sp-main} below always sparsifies \MinPl's strategy this
way, so \MaxPl's response is quantified exactly over the whole of
$\mathcal P_{\max}$, via the compact encoding of
Appendix \ref{app:sec:compact-profile}, rather than approximated.

Part~(ii) of \cref{lem:sparse-minimax} (\emph{existence}) is due to
Lipton and Young \cite{lipton1994simple} and, independently, Alth\"ofer
\cite{althofer1994sparse}: a sparse $\varepsilon$-optimal strategy of the
stated support size always exists. Part~(i) (\emph{exact soundness}), by
contrast, is not due to either source; it is the elementary observation that
any sparse strategy, optimal or not, already certifies a matching one-sided
bound on $\operatorname{val}(M)$ directly from \eqref{eq:val-two-forms}.
Both parts are stated together because the proof of \cref{thm:sp-pspace}
uses both directions.

\sparseminimax*
\begin{proof}
(i) Set $\tau\defeq\sum_{i=1}^kw_i\delta_{b_i}$, where $\delta_{b_i}$ denotes
the point mass at $b_i\in[C]$. Since $w_i\ge0$ for every $i$ and
$\sum_iw_i=1$, $\tau\in\Dist([C])$:
it assigns each $b\in[C]$ the nonnegative weight $\tau(b)=\sum_{i:b_i=b}w_i$,
and these weights sum to $1$; the $b_i$ need not be distinct, and if some
value of $b$ never occurs among $b_1,\dots,b_k$ it simply gets $\tau(b)=0$.
For any $a\in[R]$,
\[
(M\tau)_a=\sum_{b\in[C]}M(a,b)\,\tau(b)=\sum_{i=1}^kM(a,b_i)\,w_i,
\]
where the second equality regroups the sum over $b$ into the (possibly
repeated) list $b_1,\dots,b_k$, each contributing its weight $w_i$,
the quantity $\sum_{i=1}^kw_i\,M(a,b_i)$ appearing in the hypothesis. By
hypothesis this is $\ge v$ for \emph{every} $a\in[R]$, so
\[
\min_{a\in[R]}(M\tau)_a\ \ge\ v.
\]
Now, $\tau$ is one particular element of $\Dist([C])$, not
necessarily the maximizer. So it can only be a feasible competitor in the
outer maximization over $\tau$ in the first expression of
\eqref{eq:val-two-forms}: taking the maximum over \emph{all} of
$\Dist([C])$ can only match or exceed the value at this
particular $\tau$,
\[
\operatorname{val}(M)
\ \overset{\eqref{eq:val-two-forms}}{=}\ \max_{\tau'\in\Dist([C])}\min_{a\in[R]}(M\tau')_a
\ \ge\ \min_{a\in[R]}(M\tau)_a
\ \ge\ v.
\]
This proves the first half of (i). For the symmetric half, set
$\sigma\defeq\sum_{i=1}^ku_i\delta_{a_i}\in\Dist([R])$, again a
valid mixed strategy since $u_i\ge0$ and $\sum_iu_i=1$. For any
$b\in[C]$,
\[
(\sigma^{\mathsf T}M)_b=\sum_{a\in[R]}\sigma(a)\,M(a,b)=\sum_{i=1}^ku_i\,M(a_i,b)\ \le\ v
\]
by hypothesis, for every $b\in[C]$, so $\max_{b\in[C]}(\sigma^{\mathsf
T}M)_b\le v$. This time $\sigma$ is a feasible competitor in the outer
\emph{minimisation} over $\sigma$ in the second expression of
\eqref{eq:val-two-forms}, so taking the minimum over all of
$\Dist([R])$ can only match or fall below the value at this
particular $\sigma$,
\[
\operatorname{val}(M)
\ \overset{\eqref{eq:val-two-forms}}{=}\ \min_{\sigma'\in\Dist([R])}\max_{b\in[C]}(\sigma'^{\mathsf T}M)_b
\ \le\ \max_{b\in[C]}(\sigma^{\mathsf T}M)_b
\ \le\ v,
\]
proving the second half of (i). 

(ii) Let $\tau^\ast\in\Dist([C])$ attain
$\min_a(M\tau^\ast)_a=\operatorname{val}(M)\ge v+\epsilon$ (such a $\tau^\ast$
exists by compactness of $\Dist([C])$ and continuity of
$\tau\mapsto\min_a(M\tau)_a$; the inner minimum over $a$ is attained because
$[R]$ is finite). Therefore, $(M\tau^\ast)_a \ge \min_a(M\tau^\ast)_a \ge v+\epsilon$.

If $U=0$, then $M\in[0,0]^{R\times C}$, i.e.\ $M(a,b)=0$ for
every $a\in[R],b\in[C]$; consequently $(M\tau)_a=\sum_bM(a,b)\tau(b)=0$ for
every $\tau\in\Dist([C])$ and every $a\in[R]$, so
$\operatorname{val}(M)=\max_\tau\min_a(M\tau)_a=0$. The hypothesis
$\operatorname{val}(M)\ge v+\epsilon$ therefore forces $0\ge v+\epsilon$,
i.e.\ $v\le-\epsilon<0$. Taking $k=1$ (matching the general formula for $k$,
which evaluates to $\max\{1,\lceil0\rceil\}=1$ when $U=0$), any $b_1\in[C]$,
and $w_1=1$: since $M\equiv0$, $\sum_{i=1}^1w_i\,M(a,b_1)=M(a,b_1)=0$ for
every $a\in[R]$, and $0\ge v$ holds because $v<0$. So this choice already
satisfies the required inequality for every $a$, and no probabilistic
argument is needed to get it; since $M\equiv0$, the sampled value
$M(a,b_i)$ is not actually random at all, whichever $b_i$ is drawn, it is
the constant $0$.

Assume $U>0$. We prove existence by the probabilistic method: a random
empirical law succeeds with positive probability, so a fixed successful
tuple exists. The deterministic algorithm can enumerate all such tuples;
it does not need to sample from, or already know, an optimal law.
Draw $b_1,\dots,b_k\in[C]$ independently, each according to
$\tau^\ast$: that is, $\Pr[b_i=b]=\tau^\ast(b)$ for every $b\in[C]$, and the
$k$ draws $i=1,\dots,k$ do not affect one another, and set $w_i=1/k$ for
every $i$, so
$$
\hat\tau\defeq\frac1k\sum_{i=1}^k\delta_{b_i}\in\Dist([C])
$$
is the empirical distribution of the sample. This is the same kind of point-mass
mixture as the $\tau$ built in the proof of (i) above, except that here the
$b_1,\dots,b_k$ are random: the outcome of the $k$ draws, so
$\hat\tau$, and consequently $(M\hat\tau)_a$ below, is a random object rather
than a fixed one. Fix $a\in[R]$.
The variable $(M\hat\tau)_a=\frac1k\sum_iM(a,b_i)$ is an average of $k$
independent random variables, each taking values in $[0,U]$. Each of these
variables $M(a,b_i)$ has the same mean, namely
$$
\mathbb E[M(a,b_i)]=\sum_{b\in[C]}\tau^\ast(b)\,M(a,b)=(M\tau^\ast)_a
\qquad\text{for every }i,
$$
because every $b_i$ is drawn from the same distribution $\tau^\ast$; averaging
$k$ copies of this same mean therefore gives
$$
\mathbb E[(M\hat\tau)_a]=\frac1k\sum_{i=1}^k\mathbb E[M(a,b_i)]=(M\tau^\ast)_a\ \ge\ v+\epsilon,
$$
i.e.\ $(M\hat\tau)_a$ has common mean $(M\tau^\ast)_a$. Hoeffding's inequality
[Theorem 2,\cite{hoeffding1963probability}], in its standard form, bounds a sum
$S=\sum_{i=1}^kX_i$ of independent random variables $X_i\in[\alpha_i,\beta_i]$
by $\Pr[S-\mathbb E(S)\ge t]\le\exp\bigl(-2t^2/\sum_i(\beta_i-\alpha_i)^2\bigr)$
for \emph{every} $t\ge0$: this bound holds simultaneously for every
threshold $t$, so any particular value of $t$ may be substituted into it
freely; which value is useful is dictated only by what we actually want to
bound. Applying this with $X_i=M(a,b_i)\in[0,U]$ for every $i$, so
$\sum_i(\beta_i-\alpha_i)^2=kU^2$: what we want is a bound on the probability
that the \emph{average} $(M\hat\tau)_a=S/k$ deviates from its mean by at
least $\epsilon$, i.e.\ on $\Pr[S/k-\mathbb E[S/k]\ge\epsilon]$. Since $k>0$,
multiplying this inequality through by $k$ does not change which outcomes
satisfy it, so
$$
\left\{\frac Sk-\frac{\mathbb E(S)}k\ge\epsilon\right\}=\{S-\mathbb E(S)\ge k\epsilon\}
$$
Hence $\Pr[S/k-\mathbb E[S/k]\ge\epsilon]=\Pr[S-\mathbb E(S)\ge k\epsilon]$. So we can choose $t=k\epsilon$ to substitute:
doing so in the bound above gives the upper-tail bound
$$\Pr[(M\hat\tau)_a-(M\tau^\ast)_a\ge\epsilon]\le\exp(-2k\epsilon^2/U^2)$$

The mirror-image lower-tail statement, obtained identically by applying the same
bound to $-X_i\in[-U,0]$ (the same range width $U$):

$$
\Pr\bigl[(M\hat\tau)_a<(M\tau^\ast)_a-\epsilon\bigr]\le\exp\!\left(-\frac{2k\epsilon^2}{U^2}\right)
$$

Since $(M\tau^\ast)_a\ge v+\epsilon$, i.e.\ $(M\tau^\ast)_a-\epsilon\ge v$, the
event $(M\hat\tau)_a<v$ forces $(M\hat\tau)_a<(M\tau^\ast)_a-\epsilon$. So $\{(M\hat\tau)_a<v\}\subseteq
\{(M\hat\tau)_a<(M\tau^\ast)_a-\epsilon\}$, and hence, by the Hoeffding bound
above,
$$
\Pr\bigl[(M\hat\tau)_a<v\bigr]\ \le\ \Pr\bigl[(M\hat\tau)_a<(M\tau^\ast)_a-\epsilon\bigr]
\ \le\ \exp\!\left(-\frac{2k\epsilon^2}{U^2}\right).
$$
Taking a union bound over the $R$ values of $a$,
$$
\Pr\bigl[\exists a:(M\hat\tau)_a<v\bigr]
\ \le\ \sum_{a\in[R]}\Pr\bigl[(M\hat\tau)_a<v\bigr]
\ \le\ R\exp\!\left(-\frac{2k\epsilon^2}{U^2}\right)<1
$$
for the stated $k$: since $U>0$ here,
$k=\max\{1,\lceil2U^2\ln(2R)/\epsilon^2\rceil\}=\lceil2U^2\ln(2R)/\epsilon^2\rceil$
(the maximum with $1$ is automatic, since $2U^2\ln(2R)/\epsilon^2>0$ and a
ceiling of a positive number is always $\ge1$); a ceiling is always at least
the number itself, so $k\ge2U^2\ln(2R)/\epsilon^2$, i.e.\
$2k\epsilon^2/U^2\ge4\ln(2R)$; since $\exp(\cdot)$ is increasing, negating
both sides gives $\exp(-2k\epsilon^2/U^2)\le\exp(-4\ln(2R))$, and multiplying
by the positive constant $R$,
$$
R\exp\!\left(-\frac{2k\epsilon^2}{U^2}\right)\ \le\ R\exp(-4\ln(2R)).
$$
This right-hand side simplifies: since
$\exp(-4\ln(2R))=\exp\bigl(\ln((2R)^{-4})\bigr)=(2R)^{-4}=1/(16R^4)$,
$$
R\exp(-4\ln(2R))=\frac{R}{16R^4}=\frac{1}{16R^3}\ \le\ \frac1{16}\ <\ 1,
$$
using $R\ge1$ (a game always has at least one pure strategy) for the middle
inequality.

Hence with strictly positive probability the
sampled $(b_1,\dots,b_k)$ satisfies $(M\hat\tau)_a\ge v$ for every $a$
simultaneously, so such a tuple exists. Since $w_i=1/k$ for every $i$, this
tuple is exactly a choice of $b_1,\dots,b_k\in[C]$ and weights
$w_1,\dots,w_k\ge0$ summing to $1$ with $\sum_iw_iM(a,b_i)\ge v$ for every
$a\in[R]$, which is what was asserted in part (ii)'s first case. The
symmetric statement is identical, sampling from an optimal $\sigma^\ast$ and
taking a union bound over the $C$ values of $b$. The probability bound
decreases with $k$, so the proof also covers every larger integer $k$.
\end{proof}
Applied here with $U=\log_2n$ (every value arising below lies in
$[0,\log_2n]$, \cref{prop:entropy-props}(i)) and $R,C\le A^n$, so
$\ln(2R),\ln(2C)\le n\ln(2A)$, \cref{lem:sparse-minimax} gives $\epsilon$-optimal
sparse strategies of support size
\begin{equation}
\label{eq:k-def}
k(\epsilon)
=
O\!\left(1+\frac{n(\log n)^2\log(2A)}{\epsilon^2}\right),
\end{equation}
polynomial in $n,\log(2A),1/\epsilon$ despite $R,C$ being exponential in $n$.

\subsubsection{Compact Encoding of a Pure Profile}
\label{app:sec:compact-profile}
A pure profile $a\in\mathcal P_{\min}$ has an
$O(n\log(2A))$-bit description: fix an ordering of each enabled action
set and store one action index per controlled state. Because this
description is short, profiles can be enumerated one at a time, in this
order, without ever writing out or storing the full set $\mathcal
P_{\min}$. If we have no state sets, then we  have one empty profile. A tuple
of $k$ profiles takes $O(kn\log(2A))$ bits, regardless of repetitions.
\subsubsection{The Polynomial-Space Bound}
\label{app:sec:sp-main}
We separate the proof into six steps, each closing one gap between the
abstract value $V_{\mathsf{sp}}$ and an actual computation. Step~1 pins
down what a \emph{finite}-horizon version of $V_{\mathsf{sp}}$ means,
fixes a specific horizon length $T$, and bounds the error from stopping
after these $T$ rounds; every later step works within this fixed
horizon. Step~2 replaces the exact matrix game at each round, which may
need a mixed profile with exponential support to solve exactly, by a
finite empirical choice, using \cref{lem:sparse-minimax}. Step~3 turns
this into arithmetic a machine can actually execute: exact rational
operations combined with rounding to a fixed number of bits. Step~4
tracks how much error the roundings and empirical approximations of
Steps~2-3 can accumulate over the whole recursion, proving a single
uniform bound. Step~5 uses that bound to show that comparing the final
computed value against $v$ correctly decides the promise gap. Step~6
checks that the whole computation, not just its final answer, fits in
polynomial space, by bounding the bit length of every intermediate
quantity and the number of recursive frames active at once.

\begin{proof}[Proof of \cref{thm:sp-pspace}]
Let $n=|\States|$, $A=\max_{s\in\States}|\Acts(s)|$,
$e=\min\{\varepsilon_0,1\}$, and, for $n\ge2$, $L=\lceil\log_2n\rceil$; let
$\ell$ be the total bit length of the binary encoding of the arena, its
rational transition probabilities $\Trans(s,a,s')$, the rational initial
distribution $\distro$, and the rational threshold $v$. For $n=1$ compare
the identically zero value with $v$. Assume $n\ge2$ below.

\medskip\noindent\textbf{Step 1: finite horizons and their semantic meaning.}
The value $V_{\mathsf{sp}}$ is defined over infinitely many rounds, so
before any computation can begin, the infinite horizon has to be
replaced by a finite one, together with a provable bound on how much
accuracy this costs. This step does exactly that: it defines the
finite-horizon value $W_r^{\mathsf{sp}}$, bounds how far it can drift
from the true value $V_{\mathsf{sp}}$ as a function of the number of
rounds kept, and then fixes a specific horizon $T$ small enough to keep
this drift within the overall error budget.

For each integer $r\ge0$, let $W_r^{\mathsf{sp}}(\mu)$ be the value of
the $r$-round truncated game started at $\mu$, in which only the first
$r$ rounds earn a reward: \cref{thm:sp-determined} shows this value is
well defined (the upper and lower values, over strategies depending on
the full observed distribution history, coincide) and (see also the finite-horizon identification in the proof of \cref{thm:bellman}, Appendix \ref{app:bellman-proof}) satisfies
\begin{equation}
\label{eq:sp-horizon-recursion}
\begin{split}
W_0^{\mathsf{sp}}(\mu)&=0,\\
W_r^{\mathsf{sp}}(\mu)&=(1-\Discount)\Ent(\mu)
+\Discount\operatorname{val}\!
\left(\bigl[W_{r-1}^{\mathsf{sp}}(\Phi_{a,b}(\mu))\bigr]_{a,b}\right).
\end{split}
\end{equation}
Thus $W_1^{\mathsf{sp}}=(1-\Discount)\Ent$ needs no action, and $r\ge2$
uses a continuation with $r-1$ remaining rewards. The index $r$ counts
down: it is the number of reward-earning rounds still to come, not the
number already played, so $W_T^{\mathsf{sp}}$, evaluated at the start of
play, is the value of the full $T$-round truncated game, while
$W_0^{\mathsf{sp}}\equiv0$ is reached once no rounds remain. This
distinction matters directly for Step~3 below: the recursive algorithm
there is written in terms of this remaining horizon $r$, decreasing from
$T$ down to the base case $r=0$ at which no further call is made, so the
stack depth of active calls in Step~6's space bound equals the
remaining-horizon count at the point of recursion, not elapsed real
time. Indexing by elapsed time instead would need a separate counter to
detect when to stop.

The Borel Markov saddle policies that \cref{thm:sp-determined} supplies
for each $W_r^{\mathsf{sp}}$ are optimal against all distribution-history
opponents, not just Markov ones. Hence \eqref{eq:sp-horizon-recursion} is
not a recursion in which previous sampled profiles become observable:
its value depends only on the current distribution and the remaining
horizon.

\smallskip 

\emph{Monotonicity and stability of $\operatorname{val}$.}
To bound the truncation error, we first record an elementary stability
property of $\operatorname{val}$: for matrices $M,N$ of the same
dimensions, $|\operatorname{val}(M)-\operatorname{val}(N)|\le
\max_{a,b}|M_{a,b}-N_{a,b}|$, and $\operatorname{val}$ is monotone:
$M\ge N$ entrywise implies $\operatorname{val}(M)\ge\operatorname{val}(N)$.
Both follow directly from the first form of \eqref{eq:val-two-forms},
$\operatorname{val}(M)=\max_\tau\min_a(M\tau)_a$: for monotonicity, if
$M\ge N$ entrywise then $(M\tau)_a\ge(N\tau)_a$ for every $a,\tau$, so
$\min_a(M\tau)_a\ge\min_a(N\tau)_a$ for every $\tau$, and taking the max
over $\tau$ on both sides preserves the inequality; for the stability
bound, write $\Delta=\max_{a,b}|M_{a,b}-N_{a,b}|$, and let $\mathbf 1$
denote the all-ones matrix of the same dimensions as $M,N$, so that
$\Delta\mathbf 1$ is the matrix with every entry equal to $\Delta$. Then
$N-\Delta\mathbf 1\le M\le N+\Delta\mathbf 1$ entrywise, and monotonicity
applied to each side, together with
$\operatorname{val}(N\pm\Delta\mathbf1)=\operatorname{val}(N)\pm\Delta$
(adding a constant to every entry adds the same constant to every
$(M\tau)_a$ and hence to $\operatorname{val}$), gives
$\operatorname{val}(N)-\Delta\le\operatorname{val}(M)\le
\operatorname{val}(N)+\Delta$.
\smallskip 

\emph{Bounding the truncation error.}
By the paragraphs ``Finite-horizon identification'' and ``Infinite-horizon
identification and error bound'' of the proof of \cref{thm:bellman}
(\cref{app:bellman-proof}), $W_r^{\mathsf{sp}}=(\mathcal T_{\mathsf{sp}})^r0$ and
$0\le V_{\mathsf{sp}}-W_r^{\mathsf{sp}}\le\Discount^r\log_2|\States|\le\Discount^rL$
for every $r\ge0$ (equivalently, $V_{\mathsf{sp}}=\mathcal T_{\mathsf{sp}}V_{\mathsf{sp}}$,
the cancellation of the entropy terms, monotonicity and the stability bound above
give $0\le\Delta_r\le\Discount\sup\Delta_{r-1}$ for $\Delta_r=V_{\mathsf{sp}}-W_r^{\mathsf{sp}}$,
with $\Delta_0=V_{\mathsf{sp}}\in[0,L]$). In particular, for every $\mu\in\DistS$,
\begin{equation}
\label{eq:sp-truncation}
0\le V_{\mathsf{sp}}(\mu)-W_T^{\mathsf{sp}}(\mu)
\le\Discount^T L.
\end{equation}

\emph{The horizon $T$.} Since $\Discount\in(0,1)$ and $e\le1$ are fixed and $L\ge1$ (as $n\ge2$), the least positive integer $T$ with $\Discount^TL\le e/4$ exists and can be found by exact rational comparisons for $T=1,2,\ldots$.  With
$c_\Discount\defeq1/\log(1/\Discount)$, the condition is equivalent to
$T\ge c_\Discount\log(4L/e)$, so $T=\lceil c_\Discount\log(4L/e)\rceil\le
c_\Discount\log(4L/e)+1=O(1+\log(L/e))$.

\emph{Range of the finite-horizon values.} For any matrix $M$ with entries in
$[0,U]$ we have $\operatorname{val}(M)\in[0,U]$: by the first form of
\eqref{eq:val-two-forms}, $\operatorname{val}(M)=\max_\tau\min_a(M\tau)_a$, and each
$(M\tau)_a$ is a convex combination of entries in $[0,U]$. Hence, by induction on
$r$ using \eqref{eq:sp-horizon-recursion} and $\Ent\in[0,\log_2n]$
(\cref{prop:entropy-props}(i)), $W_r^{\mathsf{sp}}\in[0,\log_2n]$ for every $r$
(this is also immediate from $W_r^{\mathsf{sp}}$ being a value of a payoff in
$[0,\log_2n]$). Consequently, for every $r\ge1$ and $\mu\in\DistS$ the matrix
$\bigl[W_{r-1}^{\mathsf{sp}}(\Phi_{a,b}(\mu))\bigr]_{a,b}$ has all entries in
$[0,\log_2n]\subseteq[0,L]$, which is exactly the boundedness hypothesis of
\cref{lem:sparse-minimax} (in its part~(ii) the bound controls the deviation of
the empirical average from its mean).

 \medskip\noindent\textbf{Step 2: a finite empirical replacement for the matrix value.}
Step~1 reduced $V_{\mathsf{sp}}(\distro)$ to evaluating $W_T^{\mathsf{sp}}$
via a finite recursion, but each round of that recursion still asks for
$\operatorname{val}(M)$ of a matrix indexed by $\mathcal P_{\min}\times
\mathcal P_{\max}$, sets that can have up to $A^n$ rows and columns:
computing $\operatorname{val}(M)$ exactly would mean optimising over a
mixed strategy with possibly exponential support. This step replaces
that exact computation with a finite empirical one, $s_k(M)$, that
searches only over $k$-tuples of pure profiles; \cref{lem:sparse-minimax}
guarantees this gets within a controlled error $\alpha$ of the true
value  and shows this replacement is itself stable: nearby matrices
produce nearby empirical values, which is exactly what lets Step~4 track
error through the whole recursion later.

Put $\alpha=(1-\Discount)e/8$ and define integers
\begin{equation}
\label{eq:sampled-numeric-parameters}
J_A=1+n\left\lceil\log_2(\max\{2,A\})\right\rceil,
\qquad k=\max\left\{2,\left\lceil\frac{2L^2J_A}{\alpha^2}\right\rceil\right\}.
\end{equation}

 We show that $k$ meets the row-mixture bound
in \cref{lem:sparse-minimax}; it is polynomial in $n,\log(2A),1/e$ for
fixed $\Discount$, and its  construction needs only rational arithmetic and
integer binary lengths.

Since $|\mathcal P_{\max}|\le A^n$ (Appendix \ref{app:sec:sp-background}),
$\ln(2|\mathcal P_{\max}|)\le\ln(2A^n)=\ln2+n\ln A$. If $A\ge2$, then
$\lceil\log_2A\rceil\ge\log_2A=\ln A/\ln2$, so
$n\lceil\log_2A\rceil\ge n\ln A/\ln2>n\ln A$ (as $1/\ln2>1$), giving
$J_A=1+n\lceil\log_2A\rceil>1+n\ln A>\ln2+n\ln A$; if $A=1$, then
$\max\{2,A\}=2$ and $J_A=1+n>\ln2=\ln(2\cdot1^n)$ trivially. In
either case $J_A\ge\ln(2|\mathcal P_{\max}|)$, so
$2L^2J_A/\alpha^2\ge2L^2\ln(2|\mathcal P_{\max}|)/\alpha^2$, and $k$,
being at least $\lceil2L^2J_A/\alpha^2\rceil$, is at least the
row-mixture bound $\max\{1,\lceil2L^2\ln(2|\mathcal P_{\max}|)/\alpha^2\rceil\}$
that the row-player case of \cref{lem:sparse-minimax}(ii) requires
(with $U=L$, $C=|\mathcal P_{\max}|$, $\epsilon=\alpha$).

For the polynomial bound: $L=\lceil\log_2n\rceil=O(\log n)$, and
$J_A=1+n\lceil\log_2(\max\{2,A\})\rceil=O(n\log(2A))$. Since $\Discount$ is fixed and $e=\min\{\varepsilon_0,1\}$,
$\alpha=(1-\Discount)e/8=\Theta(e)$, so
\[
k=\max\Bigl\{2,\Bigl\lceil\frac{2L^2J_A}{\alpha^2}\Bigr\rceil\Bigr\}
=O\Bigl(\frac{(\log n)^2\cdot n\log(2A)}{e^2}\Bigr),
\]
polynomial in $n,\log(2A),1/e$. Both $\Discount$ and $e$ are themselves
fixed rational constants of the promise problem by Appendix \ref{sec:hardness};
the $1/e$ dependence is displayed explicitly to show how $k$ scales for
other fixed choices of the gap $\varepsilon_0$.

For any matrix $M$ with these profile indices, define the finite
optimisation over exactly the uniform-weight empirical laws that
\cref{lem:sparse-minimax}(ii) shows suffice (every weight equal to
$1/k$, profiles allowed to repeat):
\[
s_k(M)=\min_{(a_1,\ldots,a_k)\in\mathcal P_{\min}^k}
\max_{b\in\mathcal P_{\max}}\frac1k\sum_{i=1}^k M_{a_i,b}.
\]

The tuple represents a legitimate mixed strategy for \MinPl,
$\sigma=k^{-1}\sum_i\delta_{a_i}\in\Dist(\mathcal P_{\min})$: the uniform
distribution over the (possibly repeated) profiles $a_1,\ldots,a_k$. Call
any mixed strategy of this form, arising as the uniform distribution of
some size-$k$ sample of pure profiles, \emph{empirical}. Because $\sigma$
ranges over whole profiles rather than independent per-state choices, it
may correlate actions across a player's different controlled states,
exactly as the sampled-profile semantics permits.

 For entries in $[0,L]$  \cref{lem:sparse-minimax} gives
\begin{equation}
\label{eq:empirical-value-bound}
\operatorname{val}(M)\le s_k(M)\le\operatorname{val}(M)+\alpha.
\end{equation}
The first inequality holds because empirical mixed strategies are a
subset of all of $\Dist(\mathcal P_{\min})$. For
$\sigma=k^{-1}\sum_i\delta_{a_i}$ as above,
$(\sigma^{\mathsf T}M)_b=\sum_{a\in\mathcal P_{\min}}\sigma(a)M(a,b)
=\frac1k\sum_{i=1}^kM(a_i,b)$ for every $b$ (the same regrouping used in
the proof of \cref{lem:sparse-minimax}(i)), so
\[
s_k(M)=\min_{(a_1,\ldots,a_k)\in\mathcal P_{\min}^k}\max_b\frac1k\sum_{i=1}^kM(a_i,b)
=\min_{(a_1,\ldots,a_k)\in\mathcal P_{\min}^k}\max_b(\sigma^{\mathsf T}M)_b
=\min_{\sigma\text{ empirical}}\max_b(\sigma^{\mathsf T}M)_b,
\]
where the last minimisation ranges only over the empirical $\sigma$'s of
this uniform form. Restricting the minimisation in the second form of
\eqref{eq:val-two-forms} to this subset of $\Dist(\mathcal P_{\min})$ can
only raise its value (minimising over fewer candidates gives a result at
least as large), giving $s_k(M)\ge\operatorname{val}(M)$.

For the second inequality, apply \cref{lem:sparse-minimax}(ii) (the
row-player case) to $M$ with $\epsilon=\alpha$ and $v=\operatorname{val}(M)+\alpha$:
the hypothesis $\operatorname{val}(M)\le v-\epsilon$ holds with equality,
so the lemma supplies $a_1,\ldots,a_k\in\mathcal P_{\min}$ (repeats
allowed) with equal weights $1/k$ such that $\frac1k\sum_iM(a_i,b)\le v
=\operatorname{val}(M)+\alpha$ for every $b\in\mathcal P_{\max}$
simultaneously, provided $k$ is at least the lemma's general bound
$\lceil2U^2\ln(2|\mathcal P_{\max}|)/\alpha^2\rceil$ specialized to
$U=L$, which is valid since $M$ has entries in $[0,L]$ as shown in
Step 1: this specialization gives $k \ge \lceil2L^2\ln(2|\mathcal
P_{\max}|)/\alpha^2\rceil$. Since $J_A\ge\ln(2|\mathcal P_{\max}|)$, the
$k$ fixed in \eqref{eq:sampled-numeric-parameters} already meets this,
and the lemma's closing remark extends the guarantee to this (possibly
larger) $k$. This tuple is one feasible choice in the minimum defining
$s_k(M)$, so $s_k(M)\le\max_b\frac1k\sum_iM(a_i,b)\le\operatorname{val}(M)+\alpha$.

For any two matrices of the same dimensions, averaging, finite maxima,
and finite minima also give
\begin{equation}
\label{eq:empirical-lipschitz}
|s_k(M)-s_k(N)|\le\max_{a,b}|M_{a,b}-N_{a,b}|.
\end{equation}
To verify this, write $\Delta=\max_{a,b}|M_{a,b}-N_{a,b}|$, so
$N-\Delta\mathbf1\le M\le N+\Delta\mathbf1$ entrywise (with $\mathbf1$ the
all-ones matrix of the same dimensions). Two elementary facts about
$s_k$, proved the same way as for $\operatorname{val}$ above: it is
monotone ($M\ge N$ entrywise implies $s_k(M)\ge s_k(N)$, since for every
fixed tuple and column, $\frac1k\sum_iM_{a_i,b}\ge\frac1k\sum_iN_{a_i,b}$,
so the max over columns and then the min over tuples both preserve the
inequality), and adding a constant to every entry shifts $s_k$ by that
same constant ($s_k(N\pm\Delta\mathbf1)=s_k(N)\pm\Delta$, since shifting
every entry by $\Delta$ shifts every average $\frac1k\sum_iM_{a_i,b}$,
hence its max over $b$, hence the min over tuples, by exactly $\Delta$).
Applying monotonicity to each side of $N-\Delta\mathbf1\le M\le
N+\Delta\mathbf1$, together with these shift identities, gives
$s_k(N)-\Delta\le s_k(M)\le s_k(N)+\Delta$, i.e.\
\[
|s_k(M)-s_k(N)|\le\Delta=\max_{a,b}|M_{a,b}-N_{a,b}|.
\]

 This elementary stability property will carry child
evaluation errors through the recursion.

\medskip\noindent\textbf{Step 3: finite-precision recursive evaluation.}
Steps~1-2 replaced the infinite horizon with a finite one, and the exact
matrix game at each round with a finite empirical search over $s_k$, but
both are still stated in terms of exact real numbers: the entropy reward
$\Ent(\mu)$ is transcendental, and $s_k(M)$ as defined is an exact
minimum and maximum of real-valued sums.

 We now build a numerical recursion $\widehat W_r^{\mathsf{sp}}$, evaluated
entirely in rational arithmetic, that will approximate $W_r^{\mathsf{sp}}$
closely enough for Step~5's promise-correctness decision to be correct. To keep every
intermediate value at a bounded, fixed number of bits, which Step~6
needs for its space bound, every quantity this recursion produces is
rounded to the nearest multiple of a fixed precision $d=2^{-p}$, rather
than kept exact. Choose the least integer $p\ge1$ such that
$d=2^{-p}\le(1-\Discount)e/16$; this specific threshold is not arbitrary,
it is exactly what makes the total rounding error come out within budget
once Step~4 proves the error bound. With $p$ fixed this way, every
recursive call of $\widehat W_r^{\mathsf{sp}}$ will return a value that is
a multiple of $d$ lying in $[0,L]$, and the following rounding operation
is what enforces this at each step.

 Define the rational rounding operation
\[
\mathsf R_p(z)=2^{-p}\min\{L2^p,\max\{0,\lfloor2^p z+1/2\rfloor\}\}.
\]

For every $w\in[0,L]$ it satisfies the \emph{clipping inequality}
\[
|\mathsf R_p(z)-w|\le|z-w|+d:
\]
rounding changes a number by at most $d/2$, and clipping to $[0,L]$
cannot increase its distance from $w$. All comparisons and rounding
operations here are on rationals.

The current entropy $\Ent(\mu)$ itself need not be computed exactly: by
\cref{ass:rational} (effective dyadic evaluation, Part~2), for every
integer $p\ge1$ there is a rational $\widehat H_p(\mu)\in[0,L]$ with
$|\widehat H_p(\mu)-\Ent(\mu)|\le2^{-p}=d$, computable in time polynomial
in $\operatorname{size}(\mu)+p$. We use this $\widehat H_p$ below in place
of $\Ent$.

For rational $\mu$, define $\widehat W_0^{\mathsf{sp}}(\mu)=0$ and
\begin{equation}
\label{eq:sp-numeric-recursion}
\widehat W_r^{\mathsf{sp}}(\mu)=\mathsf R_p\!\left(
(1-\Discount)\widehat H_p(\mu)
+\Discount\min_{\vec a\in\mathcal P_{\min}^k}
\max_{b\in\mathcal P_{\max}}\frac1k\sum_{i=1}^k
\widehat W_{r-1}^{\mathsf{sp}}(\Phi_{a_i,b}(\mu))\right).
\end{equation}

This describes exactly how the nested $\min_{\vec a}\max_b\frac1k\sum_i$
of \eqref{eq:sp-numeric-recursion} is evaluated procedurally.

\begin{enumerate}
\item \emph{The $r=1$ shortcut.} Every child call is
$\widehat W_0^{\mathsf{sp}}(\cdot)\equiv0$ regardless of its argument, so
the whole $\min\max\sum$ term is $0$ without needing to try any tuple or
column at all, and
$\widehat W_1^{\mathsf{sp}}(\mu)=\mathsf R_p((1-\Discount)\widehat H_p(\mu))$
directly.

\item \emph{Outer loop, $r\ge2$.} Try every $k$-tuple
$\vec a=(a_1,\ldots,a_k)\in\mathcal P_{\min}^k$ in turn, enumerating each
of the $k$ coordinates one profile at a time as in
Appendix \ref{app:sec:compact-profile}.

\item \emph{Inner loop.} For each tuple, try every $b\in\mathcal P_{\max}$
in turn, enumerated the same way.

\item \emph{Per-pair computation.} For each fixed pair $(\vec a,b)$,
compute the $k$ child values \\$\widehat W_{r-1}^{\mathsf{sp}}(\Phi_{a_1,b}(\mu)),
\ldots,\widehat W_{r-1}^{\mathsf{sp}}(\Phi_{a_k,b}(\mu))$ one at a time,
via a fresh recursive call each, adding each to a running sum as it is
produced, so only one child value is ever held in memory at once, never
all $k$ simultaneously. The completed sum gives
$\frac1k\sum_i\widehat W_{r-1}^{\mathsf{sp}}(\Phi_{a_i,b}(\mu))$ for this
$(\vec a,b)$.

\item \emph{Max/min bookkeeping.} As each column $b$ finishes, compare
its value against a running maximum kept for the current tuple, updating
it if larger. Once every $b$ has been tried, this maximum is the value
of the inner $\max_b$ for the current $\vec a$; compare it against a
running minimum kept across tuples, updating it if smaller.

\item \emph{No caching.} If the same subproblem $(r-1,\Phi_{a_i,b}(\mu))$
arises again, whether for a different $i$, a different $b$, or a
different tuple entirely, simply recompute it from scratch by another
recursive call rather than looking it up in a table. This is what keeps
the space bound in Step~6 polynomial despite the number of distinct
subproblems being potentially exponential: memory only ever holds the
current path of nested calls, never a record of every subproblem
visited.

\item \emph{Termination.} $\mathcal P_{\min}^k$ and $\mathcal P_{\max}$
are both finite, every loop over them is bounded, and every recursive
call is made at strictly smaller $r$, so the recursion bottoms out after
at most $T$ levels.
\end{enumerate}

This computes a value depending only on $(r,\mu)$, in line with the
Markov-sufficiency recorded in Step~1 (proved in \cref{thm:sp-determined}).
Its result is the same every time the same subproblem
$(r,\mu)$ is requested, which is exactly what makes recomputing it,
rather than caching it, harmless for correctness.

\medskip\noindent\textbf{Step 4: a uniform two-sided error invariant.}
Step~3 built a numerical recursion $\widehat W_r^{\mathsf{sp}}$ that a
machine can actually execute, but executing it is not enough on its own:
every round introduces two new sources of error, from replacing
$\operatorname{val}$ by $s_k$ (up to $\alpha$) and from rounding to
precision $d$, and nothing so far bounds how these errors accumulate
over $T$ rounds of recursion. This step proves they do not compound
uncontrollably: it establishes a single, uniform bound on
$|\widehat W_r^{\mathsf{sp}}(\mu)-W_r^{\mathsf{sp}}(\mu)|$ that holds for
every round and every distribution, which is exactly what Step~5 needs
to know the final computed value is close enough to the true value to
decide the promise correctly.

We prove for every rational $\mu\in\DistS$ and $r\ge0$ that
\begin{equation}
\label{eq:sampled-numeric-error}
|\widehat W_r^{\mathsf{sp}}(\mu)-W_r^{\mathsf{sp}}(\mu)|
\le B_r,\qquad
B_0=0,\quad B_r=\Discount B_{r-1}+\Discount\alpha+2d.
\end{equation}

The base case is exact. For the inductive step let
$M_{a,b}=W_{r-1}^{\mathsf{sp}}(\Phi_{a,b}(\mu))$ and
$\widehat M_{a,b}=\widehat W_{r-1}^{\mathsf{sp}}(\Phi_{a,b}(\mu))$.
Successors are rational, so the induction hypothesis bounds each
entrywise difference by $B_{r-1}$: $\max_{a,b}|\widehat M_{a,b}-M_{a,b}| = \max_{a,b}|\widehat W_{r-1}^{\mathsf{sp}}(\Phi_{a,b}(\mu)) - W_{r-1}^{\mathsf{sp}}(\Phi_{a,b}(\mu))|
\le B_{r-1}$. By \eqref{eq:empirical-lipschitz},
\[
|s_k(\widehat M)-s_k(M)|\le\max_{a,b}|\widehat M_{a,b}-M_{a,b}|\le B_{r-1},
\qquad\text{i.e.}\qquad
s_k(M)-B_{r-1}\le s_k(\widehat M)\le s_k(M)+B_{r-1}.
\]
By \eqref{eq:empirical-value-bound}, $\operatorname{val}(M)\le
s_k(M)\le\operatorname{val}(M)+\alpha$. Combining the two: for the upper
bound, $s_k(\widehat M)\le s_k(M)+B_{r-1}\le\operatorname{val}(M)+\alpha
+B_{r-1}$; for the lower bound, $s_k(\widehat M)\ge s_k(M)-B_{r-1}\ge
\operatorname{val}(M)-B_{r-1}$ (using $s_k(M)\ge\operatorname{val}(M)$).
Together,
\[
\operatorname{val}(M)-B_{r-1}
\le s_k(\widehat M)
\le\operatorname{val}(M)+\alpha+B_{r-1}.
\]

This supplies both directions of the comparison: the algorithm built in
Step~3 cannot underestimate a continuation by more than $B_{r-1}$, or
overestimate it by more than $B_{r-1}+\alpha$. Write
$z=(1-\Discount)\widehat H_p(\mu)+\Discount\, s_k(\widehat M)$ for the
argument passed to the rounding operation, and recall
$W_r^{\mathsf{sp}}(\mu)=(1-\Discount)\Ent(\mu)+\Discount\operatorname{val}(M)$.
Subtracting,
\[
z-W_r^{\mathsf{sp}}(\mu)=(1-\Discount)\bigl[\widehat H_p(\mu)-\Ent(\mu)\bigr]
+\Discount\bigl[s_k(\widehat M)-\operatorname{val}(M)\bigr].
\]
The first bracket is at most $d$ in absolute value, since
$\widehat H_p$ approximates $\Ent$ to error $2^{-p}=d$
(\cref{ass:rational}, Part~2, as used in Step~3); the second bracket
lies in $[-B_{r-1},\alpha+B_{r-1}]$ by the two-sided bound just proved.
Since $W_r^{\mathsf{sp}}(\mu)\in[0,\log_2n]\subseteq[0,L]$ (Step~1), the
clipping inequality applies with $w=W_r^{\mathsf{sp}}(\mu)$, giving
\[
\bigl|\widehat W_r^{\mathsf{sp}}(\mu)-W_r^{\mathsf{sp}}(\mu)\bigr|
=\bigl|\mathsf R_p(z)-W_r^{\mathsf{sp}}(\mu)\bigr|
\le\bigl|z-W_r^{\mathsf{sp}}(\mu)\bigr|+d.
\]
Bounding $|z-W_r^{\mathsf{sp}}(\mu)|$ by the worse of the two directions,
$(1-\Discount)d+\Discount(B_{r-1}+\alpha)$, and adding the extra $d$ from
the clipping inequality gives
\[
\bigl|\widehat W_r^{\mathsf{sp}}(\mu)-W_r^{\mathsf{sp}}(\mu)\bigr|
\le\Discount B_{r-1}+\Discount\alpha+(1-\Discount)d+d
\le\Discount B_{r-1}+\Discount\alpha+2d=B_r,
\]
proving the induction.

The recurrence $B_0=0$, $B_r=\Discount B_{r-1}+\Discount\alpha+2d$ has the
explicit solution
\[
B_r=(\Discount\alpha+2d)\sum_{j=0}^{r-1}\Discount^j,
\]
verified by induction: at $r=0$ the sum is empty, giving $B_0=0$; if it
holds at $r-1$, then
$B_r=\Discount B_{r-1}+\Discount\alpha+2d=\Discount(\Discount\alpha+2d)
\sum_{j=0}^{r-2}\Discount^j+(\Discount\alpha+2d)
=(\Discount\alpha+2d)\bigl[\sum_{j=1}^{r-1}\Discount^j+1\bigr]
=(\Discount\alpha+2d)\sum_{j=0}^{r-1}\Discount^j$. Since all terms of the
infinite geometric series are positive, truncating it at $r-1$ terms
only decreases it, so
\[
B_r\le(\Discount\alpha+2d)\sum_{j=0}^{\infty}\Discount^j
=\frac{\Discount\alpha+2d}{1-\Discount}.
\]
Recall $\alpha=(1-\Discount)e/8$ from Step~2 and $d\le(1-\Discount)e/16$
from the choice of $p$ in Step~3. Substituting,
\[
\frac{\Discount\alpha+2d}{1-\Discount}
\le\frac{\Discount\frac{(1-\Discount)e}{8}+2\frac{(1-\Discount)e}{16}}{1-\Discount}
=\frac{\frac{(1-\Discount)e}{8}(\Discount+1)}{1-\Discount}
=\frac{(1+\Discount)e}{8}.
\]
Since $\Discount\in(0,1)$, $1+\Discount<2$, so
$\frac{(1+\Discount)e}{8}<\frac{2e}{8}=e/4$. Hence $B_r\le e/4$ for every
$r$, in particular $r=T$.

The bound is uniform in the initial distribution and the enumerated
profiles: nowhere in its derivation did $\mu$, $\nu$, or the specific
tuples chosen enter, only $r,\alpha,d,\Discount$. It also does not grow
with $k$.

\medskip\noindent\textbf{Step 5: Correctness of the decision procedure.}
Steps~1 and~4 each bounded one source of error: Step~1 bounded the cost
of truncating to $T$ rounds, and Step~4 bounded the cost of the
sparsification and rounding used to actually compute a numerical answer.
Neither step, on its own, says anything about whether comparing that
answer to $v$ decides the problem correctly. This step combines the two
error bounds into a single total, and uses the promise gap between
yes-instances and no-instances to show that the resulting comparison is
always correct on promised inputs.

Compute $z=\widehat W_T^{\mathsf{sp}}(\distro)$ and answer \textsc{Yes}
exactly when $z\le v$. By Step~4 and \eqref{eq:sp-truncation},
\[
|z-V_{\mathsf{sp}}(\distro)|
\le|z-W_T^{\mathsf{sp}}(\distro)|+|W_T^{\mathsf{sp}}(\distro)-V_{\mathsf{sp}}(\distro)|
\le e/4+e/4=e/2.
\]
Since $e=\min\{\varepsilon_0,1\}\le\varepsilon_0$ and $\varepsilon_0>0$,
$e/2\le\varepsilon_0/2<\varepsilon_0$, so $|z-V_{\mathsf{sp}}(\distro)|<\varepsilon_0$.
If $V_{\mathsf{sp}}(\distro)\le v-\varepsilon_0$, then
$z\le V_{\mathsf{sp}}(\distro)+e/2\le v-\varepsilon_0+e/2<v$. If
$V_{\mathsf{sp}}(\distro)\ge v+\varepsilon_0$, then
$z\ge V_{\mathsf{sp}}(\distro)-e/2\ge v+\varepsilon_0-e/2>v$. So the
algorithm answers \textsc{Yes} on every yes-instance and \textsc{No} on
every no-instance of the promise. The comparison $z\le v$ is between
rationals and does not require deciding equality of a transcendental
entropy expression.

\medskip\noindent\textbf{Step 6: exact bit lengths and active workspace.}
Step~5 shows that comparing $z=\widehat W_T^{\mathsf{sp}}(\distro)$ to
$v$ decides the promise problem correctly, but correctness alone does
not place the problem in $\PSPACE$: it says nothing about the resources
used to compute $z$. The algorithm of Step~3 recomputes children when
needed and stores only a single active path, rather than caching every
subproblem, which avoids holding an exponential number of values at
once; but this design choice does not by itself bound anything. Each
active frame still holds rational numbers whose bit length was never
pinned down, and the active path could in principle hold as many frames
as there are levels of recursion, a quantity not yet shown to be small.
This step closes both gaps: it bounds the exact bit length of every
rational number the algorithm ever writes down, and it bounds the
number of frames simultaneously present on the active path, showing
that their product is polynomial in $\ell$ and thereby completing the
proof of \cref{thm:sp-pspace}.

\emph{Bit length of the distribution.}
Let $D$ be the product of the denominators of all input transition
probabilities, and $D_0$ the product of the denominators of the initial
distribution's coordinates. Each individual denominator is a substring
of the $\ell$-bit input encoding, so the sum of their bit lengths is at
most $\ell$. For positive integers $a,b$ with bit lengths $p,q$,
$ab<2^p\cdot2^q=2^{p+q}$, so the bit length of a product of positive
integers is at most the sum of the bit lengths of the factors; hence
both $D$ and $D_0$ have $O(\ell)$ bits. Fix a pure profile pair $(a,b)$.
Each entry $\Trans(s,\rho_{a,b}(s),s')$ of the corresponding transition
matrix is one of the input's own rational transition probabilities, so
its denominator divides $D$ by definition of $D$; since the matrix has
only finitely many entries, all with denominator dividing $D$, the whole
matrix has common denominator dividing $D$. This allows an induction
on $t$: the base case $\mu=\distro$ has common denominator dividing
$D_0$ by definition of $D_0$. For the inductive step, suppose every
coordinate of $\mu$ has the form $m_s/(D_0D^t)$ for an integer $m_s$,
and write $\Trans(s,\rho_{a,b}(s),s')=q_{s,s'}/D$ for an integer
$q_{s,s'}$. By \eqref{eq:phi-def},
\[
\Phi_{a,b}(\mu)(s')
=\sum_{s\in\States}\frac{m_s}{D_0D^t}\cdot\frac{q_{s,s'}}{D}
=\frac{\sum_s m_sq_{s,s'}}{D_0D^{t+1}},
\]
an integer numerator over $D_0D^{t+1}$. So every coordinate of
$\Phi_{a,b}(\mu)$ again has denominator dividing $D_0D^{t+1}$, and after
$t$ updates a distribution has common denominator dividing $D_0D^t$.
Its $n$ coordinates therefore occupy $O(n(t+1)\ell)$ bits. This is a
specific denominator invariant; a polynomial number of arbitrary
rational operations alone would not suffice to prove such a bound.

\emph{Bit length of the recursion's values.}
Bounding the distribution's denominator does not account for every
rational number the algorithm manipulates: it also computes and
combines finite-precision values along the recursion
\eqref{eq:sp-numeric-recursion}. As fixed in Step~3, every recursive
call returns a multiple of $d=2^{-p}$ lying in $[0,L]$; write the
$i$-th child's return value as $j_i/2^p$ for an integer $j_i$ with
$0\le j_i\le L2^p$. For a fixed tuple and response $b$, store the
integer $\sum_i j_i$, between zero and $kL2^p$, then divide by the
common denominator $k2^p$. This takes $O(p+\log k+\log(L+1))$ bits.
Minima and maxima merely select one such rational, without introducing
new denominators. The resulting average, with denominator dividing
$k2^p$, is then combined with $(1-\Discount)\widehat H_p(\mu)$, which
has denominator $2^p$ since $\widehat H_p(\mu)$ does by construction,
and scaled by the fixed rational discount $\Discount$; this multiplies
denominators together at most once, introducing only $\Discount$'s own
fixed denominator on top of $k2^p$, so the pre-rounding value still has
polynomially many bits. Applying $\mathsf R_p$ then resets the return
denominator to exactly $2^p$. Unlike the distribution's denominator
$D_0D^t$ from above, which grows with depth, the value denominator
therefore does not compound across recursive calls.

\emph{The active workspace.}
An active frame stores the current distribution, remaining horizon, the
$k$-tuple $\vec a=(a_1,\ldots,a_k)\in\mathcal P_{\min}^k$ currently being
enumerated ($O(kn\log(2A))$ bits, one $O(n\log(2A))$-bit pure profile
per coordinate), the current column $b\in\mathcal P_{\max}$
($O(n\log(2A))$ bits), child index, running sum, and running extrema.
Rational updates, namely computing $\Phi_{a,b}(\mu)$, accumulating the
running sum, and combining the entropy and discount terms above, are
additions, multiplications, and comparisons of integers no longer than
the bit lengths already bounded; elementary arithmetic on $m$-bit
integers uses $O(m)$ auxiliary space, so these updates use space
polynomial in the lengths just established. \Cref{ass:rational} bounds only the time to compute $\widehat H_p(\mu)$,
which is polynomial in $\operatorname{size}(\mu)+p$; since a deterministic
algorithm can write to at most one new memory cell per step, a computation
running in time $t$ uses at most $O(t)$ space, so computing $\widehat H_p$
also uses only polynomial auxiliary space.

A quantity polynomial in lengths that are themselves polynomial in $\ell$ is again
polynomial in $\ell$. Summing the bounds above, a single frame occupies
$O\bigl(n(T+1)\ell+kn\log(2A)+p+\log k+\log(L+1)\bigr)$ bits: the
distribution contributes at most $O(n(T+1)\ell)$ bits since its depth
is at most $T$, the tuple and column contribute $O(kn\log(2A))$ and
$O(n\log(2A))$ bits, and the running sum, extrema, and combined value
contribute $O(p+\log k+\log(L+1))$ bits.

Having bounded the space used by a single frame, it remains to bound
how many frames coexist. On a recursive call the parent frame is
retained, but the completed child's frame is discarded before the next
child is processed. Since the recursion depth strictly decreases from
$r=T$ at the root to $r=0$ at a base case (Step~3), and only the frames
on the current path are retained, at most $T$ nontrivial frames coexist
at any time. The horizon $T$ was shown polynomial in $\ell$ in Step~1,
the sample size $k$ in Step~2, and the precision $p$ is fixed in Step~3
independently of the instance. Total space is therefore at most $T$
times a single frame's polynomial bit length, a product of two
quantities polynomial in $\ell$, hence itself polynomial in $\ell$. The
number of tuples and recomputed subproblems may be exponential; this
affects time, not the workspace bound. 
\end{proof}

\subsection{Sampled-transition semantics}
\label{app:st}
The value is unambiguous by \cref{thm:st-determined}, so the polarity
may be omitted from $\GapVal_{\mathsf{st},\varepsilon_0}$ and $V_{\mathsf{st}}$. Unlike
sampled-profile semantics, a fixed pure-profile pair no longer
determines a single successor distribution, since the arena's own
transitions may realise any of up to $n^n$ ways. We enumerate every
realisation, but sum their contributions into a single accumulator
sequentially, rather than storing their recursive subtrees, so the
number of realisations affects time, not space.

\stgapupperbound*
\begin{proof}
Let $n=|\States|$, $A=\max_{s\in\States}|\Acts(s)|$,
$e=\min\{\varepsilon_0,1\}$, and, for $n\ge2$, $L=\lceil\log_2n\rceil$; let
$\ell$ be the total bit length of the binary encoding of the arena, its
rational transition probabilities $\Trans(s,a,s')$, the rational initial
distribution $\distro$, and the rational threshold $v$. For $n=1$ compare
the identically zero value with $v$. Assume $n\ge2$ below.

Recall from \cref{sec:model} the sampled-transition objects:
$\TransReal=\States^{\States}$, so a realisation $\xi\in\TransReal$
assigns every state a successor state and $|\TransReal|\le n^n$; the
transition-realisation law
$P_{a,b}(\xi)=\prod_{s\in\States}\Trans(s,\rho_{a,b}(s),\xi(s))$ of
\eqref{eq:transition-realization-prob}, where $\rho_{a,b}$ is as in
\eqref{eq:phi-def}; and the realised-successor map
$\Psi_\xi(\mu)(s')=\sum_{s:\xi(s)=s'}\mu(s)$ of \eqref{eq:psi-def}.

By \cref{thm:st-determined}, the value $V_{\mathsf{st}}:\DistS\to[0,\log_2n]$
is the unique fixed point of the $\Discount$-contraction $\mathcal T_{\mathsf{st}}$,
defined by
\begin{equation}
(\mathcal T_{\mathsf{st}}V)(\mu)
=(1-\Discount)\Ent(\mu)+\Discount\cdot\operatorname{val}\bigl(M_V(\mu)\bigr),
\qquad
M_V(\mu)_{a,b}
\defeq\sum_{\xi\in\TransReal}P_{a,b}(\xi)\,V(\Psi_\xi(\mu)),
\label{eq:Tst-def}
\end{equation}
where $\operatorname{val}$ is as in \eqref{eq:val-two-forms}. Unlike
sampled-profile semantics, once $a,b$ are fixed there is no further
strategic choice inside $M_V(\mu)_{a,b}$: it is a fixed expectation over
up to $n^n$ realisations. \Cref{lem:sparse-minimax} reduces the number
of profiles in the outer mixture but does not change this expectation,
so we keep the entire realisation sum and evaluate its terms
sequentially rather than approximate it. Two facts make this possible
in polynomial space: realised successors only sum existing coordinates
of $\mu$, no new mass is created, and every approximate child value
shares the same dyadic denominator, so the realisation sum never
multiplies together the denominators of different realisations.

\medskip\noindent\textbf{Step 1: the exact finite-horizon recursion.}
Choose the same positive $T$ as in Appendix \ref{app:sec:sp-main}, the least
with $\Discount^TL\le e/4$. For $r$ remaining rewards,
\cref{thm:st-determined} establishes
\begin{equation}
\label{eq:st-horizon-recursion}
\begin{split}
W_0^{\mathsf{st}}&=0,\\
W_r^{\mathsf{st}}(\mu)&=(1-\Discount)\Ent(\mu)
+\Discount\operatorname{val}\!\left(
\left[\sum_{\xi\in\TransReal}P_{a,b}(\xi)
W_{r-1}^{\mathsf{st}}(\Psi_\xi(\mu))\right]_{a,b}\right).
\end{split}
\end{equation}
In particular, $W_1^{\mathsf{st}}=(1-\Discount)\Ent$ contains no
outgoing transition. \Cref{thm:st-determined} gives Borel Markov saddle
policies for these values against all distribution-history opponents,
so \eqref{eq:st-horizon-recursion} is not a game in which earlier
profiles or realisations become observable: its value depends only on
the current distribution and remaining horizon, and evaluating the same
subproblem twice does not license different information for either
player.

The finite-horizon identification $W_r^{\mathsf{st}}=(\mathcal T_{\mathsf{st}})^r0$
is also proved in the proof of \cref{thm:bellman} (Appendix \ref{app:bellman-proof},
paragraph ``Finite-horizon identification''). Every entry
$\sum_\xi P_{a,b}(\xi)W_{r-1}^{\mathsf{st}}(\Psi_\xi(\mu))$ of the continuation
matrix is a convex combination of values in $[0,\log_2n]$ (the $P_{a,b}(\xi)$ are
nonnegative and sum to $1$), so the entries lie in $[0,\log_2n]\subseteq[0,L]$ by
induction on $r$ and the range fact $\operatorname{val}(M)\in[0,U]$ for
$M\in[0,U]^{R\times C}$ of Appendix \ref{app:sec:sp-main} (Step~1). As in the
sampled-profile case, this is the boundedness hypothesis of
\cref{lem:sparse-minimax}.
The truncation bound
\begin{equation}
\label{eq:st-truncation}
0\le V_{\mathsf{st}}(\mu)-W_T^{\mathsf{st}}(\mu)\le\Discount^TL\le e/4
\qquad\text{for every }\mu\in\DistS
\end{equation}
is the error bound $0\le V_{\mathsf{st}}-W_r^{\mathsf{st}}\le\Discount^r\log_2|\States|
\le\Discount^rL$ of that same proof at $r=T$ (the argument is the one given for
$\mathsf{sp}$ in Appendix \ref{app:sec:sp-main}, Step~1, which uses only
monotonicity and stability of $\operatorname{val}$ and the shape of the recursion).

\medskip\noindent\textbf{Step 2: sparse profiles and exact weighted child sums.}
Take $\alpha,k$ from Appendix \ref{app:sec:sp-main}'s Step~2 and $p,d,\mathsf R_p$
from its Step~3. Set $\widehat W_0^{\mathsf{st}}=0$. For rational
$\mu$, first define the numerically evaluated entry
\begin{equation}
\label{eq:st-numeric-entry}
\widehat C_{r-1}(\mu)_{a,b}
=\sum_{\xi\in\TransReal}P_{a,b}(\xi)
\widehat W_{r-1}^{\mathsf{st}}(\Psi_\xi(\mu)),
\end{equation}
and then the returned value
\begin{equation}
\label{eq:st-numeric-recursion}
\widehat W_r^{\mathsf{st}}(\mu)=\mathsf R_p\!\left(
(1-\Discount)\widehat H_p(\mu)
+\Discount\min_{\vec a\in\mathcal P_{\min}^k}
\max_{b\in\mathcal P_{\max}}\frac1k\sum_i\widehat C_{r-1}(\mu)_{a_i,b}
\right).
\end{equation}
For $r=1$ the continuation is zero and all action and realisation loops
are omitted. Otherwise enumerate tuples and responses as in
Appendix \ref{app:sec:sp-main}'s Step~3. For each $a_i,b$, enumerate all maps
$\xi:\States\to\States$, compute the exact product $P_{a_i,b}(\xi)$,
recursively evaluate the one child $\Psi_\xi(\mu)$, add its weighted
value to a running sum, and discard that child's workspace once its
contribution is added. Zero-weight realisations may be skipped. Sum the
entries for all $i$, then update the maximum for the current tuple and
the outer minimum.

The order $\min_{\vec a}\max_b\sum_i\sum_\xi$ is retained throughout:
there is no maximisation over realisations and no existential choice of
representative samples, since conditional on the profiles, their
probabilities are exactly the arena's prescribed law. Each call depends
only on its rational distribution and remaining horizon, not on the
private profile or realisation history leading to it, matching Step~1's
Markov-sufficiency point.

\medskip\noindent\textbf{Step 3: a uniform two-sided error invariant.}
We prove for every rational $\mu$ and $r\ge0$ that
\[
|\widehat W_r^{\mathsf{st}}(\mu)-W_r^{\mathsf{st}}(\mu)|\le B_r,
\qquad B_0=0,\quad B_r=\Discount B_{r-1}+\Discount\alpha+2d,
\]
the same recurrence, and hence the same closed form
$B_r\le(\Discount\alpha+2d)/(1-\Discount)\le e/4$, as Appendix \ref{app:sec:sp-main}'s
Step~4. The base case is exact. Assume the bound for $r-1$, and let
$C_{r-1}(\mu)_{a,b}=\sum_\xi P_{a,b}(\xi)W_{r-1}^{\mathsf{st}}(\Psi_\xi(\mu))$
denote the exact (unrounded) counterpart of \eqref{eq:st-numeric-entry}.
For every fixed pair $a,b$,
\begin{align*}
|\widehat C_{r-1}(\mu)_{a,b}-C_{r-1}(\mu)_{a,b}|
&\le\sum_\xi P_{a,b}(\xi)
|\widehat W_{r-1}^{\mathsf{st}}(\Psi_\xi(\mu))
-W_{r-1}^{\mathsf{st}}(\Psi_\xi(\mu))|\\
&\le\sum_\xi P_{a,b}(\xi)B_{r-1}=B_{r-1},
\end{align*}
using the induction hypothesis on each term and $\sum_\xi P_{a,b}(\xi)=1$.
This is the step that prevents the error from being multiplied by
$n^n$: the coefficients $P_{a,b}(\xi)$ form a probability distribution,
so averaging over realisations cannot amplify a uniform per-realisation
error. The sum in \eqref{eq:st-numeric-entry} is exact given the dyadic
child values $\widehat W_{r-1}^{\mathsf{st}}(\Psi_\xi(\mu))$, so no
further rounding error is introduced by the realisation sum itself.

The true matrix $[C_{r-1}(\mu)_{a,b}]_{a,b}$ has entries in $[0,L]$ by
Step~1's boundedness argument, applied one level down. Applying
\eqref{eq:empirical-value-bound} and \eqref{eq:empirical-lipschitz} to
this matrix and its approximation $[\widehat C_{r-1}(\mu)_{a,b}]_{a,b}$
gives
\[
\operatorname{val}(C_{r-1})-B_{r-1}
\le s_k(\widehat C_{r-1})
\le\operatorname{val}(C_{r-1})+\alpha+B_{r-1}.
\]
Adding the current entropy introduces at most $(1-\Discount)d$ error,
and the final rounding and clipping of the complete backup costs at
most $d$ against a target in $[0,L]$, by the clipping inequality
(Appendix \ref{app:sec:sp-main}, Step~3). This mirrors Appendix \ref{app:sec:sp-main}'s
Step~4 exactly: the one-step backup error is decomposed into an entropy
term, a matrix-value term, and a final rounding/clipping term, then
combined via the triangle inequality,
\begin{align*}
|\widehat W_r^{\mathsf{st}}(\mu)-W_r^{\mathsf{st}}(\mu)|
&\le\ \underbrace{d}_{\text{final clipping}}
+\ \underbrace{(1-\Discount)d}_{\text{entropy term}}
+\ \underbrace{\Discount\bigl|s_k(\widehat C_{r-1}(\mu))-\operatorname{val}(C_{r-1}(\mu))\bigr|}_{\text{matrix-value term}}\\
&\le\ d+(1-\Discount)d+\Discount(\alpha+B_{r-1})\\
&=\ \Discount B_{r-1}+\Discount\alpha+(2-\Discount)d\\
&\le\ \Discount B_{r-1}+\Discount\alpha+2d\ =\ B_r,
\end{align*}
using $(2-\Discount)d\le2d$ since $\Discount>0$; the only difference from
Appendix \ref{app:sec:sp-main}'s Step~4 is that the matrix here is
$C_{r-1}(\mu)$, already averaged over realizations by Steps~1-2, in
place of sp's un-averaged $M$. This proves the induction.

\medskip\noindent\textbf{Step 4: correctness of the decision procedure.}
Compute $z=\widehat W_T^{\mathsf{st}}(\distro)$ and answer \textsc{Yes}
exactly when $z\le v$. By Step~3 and \eqref{eq:st-truncation},
\[
|z-V_{\mathsf{st}}(\distro)|
\le|z-W_T^{\mathsf{st}}(\distro)|+|W_T^{\mathsf{st}}(\distro)-V_{\mathsf{st}}(\distro)|
\le e/4+e/4=e/2<\varepsilon_0.
\]
If $V_{\mathsf{st}}(\distro)\le v-\varepsilon_0$, then
$z\le V_{\mathsf{st}}(\distro)+e/2\le v-\varepsilon_0+e/2<v$. If
$V_{\mathsf{st}}(\distro)\ge v+\varepsilon_0$, then
$z\ge V_{\mathsf{st}}(\distro)-e/2\ge v+\varepsilon_0-e/2>v$. So the
algorithm answers \textsc{Yes} on every yes-instance and \textsc{No} on
every no-instance of the promise.

\medskip\noindent\textbf{Step 5: exact bit lengths and active workspace.}
\emph{Bit length of the distribution.}
Let $D_0$ be the product of the denominators in the initial
distribution, and write $\distro(s)=m_s/D_0$ for integers $m_s\ge0$
summing to $D_0$. Every realised update only reassigns existing mass:
by $\Psi_\xi(\mu)(s')=\sum_{s:\xi(s)=s'}\mu(s)$, no transition
probability is multiplied into a distribution's coordinates at all,
mass is only summed. Consequently, by induction on the recursion depth,
every distribution ever passed to a recursive call has the form
$(m'_s/D_0)_s$ for integers $0\le m'_s\le D_0$ summing to $D_0$: the
base case $\distro$ has this form by definition, and if $\mu$ does,
then each coordinate of $\Psi_\xi(\mu)$ is a sum of some of the $m_s$,
still an integer between $0$ and $D_0$, over the same denominator $D_0$.
Unlike sampled-profile semantics, this denominator does not grow with
depth. It takes $O(n\ell)$ bits to store such a distribution at any
depth, since $D_0$ has $O(\ell)$ bits (it is a product of denominators
that are substrings of the $\ell$-bit input) and there are $n$
coordinates.

\emph{Bit length of the recursion's values.}
Let $D=\prod_{s\in\States}\prod_{a\in\Acts(s)}\prod_{s'\in\States}
\operatorname{den}(\Trans(s,a,s'))$, where $\operatorname{den}$ denotes a
positive input denominator. Each individual denominator is a substring
of the $\ell$-bit input, so the sum of their bit lengths is at most
$\ell$, and hence $D$ has $O(\ell)$ bits by the same product-bit-length
fact used in Appendix \ref{app:sec:sp-main}, Step~6. A product $P_{a,b}(\xi)$
selects one transition entry for each source state, so its unreduced
denominator is a subproduct of the factors defining $D$; therefore
$w_{a,b,\xi}=D\times P_{a,b}(\xi)$ is a nonnegative integer, computable
from the individual factors' numerators and denominators by elementary
integer arithmetic, in time and space polynomial in the bit lengths
already bounded, and $\sum_{\xi\in\TransReal}w_{a,b,\xi}=D$.
If a child returns $j_\xi/2^p$, then
\[
\widehat C_{r-1}(\mu)_{a,b}
=\frac{\sum_\xi w_{a,b,\xi}j_\xi}{D\times 2^p},
\qquad
0\le\sum_\xi w_{a,b,\xi}j_\xi\le D\times L\times 2^p,
\]
since every summand is nonnegative and the total weight
$\sum_\xi w_{a,b,\xi}=D$ times the largest possible child value $L2^p$
bounds the sum. All partial sums obey the same bound, so the whole
realisation loop needs an integer accumulator with $O(\ell+p+\log(L+1))$
bits: summing exponentially many terms therefore costs time, one
addition per realisation, but not space, since the accumulator's bit
length never depends on how many realisations have been added so far.
The empirical average over $k$ rows has common denominator $kD2^p$ and
numerator at most $kDL2^p$, adding only $O(\log k)$ bits on top.
Computing a new entry reuses this accumulator; it does not multiply
together the denominators of different realisations or different rows.
Maxima and minima select existing rational values without introducing
new denominators, and the final rounding resets the returned value to
denominator $2^p$, exactly as in Appendix \ref{app:sec:sp-main}, Step~6.

\emph{The active workspace.}
A realisation $\xi:\States\to\States$ is enumerated as an $n$-digit
base-$n$ counter, one digit per state recording its image, using
$O(n\log n)$ bits. An active frame stores the current distribution
($O(n\ell)$ bits, above), the current $k$-tuple and column
($O(kn\log(2A))$ and $O(n\log(2A))$ bits, as in
Appendix \ref{app:sec:sp-main}, Step~6), the current realisation counter
($O(n\log n)$ bits), and the bounded accumulators and extrema
($O(\ell+p+\log(L+1)+\log k)$ bits, from above). All local arithmetic,
including the entropy routine, uses only polynomial auxiliary space in
these lengths, by the same two-part argument as Appendix \ref{app:sec:sp-main},
Step~6: elementary arithmetic on integers of already-bounded bit length
uses space polynomial in that length, and \cref{ass:rational} bounds
only $\widehat H_p$'s time, but since a deterministic algorithm can
write to at most one new memory cell per step, a computation running in
time $t$ uses at most $O(t)$ space, so the entropy routine also uses
only polynomial auxiliary space. On a recursive call the parent frame is
retained, but each realisation's child frame is discarded once its
contribution is added to the running sum, and each completed child in
the tuple-and-column enumeration is discarded before the next is
processed. Since the recursion depth strictly decreases from $r=T$ at
the root to $r=0$ at a base case, and only the frames on the current
path are retained, at most $T$ nontrivial frames coexist at any time.
The horizon $T$ was shown polynomial in $\ell$ in Appendix \ref{app:sec:sp-main}'s
Step~1, the sample size $k$ in its Step~2, and the precision $p$ is
fixed in its Step~3 independently of the instance. Total space is
therefore at most $T$ times a single frame's polynomial bit length, a
product of two quantities polynomial in $\ell$, hence itself polynomial
in $\ell$. The number of realisations, tuples, and recomputed
subproblems may be exponential; this affects time, not the workspace
bound. 
\end{proof}

\section{The One Player Case}
\label{sec:onep}
This section focuses on the single-decision-maker case, $\States_{\max}=\emptyset$.
We begin with \cref{lem:pure-suffices}, used for the upper-bound results : when there is a single
decision-maker, optimising over the full mixed (or, for the averaged
semantics, continuous statewise-mixed) strategy space is equivalent to
optimising over pure profiles alone.

Fix $f\in C(\DistS)$, the space of continuous real-valued functions on
$\DistS$. For $\Sem\in\{\mathsf{sp},\mathsf{st}\}$ define
\[
\begin{gathered}
M_f^{\mathsf{sp}}(\mu)_{a,b}=f(\Phi_{a,b}(\mu)),\qquad
M_f^{\mathsf{st}}(\mu)_{a,b}=\sum_\xi P_{a,b}(\xi)f(\Psi_\xi(\mu)),\\
(\mathcal T_\Sem f)(\mu)=(1-\Discount)\Ent(\mu)+\Discount\operatorname{val}(M_f^\Sem(\mu)),
\end{gathered}
\]
where $\operatorname{val}(\cdot)$ is the common minimax value of a finite
zero-sum matrix game, as given by the finite minimax theorem in
\eqref{eq:matrix-minimax-value}; the proof below also uses the equivalent
one-sided characterisation of \eqref{eq:val-two-forms}, which lets either
player's opponent be restricted to a pure best response without changing
the value.
For the averaged semantics define separately
\begin{align*}
(\mathcal T^+_{\mathsf{av}}f)(\mu)
&=(1-\Discount)\Ent(\mu)+\Discount\min_{x\in\mathcal X}
\max_{y\in\mathcal Y}f(\DistOp_{x,y}(\mu)),\\
(\mathcal T^-_{\mathsf{av}}f)(\mu)
&=(1-\Discount)\Ent(\mu)+\Discount\max_{y\in\mathcal Y}
\min_{x\in\mathcal X}f(\DistOp_{x,y}(\mu)).
\end{align*}
When $\States_{\max}=\emptyset$, $\mathcal P_{\max}$ contains only the
empty (trivial) profile, which we call $b_0$, and $\mathcal Y$ is the
single point $\{y_0\}$; write $\Phi_a:=\Phi_{a,b_0}$,
$P_a(\xi):=P_{a,b_0}(\xi)$, and $\DistOp_x:=\DistOp_{x,y_0}$ for
$a\in\mathcal P_{\min}$, $x\in\mathcal X$. Consequently
$\max_{y\in\mathcal Y}$ is vacuous, so
$\mathcal T^+_{\mathsf{av}}=\mathcal T^-_{\mathsf{av}}$; write
$\mathcal T_{\mathsf{av}}$ for their common value,
\begin{equation}
(\mathcal T_{\mathsf{av}}f)(\mu)=(1-\Discount)\Ent(\mu)+\Discount\min_{x\in\mathcal X}f(\DistOp_x(\mu)).
\label{eq:Tav-def}
\end{equation}
\puresuffices*

\begin{proof}

\emph{Sampled semantics.} Since $W_{t+1}=\mathcal T_\Sem W_t$, unfolding
the definition of $\mathcal T_\Sem$ gives
$W_{t+1}(\mu)=(1-\Discount)\Ent(\mu)+\Discount\operatorname{val}(M_{W_t}^\Sem(\mu))$,
where $M_{W_t}^\Sem(\mu)$ denotes the matrix $M_f^\Sem(\mu)$ defined
above, evaluated at $f=W_t$. With $\States_{\max}=\emptyset$,
$\mathcal P_{\max}=\{b_0\}$, so this matrix has a single column; writing
$M_a:=M_{W_t}^\Sem(\mu)_{a,b_0}$, the definition of $M_f^\Sem$ gives
$M_a=W_t(\Phi_a(\mu))=C^{\mathsf{sp}}_t(\mu,a)$ for $\Sem=\mathsf{sp}$ and
$M_a=\sum_\xi P_a(\xi)W_t(\Psi_\xi(\mu))=C^{\mathsf{st}}_t(\mu,a)$ for
$\Sem=\mathsf{st}$. By \eqref{eq:val-two-forms},
$\operatorname{val}(M_{W_t}^\Sem(\mu))=\min_{\sigma\in\Dist(\mathcal
P_{\min})}\sigma^{\mathsf T}M_{W_t}^\Sem$,  a linear functional of $\sigma$,
minimised at a vertex of $\Dist(\mathcal P_{\min})$, i.e.\ at a pure $a$,
for both $\Sem=\mathsf{sp}$ and $\Sem=\mathsf{st}$.

\smallskip
\emph{Averaged-distribution semantics.} We show by induction on $t$ that
$W_t$ is concave on $\DistS$. The base case $W_0\equiv0$ is trivially
concave. For the inductive step, suppose $W_t$ is concave. For fixed
$x\in\mathcal X$, $\mu\mapsto W_t(\DistOp_x(\mu))$ is concave, being $W_t$
composed with the map $\mu\mapsto\DistOp_x(\mu)$, which is linear in $\mu$
for fixed $x$ by \eqref{eq:distribution-update}. The pointwise infimum of
any family of concave functions is concave (dual to the fact that a
pointwise supremum of convex functions is convex), so
$\mu\mapsto\inf_{x\in\mathcal X}W_t(\DistOp_x(\mu))$ is concave, and hence so
is $W_{t+1}=(1-\Discount)\Ent+\Discount\inf_x W_t(\DistOp_x(\cdot))$, being a
sum of two concave functions; this closes the induction. Now fix $\mu$ and
consider $x\mapsto W_t(\DistOp_x(\mu))$: $\DistOp_x(\mu)$ is linear in $x$
for fixed $\mu$ (again by \eqref{eq:distribution-update}), so this is a
concave function of $x$, since $W_t$ is concave.

Every point $x\in\mathcal X$ is exactly the convex combination of its
vertices given by its induced profile law $\widehat x$
(\eqref{eq:induced-profile-laws}): $x=\sum_a\widehat x(a)\,a$, where
$\widehat x(a)=\prod_sx(s)(a(s))$ is the probability of the pure
profile $a$ under independently sampling each state's action from $x$.
Linearity of $\DistOp_x(\mu)$ in $x$ (\eqref{eq:distribution-update})
gives
\[
\DistOp_x(\mu)=\sum_a\widehat x(a)\DistOp_a(\mu)=\sum_a\widehat x(a)\Phi_a(\mu),
\]
using $\DistOp_a=\Phi_a$ as above. Since $W_t$ is concave, Jensen's
inequality gives
\[
W_t(\DistOp_x(\mu))=W_t\Bigl(\sum_a\widehat x(a)\Phi_a(\mu)\Bigr)
\ge\sum_a\widehat x(a)W_t(\Phi_a(\mu)).
\]
Since the $\widehat x(a)$ are nonnegative and sum to $1$, this weighted
average of the values $W_t(\Phi_a(\mu))$ is at least their smallest
term:
\[
\sum_a\widehat x(a)W_t(\Phi_a(\mu))\ge\min_aW_t(\Phi_a(\mu)).
\]
Combining the two,
\[
W_t(\DistOp_x(\mu))
\ge\sum_a\widehat x(a) W_t(\Phi_a(\mu))
\ge\min_a W_t(\Phi_a(\mu)).
\]

Its vertices are exactly the pure profiles; since each pure profile $a$,
viewed as the degenerate point $x=a\in\mathcal X$, satisfies
$\DistOp_a=\Phi_a$ (as above), the minimum over all of $\mathcal X$ is at
most its value at this particular point:
$\min_{x\in\mathcal X}W_t(\DistOp_x(\mu))\le
W_t(\DistOp_a(\mu))=W_t(\Phi_a(\mu))$ for every pure $a$. Taking the
minimum over $a$ on the right-hand side gives
$\min_{x\in\mathcal X}W_t(\DistOp_x(\mu))\le\min_aW_t(\Phi_a(\mu))$.

The Jensen inequality above, $W_t(\DistOp_x(\mu))\ge\min_aW_t(\Phi_a(\mu))$,
holds for every $x\in\mathcal X$, and its right-hand side does not depend
on $x$; taking the minimum over $x\in\mathcal X$ of both sides therefore
gives $\min_{x\in\mathcal X}W_t(\DistOp_x(\mu))\ge\min_aW_t(\Phi_a(\mu))$.
Combined with the reverse inequality just shown,
$\min_{x\in\mathcal X}W_t(\DistOp_x(\mu))=\min_aW_t(\Phi_a(\mu))$, so the
infimum over all of $\mathcal X$ is attained at a pure profile.

Since $\mathcal P_{\min}$ is finite, the minimum in each step is attained;
picking a minimising $a$ at each step of the recursion yields, from any
starting distribution, a witnessing sequence of pure actions achieving
$W_{t+1}$.
\end{proof}

\begin{corollary}[One-player averaged and sampled-profile values coincide]
\label{cor:av-sp-coincide}
If $\States_{\max}=\emptyset$, then $V_{\mathsf{av}}=V_{\mathsf{sp}}$.
\end{corollary}
\begin{proof}
Both value functions arise as limits of finite-horizon backups from the same
base case $W_0\equiv0$:
$W_{t+1}^{\mathsf{av}}=(1-\Discount)\Ent+\Discount\min_{x\in\mathcal
X}W_t^{\mathsf{av}}(\DistOp_x(\cdot))$ and
$W_{t+1}^{\mathsf{sp}}=(1-\Discount)\Ent+\Discount\min_{a\in\mathcal
P_{\min}}W_t^{\mathsf{sp}}(\Phi_a(\cdot))$. The proof of
\cref{lem:pure-suffices} shows, at every step of this recursion, that
$\min_{x\in\mathcal X}W_t(\DistOp_x(\mu))=\min_{a\in\mathcal
P_{\min}}W_t(\Phi_a(\mu))$ whenever $W_t$ is concave, and that concavity is
preserved from step to step. So if $W_t^{\mathsf{av}}=W_t^{\mathsf{sp}}$, the
two recursions apply the same minimisation to the same function and produce
the same $W_{t+1}$; since $W_0^{\mathsf{av}}=W_0^{\mathsf{sp}}\equiv0$,
induction on $t$ gives $W_t^{\mathsf{av}}=W_t^{\mathsf{sp}}$ for every
$t\ge0$. Both sequences converge uniformly, at the same geometric rate, to
their respective infinite-horizon values as $t\to\infty$
(\eqref{eq:finite-infinite-error} for $\mathsf{av}$,
\eqref{eq:sampled-value-limit} for $\mathsf{sp}$), and a sequence has at
most one limit, so
$V_{\mathsf{av}}=\lim_tW_t^{\mathsf{av}}=\lim_tW_t^{\mathsf{sp}}=V_{\mathsf{sp}}$.
\end{proof}

\subsection{Upper Bounds for One Player}
\label{app:uponep}
Since $\States_{\max}=\emptyset$, $\mathcal P_{\max}$ contains only the
empty profile $b_0$, so $\StratMax^{\mathsf{av}}$ contains only the
single strategy $\tau_0$ that plays $b_0$ at every stage, regardless of
history. Consequently $\sup_{\stratmax\in\StratMax^{\mathsf{av}}}$ in
the definitions of both $V^+_{\mathsf{av}}$ and $V^-_{\mathsf{av}}$
ranges over the one strategy $\tau_0$ and may be dropped, so
\[
V^+_{\mathsf{av}}(\distr)=\inf_{\stratmin\in\StratMin^{\mathsf{av}}}
\mathsf{Obj}^{\mathsf{av}}_{\Discount}(\stratmin,\tau_0;\distr)
=V^-_{\mathsf{av}}(\distr)
\]
for every $\distr\in\DistS$. As with $\mathsf{sp}$ and $\mathsf{st}$
(\cref{thm:sp-determined,thm:st-determined}), we therefore omit the
polarity in the one-player case and write $V_{\mathsf{av}}$,
$\GapVal_{\mathsf{av},\varepsilon_0}$.
\begin{lemma}[One-player averaged gap value lies in $\NP$]
\label{cor:1p-etr}
\label{cor:1p-av-np}
For fixed rational $\Discount\in(0,1)$ and $\varepsilon_0>0$, if
$\States_{\max}=\emptyset$ then $\GapVal_{\mathsf{av},\varepsilon_0}\in\NP$.
\end{lemma}
\begin{proof}
Recall that a promise problem lies in $\NP$ if there is a polynomial-length
certificate and a polynomial-time verifier such that every yes-instance
has some certificate the verifier accepts (\emph{completeness}) and every
no-instance has every certificate rejected by the verifier
(\emph{soundness}); we exhibit exactly these four ingredients.

Since $\States_{\max}=\emptyset$, every state lies in $\States_{\min}$,
and by \cref{lem:pure-suffices}, a finite-horizon optimum is attained
by pure profiles, so every trajectory considered here is pure. So by
\eqref{eq:phi-def},\eqref{eq:distribution-update} the successor map
$\Upsilon_{a,b} = \Phi_{a,b}$ does not depend on $b$; write
$\Phi_a(\mu)(s')\defeq\sum_{s\in\States}\mu(s)\Trans(s,a(s),s')$ for
this common value.

Let $e=\min\{\varepsilon_0,1\}$ and $L=\lceil\log_2n\rceil$, and let $T$
be as in the proof of \cref{thm:etr-ub} (Step 1): the least positive
integer with $\Discount^TL\le e/4$. Thus $U_T(\distro)=\min_{\vec a}U_T^{\vec
a}(\distro)$, where $\vec a=(a_0,\ldots,a_{T-2})$ ranges over sequences
of pure profiles, $\mu_0=\distro$, $\mu_{t+1}=\Phi_{a_t}(\mu_t)$, and
$U_T^{\vec a}(\distro)=(1-\Discount)\sum_{t=0}^{T-1}\Discount^t\Ent(\mu_t)$.
The certificate is such a sequence $\vec a$, with $O(Tn\log(2A))$ bits by
Appendix \ref{app:sec:compact-profile}.

Choose the least $p\ge1$ with $2^{-p}\le e/8$. The verifier computes
$z^{\vec a}=(1-\Discount)\sum_{t=0}^{T-1}\Discount^t\widehat H_p(\mu_t)$
along the guessed $\vec a$ and accepts iff $z^{\vec a}\le v$. Since
$(1-\Discount)\sum_{t=0}^{T-1}\Discount^t\le1$ and each term satisfies
$|\widehat H_p(\mu_t)-\Ent(\mu_t)|\le2^{-p}$,
\[
|z^{\vec a}-U_T^{\vec a}(\distro)|\le2^{-p}\le e/8
\]
for every $\vec a$. By \eqref{eq:finite-infinite-error},
$0\le V^+_{\mathsf{av}}(\distro)-U_T(\distro)\le e/4$; since
$V^+_{\mathsf{av}}=V_{\mathsf{av}}$ was shown above, this reads
$0\le V_{\mathsf{av}}(\distro)-U_T(\distro)\le e/4$, so for
the optimal $\vec a^*$,
\[
|z^{\vec a^*}-V_{\mathsf{av}}(\distro)|
\le|z^{\vec a^*}-U_T(\distro)|+|U_T(\distro)-V_{\mathsf{av}}(\distro)|
\le e/8+e/4=3e/8,
\]
and for every $\vec a$, since $U_T^{\vec a}(\distro)\ge U_T(\distro)$,
\[
z^{\vec a}\ge U_T^{\vec a}(\distro)-e/8\ge U_T(\distro)-e/8
\ge V_{\mathsf{av}}(\distro)-e/4-e/8=V_{\mathsf{av}}(\distro)-3e/8.
\]
Since $e\le\varepsilon_0$, $3e/8<\varepsilon_0$. If
$V_{\mathsf{av}}(\distro)\le v-\varepsilon_0$, the certificate
$\vec a^*$ gives $z^{\vec a^*}\le v-\varepsilon_0+3e/8<v$, so it is
accepted: completeness holds. If $V_{\mathsf{av}}(\distro)\ge v+\varepsilon_0$, every
certificate $\vec a$ gives $z^{\vec a}\ge v+\varepsilon_0-3e/8>v$, so
every certificate is rejected: soundness holds. Computing each $\mu_t$
takes polynomial time and computing each $\widehat H_p(\mu_t)$ takes
polynomial time by \cref{ass:rational}, so verification is
polynomial-time. Together with the certificate's polynomial length,
this establishes all four ingredients recalled above, so
$\GapVal_{\mathsf{av},\varepsilon_0}\in\NP$.
\end{proof}

\begin{lemma}[One-player sampled-profile gap value lies in $\NP$]
\label{cor:1p-sp-np}
For fixed rational $\Discount\in(0,1)$ and $\varepsilon_0>0$, if
$\States_{\max}=\emptyset$ then $\GapVal_{\mathsf{sp},\varepsilon_0}\in\NP$.
\end{lemma}
\begin{proof}
With one player the matrix has one column, so $\operatorname{val}(M)
=\min_\sigma\sigma^{\mathsf T}M$ minimises a linear functional of
$\sigma$ over the probability simplex on $\mathcal P_{\min}$. A linear
functional over a simplex attains its minimum at an extreme point, and
the simplex's extreme points are exactly the point masses on pure
profiles, so the minimum is attained at some pure profile $a$.
Consequently a finite-horizon optimum follows a deterministic sequence
of $T-1$ pure profiles.

Since $\States_{\max}=\emptyset$, every state lies in $\States_{\min}$,
so by \eqref{eq:phi-def} the successor map $\Phi_{a,b}$ does not depend
on $b$; write $\Phi_a(\mu)(s')\defeq\sum_{s\in\States}\mu(s)\Trans(s,a(s),s')$
for this common value, exactly as in \cref{cor:1p-av-np}. Under
sampled-profile semantics a pure profile has nothing to sample: with no
genuinely mixed action to aggregate, the distribution is updated by the
same map $\Phi_a$ used under averaged-distribution semantics, so a
fixed pure-profile trajectory's distribution sequence and entropy
values are identical under either semantics. The polynomial-time
verifier and the promise margins are therefore exactly those of
\cref{cor:1p-av-np}. This proves $\NP$ membership.
\end{proof}

\begin{lemma}[One-player sampled-transition gap value]
\label{cor:1p-st-pspace}
For fixed rational $\Discount\in(0,1)$ and $\varepsilon_0>0$, if
$\States_{\max}=\emptyset$ then $\GapVal_{\mathsf{st},\varepsilon_0}\in\PSPACE$.
\end{lemma}
\begin{proof}
With $\States_{\max}=\emptyset$, every continuation matrix
$M=[\sum_\xi P_{a,b}(\xi)W_{r-1}^{\mathsf{st}}(\Psi_\xi(\mu))]_{a,b}$
has a single column; write $M_a$ for its entry at the trivial column,
matching \cref{lem:pure-suffices}'s notation. Then
$\operatorname{val}(M)=\min_\sigma\sigma^{\mathsf T}M$ minimises a
linear functional of $\sigma$ over the probability simplex on
$\mathcal P_{\min}$; this minimum is attained at an extreme point of
the simplex, that is, at some pure profile $a^*$, giving
$\operatorname{val}(M)=M_{a^*}$. Consequently, for the empirical value
$s_k(M)=\min_{(a_1,\ldots,a_k)}\frac1k\sum_iM_{a_i}$ of
Appendix \ref{app:sec:sp-main}'s Step~2, setting every $a_i=a^*$ already attains
$s_k(M)=M_{a^*}=\operatorname{val}(M)$ exactly, for every $k\ge1$: no
averaging over several profiles is needed, and no sparsification error
remains. In particular $\alpha=0$ and $k=1$ suffice in
Appendix \ref{app:sec:sp-main}'s Step~2, and the response loop over
$\mathcal P_{\max}$ in its Steps~2 and 3 is omitted, since
$\mathcal P_{\max}$ has only the single trivial column.

Keep the exact sum over all realisations (Step~2 above) and the
fixed-precision recursion \eqref{eq:st-numeric-recursion} unchanged.
With $\alpha=0$, the error recurrence of Step~3 becomes
$B_r=\Discount B_{r-1}+2d$, so its general closed form
$B_r\le(\Discount\alpha+2d)/(1-\Discount)$ specializes to
$B_r\le2d/(1-\Discount)\le e/4$, so \eqref{eq:st-truncation} and Step~4's
argument go through unchanged and prove promise correctness. The
bit-length argument of Step~5 does not depend on the specific value of
$k$, beyond $k$ itself being polynomial in $\ell$, which $k=1$ trivially
is, so the same space bound applies. Unlike the one-player $\mathsf{sp}$
case (\cref{cor:1p-sp-np}), a pure choice here does not produce a
deterministic successor, since the arena's own transitions still
realise stochastically; so this argument does not yield a
polynomial-length trajectory certificate or an $\NP$ upper bound, only
the polynomial-space bound above.
\end{proof}

\subsection{Lower Bounds}
\label{app:lower}
We next give the lower bound once, in a form that can be cited from all
three semantic subsections. The argument runs in three steps. We start
from a problem already known to be $\NP$-hard: distinguishing, for a
family of triples over some universe, an instance that can be
partitioned into disjoint triples from one where every such partition
falls well short of covering the universe ~\cite{petrank1994gap}. We
first convert this into an equally hard question about Set-Cover:
distinguishing an instance with a small, exact cover from one where
every cover needs substantially more sets (\cref{lem:bounded-cover-gap}).
We then feed that Set-Cover instance into the gadget described next. It
is built so that its one-step entropy $\Ent(\distr_1)$ comes out small
exactly when the chosen cover is small, and larger otherwise. Because
every state the gadget reaches after one step stays fixed forever after,
this one-step gap passes through unchanged to the infinite-horizon
objective.

We restrict this reduction to deterministic arenas for two reasons. First, on
a deterministic arena, a pure profile leaves nothing for sampled-profile
semantics to sample and nothing for sampled-transition semantics to realise,
so all three semantics agree exactly on the resulting trajectory; this is
what lets a single construction transfer hardness to all three semantics at
once, rather than requiring three separate reductions. Second, since every
deterministic game is also a (degenerate) general entropy stochastic game,
hardness proved on this restricted family is at least as strong as hardness
proved on the general one; restricting to it only makes the construction
simpler.

The rest of this subsection assembles the theorem from four facts, each
proved once and then combined. Appendix \ref{sec:set-cover-gadget} defines the
game construction, which turns any Set-Cover instance into a
single-player entropy game with one absorbing state per available set.
Appendix \ref{sec:collapsing-entropy} proves, in \cref{lem:set-cover-collapse},
that on this game the value under all three semantics is the same single
number: the entropy of the distribution reached after one step, so a
gap in that one number is automatically a gap under every semantics at
once. Appendix \ref{sec:entropy-routing} proves, in \cref{lem:set-cover-entropy},
a precise two-sided bound on that one-step entropy in terms of how many
sets the routing uses, so a small cover gives low entropy and a cover
forced to use more sets gives provably higher entropy.
Appendix \ref{sec:bounded-cover-gap} proves, in \cref{lem:bounded-cover-gap}, that
distinguishing a small-cover Set-Cover instance from a large-cover one is
itself $\NP$-hard, by reduction from Petrank's perfect-completeness gap
for bounded-occurrence three-dimensional matching~\cite{petrank1994gap}:
any sufficiently small cover can be converted, by repeatedly discarding a
triple that overlaps another, into an almost-as-small disjoint matching,
so a cover much smaller than the claimed gap would yield a matching
larger than Petrank's promise allows.
 Appendix \ref{sec:hardness-reduction} then combines all
three: feeding the hard instances from \cref{lem:bounded-cover-gap}
through the gadget of Appendix \ref{sec:set-cover-gadget}, and reading off the
resulting entropy gap via \cref{lem:set-cover-collapse,lem:set-cover-entropy},
turns the $\NP$-hardness of telling the two Set-Cover cases apart into
the $\NP$-hardness of telling the two entropy values apart, which is
the theorem.

\subsubsection{The Set-Cover gadget}
\label{sec:set-cover-gadget}
\begin{definition}[Deterministic entropy game]
\label{def:deterministic-eg}
A game is \emph{deterministic} if for every state-action pair
$s,a$, $\Trans(s,a)=\delta_{s'}$ for some $s'$: every action, at every state, leads
with certainty to a single successor state.
\end{definition}

\textbf{Set-Cover problem:} Given a finite set, the \emph{universe} $U$,
and a collection of subsets $S_1,\ldots,S_M\subseteq U$, the goal is to choose as few of
the $S_j$ as possible whose union is all of $U$ (a \emph{cover}); deciding whether a
cover of a given size exists is a classical $\NP$-hard
problem~\cite{karp2009reducibility}.

\begin{figure}[t]
\centering
\begin{tikzpicture}[>=Latex, thick, every node/.style={circle,draw,minimum size=7mm,inner sep=0pt}]
  \node (e1) at (0,4.5) {$e_1$};
  \node (e2) at (0,3) {$e_2$};
  \node (e3) at (0,1.5) {$e_3$};
  \node (e4) at (0,0) {$e_4$};
  \node (f1) at (5,3.75) {$f_1$};
  \node (f2) at (5,2.25) {$f_2$};
  \node (f3) at (5,0.75) {$f_3$};
  \draw[->] (e1) -- (f1);
  \draw[->] (e2) -- (f1);
  \draw[->] (e2) -- (f2);
  \draw[->] (e3) -- (f2);
  \draw[->] (e3) -- (f3);
  \draw[->] (e4) -- (f3);
  \foreach \f in {f1,f2,f3} {
    \draw[->] (\f) to [out=45,in=-45,looseness=6] (\f);
  }
  \node[draw=none] at (0,5.1) {$\{e_u\}_{u\in U}$};
  \node[draw=none] at (5,4.35) {$\{f_j\}_{j\le M}$};
  \node[draw=none,align=center] at (-1.6,2.25) {$\distro$\\uniform};
\end{tikzpicture}
\caption{The Set-Cover gadget (Construction~\ref{constr:set-cover}) for the small
instance $U=\{1,2,3,4\}$, $S_1=\{1,2\}$, $S_2=\{2,3\}$, $S_3=\{3,4\}$. Left: one
state $e_u$ per universe element $u\in U$, together carrying the initial
distribution $\distro$, uniform over all $m=|U|$ of them. Right: one state $f_j$
per set $S_j$, each absorbing (self-loop). \MinPl's only choice, at each $e_u$, is
which containing set's state to move to: state $e_2$, for instance, has two
available actions, one to $f_1$ (since $2\in S_1$) and one to $f_2$ (since
$2\in S_2$), while $e_1$ and $e_4$ each have only one, being covered by a single
set. Concentrating many elements' mass onto few $f_j$ states lowers the entropy of
$\distr_1$; spreading it across many raises it, exactly like choosing a small
versus a wasteful cover.}
\label{fig:set-cover-gadget}
\end{figure}
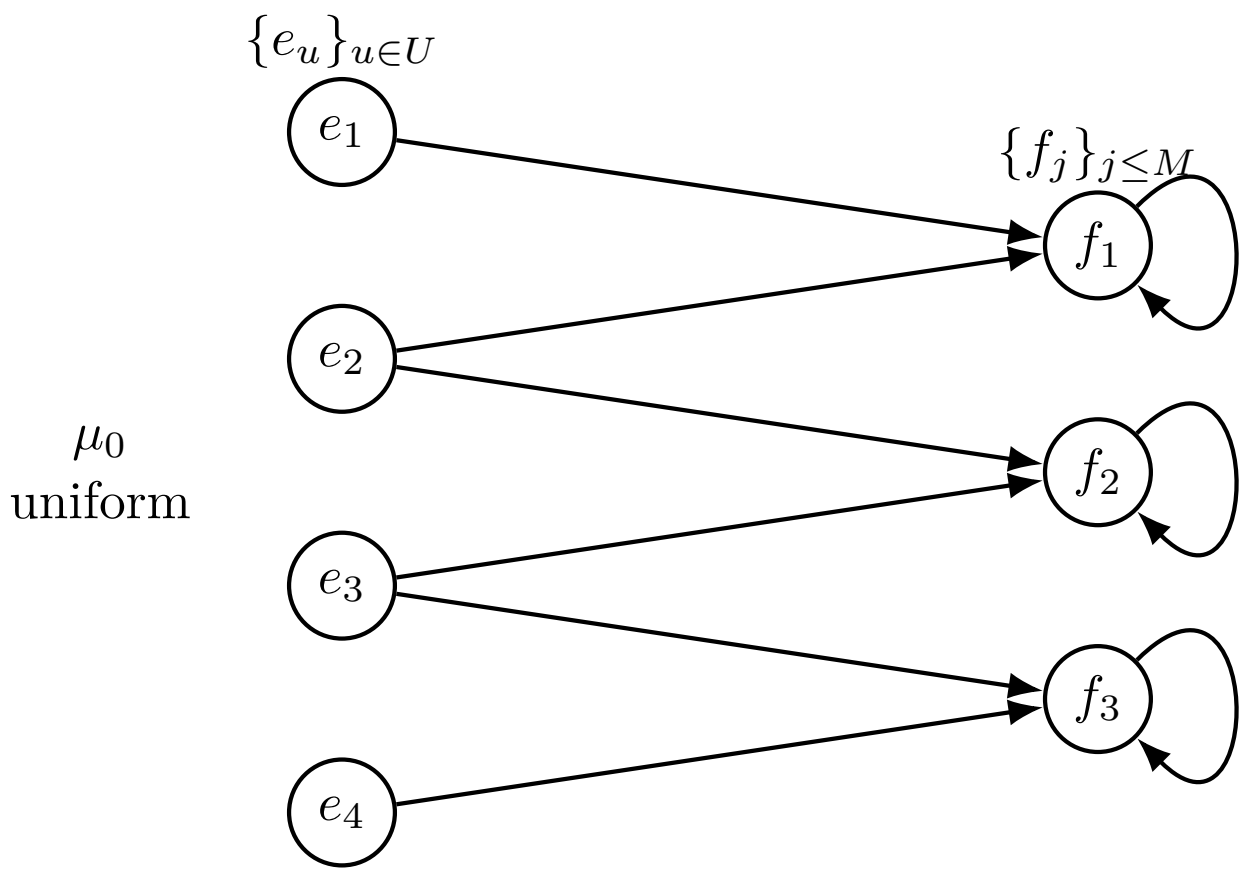

\begin{construction}[Set-Cover gadget]
\label{constr:set-cover}
Given a universe $U$, $|U|=m$, and sets $S_1,\ldots,S_M\subseteq U$ each of size at
most $B$ such that $\bigcup_{j=1}^M S_j = U$. Build the deterministic, single-player entropy game $\mathcal E$
with $\States_{\max}=\emptyset$, $\States_{\min}=\{e_u : u\in U\}\cup\{f_j : j\le M\}$,
$\distro$ uniform on $\{e_u\}_{u\in U}$ (so $\distro(e_u)=1/m$), and, for each $u$, one
action per set $S_j\ni u$ at $e_u$, deterministically leading to $f_j$; each $f_j$ is
absorbing (its unique available action self-loops). \MinPl (the only player) chooses,
possibly with mixed strategies, an action at each $e_u$.
\end{construction}
\cref{fig:set-cover-gadget} illustrates the construction.

\subsubsection{Collapsing to one-step entropy}
\label{sec:collapsing-entropy}
Since $\States_{\max}=\emptyset$, \MaxPl has only the trivial strategy
$\tau_0$; write $\mathsf{Obj}^\Sem_\Discount(\stratmin;\distr)
:=\mathsf{Obj}^\Sem_\Discount(\stratmin,\tau_0;\distr)$ for the resulting
one-argument form. With $\stratmax$ trivial, the upper and lower values
$\inf_{\stratmin}\sup_{\stratmax}$ and $\sup_{\stratmax}\inf_{\stratmin}$
both collapse to $\inf_{\stratmin}$ alone, so
$\inf_{\stratmin}\mathsf{Obj}^{\Sem}_{\Discount}(\stratmin;\distro)$ is
exactly the one-player value already in use: the polarity-free
$V_\Sem(\distro)$, for every $\Sem\in\{\mathsf{av},\mathsf{sp},\mathsf{st}\}$.
For $\mathsf{av}$ the upper and lower values coincide in this
one-player case by the argument given just before \cref{cor:1p-av-np};
for $\mathsf{sp}$ and $\mathsf{st}$ they already coincide in general,
with or without a lone decision-maker
(\cref{thm:sp-determined,thm:st-determined}). This is the value used in
\cref{cor:1p-av-np,cor:1p-sp-np,cor:1p-st-pspace}. By
\cref{lem:pure-suffices}, restricting \MinPl to pure profiles already
suffices to compute this value, and shows the infimum is in fact
attained, under every semantics, so the next lemma is stated directly
for a pure profile.
\begin{lemma}
\label{lem:set-cover-collapse}
For $\mathcal E$ as in Construction~\ref{constr:set-cover}, any semantics
$\Sem\in\{\mathsf{av},\mathsf{sp},\mathsf{st}\}$, and any pure profile $a$ of
\MinPl (one action per state $e_u$),
$$
\mathsf{Obj}_\gamma \defeq \mathsf{Obj}^{\Sem}_{\Discount}(a;\distr_0)
= (1-\gamma)\log_2 m + \gamma\cdot\Ent(\distr_1),
$$
independent of $\Sem$.
\end{lemma}
\begin{proof}
\emph{The entropy trajectory takes only two values.} The entropy trajectory only ever takes two values: $\log_2 m$ at time $0$, then
$\Ent(\distr_1)$ forever after, since every $f_j$ is absorbing. So the infinite
discounted sum defining $\mathsf{Obj}_\Discount$ collapses to a weighted average of just
these two fixed numbers, not a sum over a changing trajectory.

\emph{Computing $\Ent(\distr_0)$.} $\distro$ is uniform on the $m$ states $\{e_u\}_{u\in U}$ by construction,
and this does not depend on \MinPl's strategy at all. \MinPl only has actions
available at the $e_u$ states, and choosing one of them determines the transition
\emph{into} $\distr_1$, not $\distro$ itself; $\distro$ is fixed data of the
instance, not something any strategy can influence. So, by
Proposition~\ref{prop:entropy-props}(i), the uniform distribution on $m$ states has
entropy exactly $\log_2 m$, giving $\Ent(\distr_0)=\log_2 m$ regardless of the
strategy played.

\emph{Computing $\Ent(\distr_t)$ for $t\ge1$.} Every state $f_j$ is absorbing, meaning its only available action
self-loops back to $f_j$ with certainty (Construction~\ref{constr:set-cover}). Once
the first round of play moves all the mass from the $e_u$ states onto the $f_j$
states, producing $\distr_1$, every state in the support of $\distr_1$ is absorbing,
so no later round of play can move any mass anywhere else: $\distr_t=\distr_1$ for
every $t\ge1$. The entropy trajectory is therefore exactly
$\log_2m,\Ent(\distr_1),\Ent(\distr_1),\ldots$, constant from time $1$ onward.

\emph{Why the trajectory is the same under all three semantics.}
Sampled-profile semantics differs from averaged-distribution semantics
only in how a genuinely mixed action is aggregated, and
sampled-transition semantics differs from both only in how a genuinely
stochastic transition is realised. For the pure profile $a$ used here,
the corresponding degenerate statewise mixed action has $x(s)=\delta_{a(s)}$
for every $s\in\States_{\min}$, so its induced profile law
(\cref{eq:induced-profile-laws}) is $\widehat x(a')=\prod_{s\in\States_{\min}}
\delta_{a(s)}(a'(s))$, which equals $1$ if $a'=a$ and $0$ otherwise; that
is, $\widehat x=\delta_a$, the point mass on $a$. Sampling a profile from
$\widehat x$, as sampled-profile semantics does, therefore always returns
$a$ itself: there is nothing to sample, exactly as used in
\cref{cor:1p-sp-np}. Likewise, by \cref{def:deterministic-eg} every
$\Trans(s,a)$ is already a point mass, so sampled-transition semantics
has no randomness to realise either. Hence $\distr_1$, and every
subsequent $\distr_t=\distr_1$ for $t\ge1$, is identical regardless of
$\Sem$.

\emph{Assembling $\mathsf{Obj}_\gamma$.} Substituting this trajectory into the definition of $\mathsf{Obj}_\gamma$ and
splitting off the $t=0$ term,
\begin{align*}
  \mathsf{Obj}_{\gamma}=(1-\gamma)\sum_{t\ge0}\gamma^t\Ent(\distr_t)
    &=(1-\gamma)\Ent(\distr_0)+(1-\gamma)\Ent(\distr_1)\sum_{t\ge1}\gamma^t\\
    &=(1-\gamma)\log_2 m+\gamma\,\Ent(\distr_1),
\end{align*}
using $\sum_{t\ge1}\gamma^t=\gamma\sum_{t\ge0}\gamma^t=\dfrac{\gamma}{1-\gamma}$, the
standard geometric series shifted by one term.
\end{proof}

\begin{remark}[Reducing $\mathsf{Obj}_\gamma$-hardness to $\Ent(\distr_1)$-hardness]
\label{rem:disc-substitution}
Since $\gamma>0$ and $\log_2 m$ are fixed, $\mathsf{Obj}_\gamma=(1-\gamma)\log_2
m+\gamma\,\Ent(\distr_1)$ is an increasing affine function of $\Ent(\distr_1)$. A routing minimising $\Ent(\distr_1)$ also
minimises $\mathsf{Obj}_\gamma$, and an $\varepsilon$-gap between two values of
$\Ent(\distr_1)$ becomes a $\gamma\varepsilon$-gap between the corresponding values of
$\mathsf{Obj}_\gamma$. We therefore analyse $\min\Ent(\distr_1)$ once and read off
$\mathsf{Obj}_\gamma$ at the end.
\end{remark}

\subsubsection{Entropy as a function of the routing}
\label{sec:entropy-routing}

The next lemma shows how
$\Ent(\distr_1)$ depends on the routing, in terms of how many sets are used and how
evenly the universe's elements are spread across them.
\begin{lemma}[Entropy of a routing]
\label{lem:set-cover-entropy}
Write $\distr_1(f_j)=c_j/m$ for the (integer) load $c_j\in\{0,\ldots,B\}$ routed to
$S_j$ (so $\sum_j c_j=m$), and $s=|\{j : c_j>0\}|$ for the number of sets used, which
is necessarily a set cover of $U$. Then, for this integer $s$,
$
E^*(s) - \frac{B\log_2 B}{m}\ \le\ \Ent(\distr_1)\ \le\ \log_2 s
$
where
$E^*(s)\ \defeq\ \log_2 m - \frac{(m-s)\,B\log_2 B}{m(B-1)}$
and $E^*$ is strictly increasing in $s$. In general, $\Ent(\distr_1)\ge\log_2(m/B)$ with equality iff $B$ divides $m$ and
$s=m/B$, then $\Ent(\distr_1) = E^*(m/B)=\log_2(m/B)$.
\end{lemma}
\begin{proof}
\emph{Intuition.} The load cap $B$ stops any single $f_j$ from
absorbing more than a $B/m$ share of the mass, so using \emph{fewer} sets forces
\emph{more} mass onto each of the remaining ones; entropy is minimised exactly when
this forcing is as extreme as possible, every used set loaded to its cap $B$ (uniform
on $m/B$ states) when that many sets suffice, and it can only rise if more sets than
necessary are used, since that spreads the same total mass thinner.

$$\Ent(\distr_1) = -\sum_{j:c_j>0}\frac{c_j}{m}\log_2\frac{c_j}{m} = \sum_{j:c_j>0}\frac{c_j}{m} [\log_2m - \log_2c_j] = \log_2m-\frac1m\sum_{j:c_j>0}c_j\log_2c_j$$
Minimizing $\Ent(\distr_1)$ at fixed support
size $s$ is equivalent to \emph{maximising} $F(c)\defeq\sum_jc_j\log_2c_j$ over the
polytope $P_s=\{c\in[1,B]^s:\sum_jc_j=m\}$, the possible loads when exactly $s$ sets
are used (we first relax integrality; the correction for integer loads is bounded
below).

\textbf{Upper bound.} By concavity of $\Ent$ restricted to distributions supported on
a fixed set of size $s$ (Proposition~\ref{prop:entropy-props}(i)), $\Ent(\distr_1)\le\log_2s$
always, with equality iff $\distr_1$ is uniform on its support.

\textbf{Basic lower bound and the equality case.} $F$ is a sum of the convex function
$x\mapsto x\log_2x$ applied coordinatewise, and every $c_j\le B$ caps every atom's mass
at $\|\distr_1\|_\infty\le B/m$; the elementary bound $\Ent(\nu)\ge-\log_2\|\nu\|_\infty$
(equality iff $\nu$ is uniform on its support) then gives $\Ent(\distr_1)\ge\log_2(m/B)$ for
\emph{every} routing, regardless of $s$. Equality forces $\|\distr_1\|_\infty=B/m$ exactly,
i.e.\ every used set loaded to exactly $B$; since loads sum to $m$, this is only possible
when $s=m/B$ (if $B$ divides $m$; otherwise the bound is strict), a \emph{balanced exact
cover}.

\textbf{Refined bound via the vertex structure of $P_s$.} Fix $s$ and consider
$\max_{c\in P_s}F(c)$. Since $F$ is convex and $P_s$ is a box intersected with one
hyperplane, the maximum is attained at a vertex, and every vertex of such a polytope has
at most one coordinate strictly inside $(1,B)$, the rest sitting at a bound; otherwise two interior coordinates could be perturbed by $\pm\varepsilon$ while preserving feasibility, contradicting extremality. Writing
$q=\lfloor(m-s)/(B-1)\rfloor$ for the number of coordinates pinned to $B$ (each set
starts from a baseline load of $1$ and absorbs excess mass $m-s$ in units of $B-1$ until
exhausted), $\rho=1+(m-s)-q(B-1)\in[1,B]$ for the one fractional coordinate, and the
remaining $s-1-q$ coordinates at $1$, the true (integer-load) optimum satisfies
\[
  F^\star(s) = qB\log_2B+\rho\log_2\rho,\qquad
  \Ent(\distr_1)\ \ge\ \log_2m-\tfrac1mF^\star(s),
\]
exactly, since the integer feasible region is a subset of the relaxed polytope $P_s$.

Since $\rho \in [1,B]$, $\rho\log_2\rho \le B\log_2B$. Then,

\[
F^\star(s) \le qB\log_2B+B\log_2B = (q+1)B\log_2B \le ((m-s)/(B-1)+1)B\log_2B
\]

Hence,
\[
    \Ent(\distr_1)\ \ge \log_2m-\tfrac1m\left(\left(\frac{m-s}{B-1}+1\right)B\log_2B\right) = E^*(s)-\frac{B\log_2B}{m}
\]
where $E^*(s)\ =\ \log_2 m - \frac{(m-s)\,B\log_2 B}{m(B-1)}$.

Hence, $E^*$ is increasing in $s$, matching the monotonicity claimed in the lemma. (For $B=2$ the
correction vanishes identically, since $B-1=1$ divides every integer exactly; the slack
term is only needed for $B\ge3$.)

\textbf{Consistency check.} At $s=m/B$, $m-s=m(B-1)/B$, so
$\frac{(m-s)B\log_2B}{m(B-1)}=\log_2B$ exactly and $E^*(m/B)=\log_2m-\log_2B=\log_2(m/B)$,
recovering the basic lower bound above \emph{exactly}, with no rounding slack, as it
must, since $s=m/B$ with all loads at $B$ is already an integer vertex ($q=s$).

\textbf{Worked example.} Take $m=6$, $B=2$. A perfect-matching cover
$\{S_1,S_2,S_3\}=\{\{1,2\},\{3,4\},\{5,6\}\}$ achieves $s=m/B=3$ with every load exactly
$2$, so $\distr_1$ is uniform on $3$ atoms and $\Ent(\distr_1)=\log_23\approx1.585$, matching
$E^*(3)=\log_26-\log_22=\log_23$ exactly. If instead the available sets force $s=4$ sets
to be used, the entropy-minimising loads are $(2,2,1,1)$ (the vertex above, with
$q=\lfloor(6-4)/1\rfloor=2$ sets at $B=2$ and one fractional/remaining coordinate each at
$1$), giving
\[
  \Ent(\distr_1)=-2\cdot\tfrac26\log_2\tfrac26-2\cdot\tfrac16\log_2\tfrac16\approx1.918,
\]
matching $E^*(4)=\log_26-\frac{(6-4)\cdot2\log_22}{6\cdot1}=\log_26-\tfrac23\approx1.918$;
the two match exactly here, rather than merely up to the $B\log_2B/m$
slack, because $B=2$, for which that correction vanishes identically,
as noted above. This need not happen for $B\ge3$: the general
conclusion is not that $\Ent(\distr_1)$ equals $E^*(s)$, but that both
the achieved minimum entropy and its lower bound $E^*(s)-B\log_2B/m$
increase with $s$, since $E^*$ is increasing in $s$ and the subtracted
term $B\log_2B/m$ does not depend on $s$. Thus being forced from a
minimum $3$-set cover up to a $4$-set cover provably costs at least
$E^*(4)-E^*(3)\approx0.333$ bits of entropy in this lower bound, for
every $B$, exactly the mechanism Theorem~\ref{thm:set-cover-hard}
exploits: an inapproximability gap in cover size becomes, via this
lemma, a fixed additive gap in entropy.
\end{proof}

\subsubsection{A perfect-completeness Set-Cover gap}
\label{sec:bounded-cover-gap}
Generic approximation hardness for Set-Cover would not suffice for the
theorem below: the entropy calculation needs a yes-case that partitions
the universe into maximum-size sets, not merely an approximately-optimal
cover.

\begin{lemma}[A bounded-Set-Cover gap with perfect completeness]
\label{lem:bounded-cover-gap}
There is a fixed rational $c>0$ with the following property. Consider an
explicitly given set system consisting of a nonempty universe $U$ of size
$m$, with $3\mid m$, and sets $S_1,\ldots,S_M\subseteq U$, each of size at
most $B=3$, with $\bigcup_jS_j=U$ (so some cover of $U$ exists). It is
$\NP$-hard to distinguish:
\begin{enumerate}[(i)]
\item the case where $U$ can be partitioned into $m/3$ of the given sets,
each of size exactly $3$;
\item the case where every cover of $U$ by the given sets uses at least
$(1+c)m/3$ of them.
\end{enumerate}
\end{lemma}
\begin{proof}
Use the perfect-completeness gap for bounded-occurrence
three-dimensional matching~\cite[Theorem 4.4]{petrank1994gap}. In the form
needed here, there is a fixed rational $\eta\in(0,1)$, independent of the
instance, such that it is $\NP$-hard to distinguish a triple system with a
perfect matching (a partition of the universe into disjoint triples) from
one whose every disjoint matching covers at most a $(1-\eta)$ fraction of
the universe. The cited hard-gap statement counts elements occurring exactly
once in a chosen subfamily; a disjoint matching is a special case, so its
soundness implies this version. Taking three disjoint copies if necessary
makes the universe size $m=3k$ without changing either ratio; write $k$ for
this quotient (the number of triples in a perfect matching, when one
exists).

Regard the triples as available sets and add every singleton. This ensures
that the union is the whole universe and preserves the bound $B=3$. A
perfect matching supplies the yes-case partition, and no cover can use
fewer than $k$ sets because each set has size at most three.

For soundness, consider any cover consisting of $s_3$ triples and $s_1$
singletons, with $s=s_3+s_1$. If $U_3$ is the union of its triples, every
element of the universe outside $U_3$ must be covered by some singleton
(triples and singletons are the only available sets), so at most $s_1$
elements lie outside $U_3$: $|U_3|\ge m-s_1$. Their excess incidence count
is $R\defeq3s_3-|U_3|$ (the total multiplicity, $3s_3$, of elements across
the $s_3$ triples, minus the number of distinct elements they cover); since
$|U_3|\ge m-s_1$, $R\le3s_3+s_1-m$.

Repeatedly delete a triple that intersects another remaining triple, until
the remaining triples are pairwise disjoint. Each deletion decreases $R$,
recomputed on the remaining triples, by at least one: the deleted triple
shares at least one element with another remaining triple, so at most two
of its three elements are covered by no other remaining triple, and
removing it can only decrease the union of the remaining triples by that
many $(|U'_3|\ge |U_3|-2)$, while the triple count drops by exactly one $(s'_3 = s_3-1)$; a short computation
gives $R' = 3s'_3-|U'_3| \le 3s_3-3-|U_3|+2 = R-1$. Since $R\ge0$ throughout and strictly decreases at each
step, at most $R$ deletions occur before the remaining triples are
disjoint, and at least $s_3-R$ triples remain. Using $R\le3s_3+s_1-m$,
\[
s_3-R\ \ge\ m-2s_3-s_1\ \ge\ m-2(s_3+s_1)\ =\ m-2s\ =\ 3k-2s.
\]

The remaining triples form a disjoint matching drawn from the original
triple system, of size at least $3k-2s$. In a no-instance, every such
matching has size at most $(1-\eta)k$ (a matching of $j$ triples covers
$3j$ elements, and $3j\le(1-\eta)\cdot3k$ gives $j\le(1-\eta)k$). Therefore
$3k-2s\le(1-\eta)k$, i.e.\ $s\ge(1+\eta/2)k$. Setting $c\defeq\eta/2>0$, a
fixed rational independent of the instance, gives $s\ge(1+c)k=(1+c)m/3$,
proving soundness. The transformation, including the added singletons, is
computable in time polynomial in $m$, proving the lemma.
\end{proof}

\subsubsection{The hardness reduction}
\label{sec:hardness-reduction}

By the same reasoning used to prove \cref{lem:set-cover-collapse}, but without
restricting to a pure profile: since $\distro$ is fixed data of the instance,
not a choice made by any strategy, and since every state in the support of
the resulting $\RandDistr_1$ is absorbing, the entropy trajectory under
\emph{any} $\stratmin\in\StratMin^\Sem$, of any memory or randomization,
still takes only two values, $\log_2m$ at time $0$ and $\Ent(\RandDistr_1)$
forever after, so
$$
\mathsf{Obj}^\Sem_\Discount(\stratmin;\distro)
=(1-\Discount)\log_2m+\Discount\,\mathbb E\bigl[\Ent(\RandDistr_1)\bigr],
$$
 In
particular the objective depends on $\stratmin$ only through this one
expectation, not on anything else the strategy does.

Moreover $\distro$ is fixed before play begins, so there is no history prior
to \MinPl's first move: a history-dependent strategy's action at time $0$ is
just some (possibly randomized) choice of action at each $e_u$, exactly as
available to a memoryless strategy, and nothing said above used purity or
absence of memory to reach this point. By \cref{lem:pure-suffices}, applied
at $\distro$, this one-shot minimization over randomized choices is attained,
under every semantics, by a pure profile $a$, i.e.\ a fixed assignment of one
action to each $e_u$ with no time or history dependence at all, which is
precisely a memoryless strategy. Hence
$$
\inf_{\stratmin\in\StratMin^\Sem}\mathsf{Obj}^\Sem_\Discount(\stratmin;\distro)
=\min_{a}\ \mathsf{Obj}^\Sem_\Discount(a;\distro),
$$
the minimum on the right taken over pure profiles $a$, which is exactly the
quantity \cref{lem:set-cover-collapse} computes. The hardness result below,
being a statement about this infimum, therefore already applies with \MinPl
restricted to memoryless strategies.

\onephardness*

\begin{proof}
\emph{Intuition.} Lemma~\ref{lem:set-cover-entropy} says that using more sets than
the minimum cover costs entropy: spreading the same mass over extra sets can only raise
$\Ent(\distr_1)$. So if a Set-Cover instance is hard to approximate, meaning its minimum
cover jumps by a constant factor when no small cover exists, that same jump forces a gap
onto $\min\Ent(\distr_1)$. E.g. on $6$
elements with $B=2$: a minimum $3$-set cover gives $\Ent\approx1.585$, while being forced
to $4$ sets costs $\approx0.333$ more bits.

\emph{Reducing to deterministic routings.} For a randomised strategy, the minimum cover
strategies is attained at a deterministic (pure) routing under every semantics, by
\cref{lem:pure-suffices}; we therefore analyse deterministic
routings, each of which sends every $u$ to exactly one $S_j\ni u$. By
Lemma~\ref{lem:set-cover-collapse} it suffices to prove both claims for $\Ent(\distr_1)$
in Construction~\ref{constr:set-cover}; the discounted case then follows by the affine
substitution noted in the main text (which maps a fixed additive gap $\varepsilon_0$ in $\Ent(\distr_1)$ to
a fixed additive gap $\gamma\varepsilon_0$ in $\mathsf{Obj}_\gamma$, and an exact
threshold to an exact threshold). Construction~\ref{constr:set-cover} is computable in
time polynomial in $m+M$.

\emph{Turning the Set-Cover gap into an entropy gap.} Fix $B=3$. The
theorem only requires \emph{some} fixed set-size bound with a fixed gap,
not a bound holding for every $B$, so it suffices to exhibit one, and
$B=3$ is what Lemma~\ref{lem:bounded-cover-gap} supplies: a fixed
rational $c>0$ such that it is $\NP$-hard to distinguish a set system on
a universe of size $m$ (all sets of size at most $3$) that admits an
exact partition into $m/3$ triples from one where every set cover needs
at least $(1+c)m/3$ sets. This constant $c$ comes from
Lemma~\ref{lem:bounded-cover-gap}'s reduction from
bounded-occurrence three-dimensional matching~\cite[Theorem 4.4]{petrank1994gap}, not
from the entropy game; all we use here is that it is fixed, positive,
and independent of $m$.

In the first case,
routing every $u$ through a set of a minimum cover of size $m/B$ gives every load exactly
$B$, so $\min\Ent(\distr_1)=\log_2(m/B)$ by the equality case of
Lemma~\ref{lem:set-cover-entropy}. In the second, every routing's support size $s$
satisfies $s\ge(1+c)m/B$ (any routing's support is a cover), so by
Lemma~\ref{lem:set-cover-entropy},
\begin{align*}
  \Ent(\distr_1) &\ge\ E^*\!\big((1+c)m/B\big)-\frac{B\log_2B}{m}\\
  &=\ \log_2(m/B)+\frac{c\log_2B}{B-1}-\frac{B\log_2B}{m}
\end{align*}
For $m$ larger than a constant depending only on $B,c$ $\left(m\ge\frac{2B(B-1)}{c}\right)$, the correction term $O(\log_2B/m)$
term is below half of the fixed constant $\frac{c\log_2B}{B-1}$, giving
$\Ent(\distr_1)\ge\log_2(m/B)+\frac{c\log_2B}{4(B-1)}+\varepsilon_0$ for
$\varepsilon_0\defeq\frac{c\log_2B}{4(B-1)}>0$, a constant independent of the instance.
Taking $v=\log_2(m/B)+\frac{c\log_2B}{4(B-1)}$, this is exactly a valid $\varepsilon_0$-gap promise separating
the two cases: $\min\Ent(\distr_1)=\log_2(m/B)\le v-\varepsilon_0$ in the first case and $\min\Ent(\distr_1)\ge v+\varepsilon_0$
in the second. This
gives an $\NP$-hard many-one reduction for $\min\Ent(\distr_1)$, at threshold
$v=\log_2(m/B)+\frac{c\log_2B}{4(B-1)}$ and gap $\varepsilon_0 = \frac{c\log_2B}{4(B-1)}$, and it transfers to $\mathsf{Obj}_\gamma$ by the substitution of
Remark~\ref{rem:disc-substitution}.

\emph{Extending to $\mathsf{Obj}_\gamma$:} For a fixed $c>0$, taking $v^\ast = (1-\gamma)\log_2m+\gamma\cdot\left(\log_2(m/B)+\frac{c\log_2B}{4(B-1)}\right)$ and $\varepsilon^\ast = \gamma \cdot \frac{c\log_2B}{4(B-1)}$ separates the two cases for $\mathsf{Obj}_\gamma$.

\emph{Rational threshold and gap.} The problem $\GapVal_{\Sem,\varepsilon_0}$ asks for a rational threshold and a rational gap, whereas $v^\ast$ and $\varepsilon^\ast$ are irrational in general (they involve $\log_2B$ and $\log_2(m/B)$). Fix a rational $\varepsilon_1$ with $0<\varepsilon_1<\varepsilon^\ast/2$; it depends only on $\gamma$ and $c$, not on the instance. Using the logarithm series from the proof of \cref{ass:rational}, compute in time polynomial in $m$ a rational $v'$ with $|v'-v^\ast|<\varepsilon^\ast/2$. In the first case the value is at most $v^\ast-\varepsilon^\ast<v'-\varepsilon_1$, and in the second it is at least $v^\ast+\varepsilon^\ast>v'+\varepsilon_1$. Hence $(v',\varepsilon_1)$ is a rational threshold and gap that separate the two cases, and $\GapVal_{\Sem,\varepsilon_1}$ is $\NP$-hard for every $\Sem$. Instances with $m$ below the constant $2B(B-1)/c$ are solved by brute force, so they do not affect hardness.

\medskip

\emph{Exact $\NP$-hardness (\cref{thm:set-cover-exact}).} The exact problem needs a
\emph{rational} threshold, whereas $\log_2(m/B)$ is generally irrational, so we
do not use it as the threshold. Let $M\defeq\log_2(m/B)+\varepsilon_0$ be the
midpoint of the promise gap above, so that yes-instances have
$\min\Ent(\distr_1)\le M-\varepsilon_0$ and no-instances have
$\min\Ent(\distr_1)\ge M+\varepsilon_0$. Using the rapidly convergent series
for the logarithm from the proof of \cref{ass:rational}, compute in time polynomial in
$m$ a rational $v'$ with $|v'-M|<\varepsilon_0/2$ (here $\varepsilon_0$ is a fixed rational
constant, or any rational constant below the fixed gap). Then
$M-\varepsilon_0<v'<M+\varepsilon_0$, hence
$\min\Ent(\distr_1)\le v'$ on yes-instances and
$\min\Ent(\distr_1)> v'$ on no-instances. Thus the exact comparison
``$\min\Ent(\distr_1)\le v'$'' decides the same $\NP$-hard cover-size question, with the
rational threshold $v'$. For $\mathsf{Obj}_\gamma$ the same argument is applied to the
midpoint $(1-\gamma)\log_2m+\gamma M$ with tolerance $\gamma\varepsilon_0/2$. The gap
statement is thus the stronger of the two, and exact hardness follows from it.
\end{proof}

\section{Comparing the semantics}
\label{app:semantic-comparison}
This appendix proves \cref{thm:semantic-value-dominance} and gives the examples showing that $V_{\mathsf{sp}}$ and $V_{\mathsf{st}}$ are incomparable. Below, ``pure'' strategy has the meaning of \cref{sec:model}: it selects one pure profile.

\begin{remark}[Deterministic transitions]
If every enabled transition is deterministic, then for every profile pair $(a,b)$ the realisation $\xi$ is determined, so $\Psi_\xi=\Phi_{a,b}$ and the sampled-profile and sampled-transition updates coincide for every profile law; hence $V_{\mathsf{sp}}=V_{\mathsf{st}}$ in such arenas. Deterministic transitions alone do not give equality with the averaged values.
\end{remark}

\begin{remark}[Pure strategies]
\label{rem:pure-coincidence}
For the same pair of pure Borel distribution-history strategies, the
averaged-distribution and sampled-profile path laws, and hence their discounted
payoffs, coincide, even with stochastic arena transitions. Indeed, a pure
strategy selects one pure profile at each history. Its averaged prescription is
the family of Dirac actions at that profile, and its sampled prescription is
the Dirac law on it. At any common history, with selected pair $(a,b)$, both
next-state laws equal $\delta_{\Phi_{a,b}(\mu)}$, because
$\DistOp_{a,b}=\Phi_{a,b}$. Starting from the same distribution, induction
shows that every finite prefix agrees, so all later selected profiles agree as
well, even if the strategies depend on the whole distribution history. With
deterministic transitions as well, the sampled-transition law agrees too (previous
remark), so all three path laws and payoffs coincide for pure strategies.
These are equalities of induced laws for specified strategies, not an
identification of all admissible averaged and sampled strategies: in
\cref{thm:nondetermined} the game is deterministic, but the averaged upper value
is one, the averaged lower value is $(1-\Discount)+\Discount h(1/4)$, and the
two sampled values equal $(1-\Discount)+\Discount/2$.
\end{remark}

\subsection{Open-loop comparison of the three semantics}
\label{app:cmp-openloop}
\begin{theorem}[Open-loop entropy ordering]
\label{thm:open-loop-order}
Let $(\sigma_t,\tau_t)_{t\ge0}$ be prescribed profile laws depending on
time only, not on observed distributions. Under averaging use their
statewise marginals $x_t,y_t$; under both sampled semantics use the laws
$\sigma_t,\tau_t$ themselves, with fresh independent draws each round.
Write $\mu_t^{\mathsf{av}}$ for the deterministic averaged trajectory.
For every $t\ge0$,
\[
\Ent(\mu_t^{\mathsf{av}})
\ge\mathbb E^{\mathsf{sp}}[\Ent(\RandDistr_t)]
\ge\mathbb E^{\mathsf{st}}[\Ent(\RandDistr_t)].
\]
Consequently the same weak ordering holds for their normalised discounted
payoffs. No independence between a player's different state coordinates
is required. The inequalities need not be strict.
\end{theorem}

\begin{proof}
Throughout, treat a distribution $\mu\in\DistS$ as a row vector in
$\mathbb R^{1\times\States}$, so that a stochastic matrix acts on it by
right multiplication. For each $j\ge0$, draw a profile pair $(A_j,B_j)$
with law $\sigma_j\times\tau_j$, independently across $j$; conditional
on all of $(A_j,B_j)_{j\ge0}$, draw realisations $\xi_j\sim
P_{A_j,B_j}$, independently across $j$. For a fixed time $t$ only
finitely many of these draws are ever used. Define two processes from
this single common experiment: $\RandDistr_0^{\mathsf{sp}}=\RandDistr_0^{\mathsf{st}}=\distro$,
and, for $j\ge0$,
\[
\RandDistr_{j+1}^{\mathsf{sp}}=\Phi_{A_j,B_j}(\RandDistr_j^{\mathsf{sp}}),
\qquad
\RandDistr_{j+1}^{\mathsf{st}}=\Psi_{\xi_j}(\RandDistr_j^{\mathsf{st}}).
\]
Since both use the same profile draws $(A_j,B_j)_{j\ge0}$, and since
$\sigma_j,\tau_j$ are open-loop (do not depend on the observed
distribution), each has exactly the marginal law of the corresponding
semantics' distribution process: this is immediate from
\eqref{eq:sp-update} and \eqref{eq:st-update}, applied round by round.
So it suffices to prove the stated inequalities for
$\RandDistr_t^{\mathsf{sp}},\RandDistr_t^{\mathsf{st}}$ as just
constructed.

Let $K_{a,b}$ be the stochastic matrix with entries
$(K_{a,b})_{s,s'}=\Trans(s,\rho_{a,b}(s),s')$, and let $R_\xi$ be the
$\{0,1\}$-valued matrix with $(R_\xi)_{s,s'}=\mathbf1_{\{\xi(s)=s'\}}$. By
\eqref{eq:phi-def} and \eqref{eq:psi-def},
$\Phi_{a,b}(\mu)=\mu K_{a,b}$ and $\Psi_\xi(\mu)=\mu R_\xi$ for every
$\mu$, so
\begin{equation}
\label{eq:open-loop-exact-processes}
\RandDistr_t^{\mathsf{sp}}=\distro K_{A_0,B_0}\cdots K_{A_{t-1},B_{t-1}},
\qquad
\RandDistr_t^{\mathsf{st}}=\distro R_{\xi_0}\cdots R_{\xi_{t-1}},
\end{equation}
both exactly, as products of random matrices. Since the coordinates of
$\xi$ are independent given $(A,B)$ with $\xi(s)\sim\Trans(s,\rho_{A,B}(s),\cdot)$,
$\mathbb E[R_{\xi_j}\mid A_j=a,B_j=b]=K_{a,b}$ for every $j,a,b$.

We first show, by induction on $t$, that
$\mathbb E[\RandDistr_t^{\mathsf{st}}\mid\mathcal A_t]=\distro
K_{A_0,B_0}\cdots K_{A_{t-1},B_{t-1}}$, where $\mathcal
A_t=(A_0,B_0,\ldots,A_{t-1},B_{t-1})$. The base case $t=0$ is the
equality $\distro=\distro$. For the step, write
$\RandDistr_{t+1}^{\mathsf{st}}=\RandDistr_t^{\mathsf{st}}R_{\xi_t}$.
The profile pairs are independent across time, so $(A_t,B_t)$ is
independent of $\mathcal A_t$; and, conditional on all profiles,
$\xi_t$ is independent of $\xi_0,\ldots,\xi_{t-1}$, hence of
$\RandDistr_t^{\mathsf{st}}$, which is a function of
$\xi_0,\ldots,\xi_{t-1}$ and $\mathcal A_t$ alone. So, conditional on
$\mathcal A_{t+1}=(\mathcal A_t,A_t,B_t)$, the row vector
$\RandDistr_t^{\mathsf{st}}$ and the matrix $R_{\xi_t}$ are
independent, and their product's conditional mean factors entrywise:
\[
\mathbb E\bigl[\RandDistr_t^{\mathsf{st}}R_{\xi_t}\mid\mathcal A_{t+1}\bigr]
=\mathbb E\bigl[\RandDistr_t^{\mathsf{st}}\mid\mathcal A_{t+1}\bigr]\cdot
\mathbb E\bigl[R_{\xi_t}\mid\mathcal A_{t+1}\bigr].
\]
The first factor equals $\mathbb E[\RandDistr_t^{\mathsf{st}}\mid\mathcal
A_t]$, since conditioning further on the independent pair $(A_t,B_t)$
does not change it, which is $\distro
K_{A_0,B_0}\cdots K_{A_{t-1},B_{t-1}}$ by the induction hypothesis. The
second factor is $\mathbb E[R_{\xi_t}\mid A_t,B_t]=K_{A_t,B_t}$, since
$\xi_t$'s law depends only on $(A_t,B_t)$. Multiplying proves the
inductive step. Combined with \eqref{eq:open-loop-exact-processes},
this gives
\begin{equation}
\label{eq:open-loop-conditional-mean}
\mathbb E[\RandDistr_t^{\mathsf{st}}\mid\mathcal A_t]
=\distro K_{A_0,B_0}\cdots K_{A_{t-1},B_{t-1}}
=\RandDistr_t^{\mathsf{sp}}.
\end{equation}

Next, define the marginal action at $s\in\States_{\min}$ by
$x_j(s)(\alpha)=\sum_{a:a(s)=\alpha}\sigma_j(a)$, and $y_j$ analogously
from $\tau_j$ at $s\in\States_{\max}$, and set $\overline
K_j\defeq\sum_{a,b}\sigma_j(a)\tau_j(b)K_{a,b}$. At a Min state $s$,
every $K_{a,b}$'s row depends on $a,b$ only through $a(s)$, so
$(\overline K_j)_{s,\cdot}=\sum_a\sigma_j(a)\Trans(s,a(s),\cdot)=\sum_\alpha
x_j(s)(\alpha)\Trans(s,\alpha,\cdot)$, and symmetrically at a Max
state; comparing with \eqref{eq:distribution-update} identifies
$\overline K_j$ exactly with the matrix of $\DistOp_{x_j,y_j}$, using
only the one-state marginals $x_j(s),y_j(s)$, not any correlation
$\sigma_j,\tau_j$ may impose across different states. Consequently
$\mu_{j+1}^{\mathsf{av}}=\DistOp_{x_j,y_j}(\mu_j^{\mathsf{av}})=\mu_j^{\mathsf{av}}\overline
K_j$, so by induction on $j$, $\mu_t^{\mathsf{av}}=\distro\overline
K_0\cdots\overline K_{t-1}$ exactly.

We now show, again by induction on $t$, that
$\mathbb E[\RandDistr_t^{\mathsf{sp}}]=\distro\overline K_0\cdots\overline
K_{t-1}$. The base case is trivial. For the step,
$\RandDistr_{t+1}^{\mathsf{sp}}=\RandDistr_t^{\mathsf{sp}}K_{A_t,B_t}$
with $\RandDistr_t^{\mathsf{sp}}$ a function of $\mathcal A_t$
independent of $(A_t,B_t)$, so
\[
\mathbb E[\RandDistr_{t+1}^{\mathsf{sp}}]
=\mathbb E[\RandDistr_t^{\mathsf{sp}}]\cdot\mathbb E[K_{A_t,B_t}]
=\bigl(\distro\overline K_0\cdots\overline K_{t-1}\bigr)\overline K_t,
\]
using $\mathbb E[K_{A_t,B_t}]=\sum_{a,b}\sigma_t(a)\tau_t(b)K_{a,b}=\overline
K_t$, which proves the step. Combined with $\mu_t^{\mathsf{av}}=\distro\overline
K_0\cdots\overline K_{t-1}$ above, this gives
\begin{equation}
\label{eq:open-loop-unconditional-mean}
\mathbb E[\RandDistr_t^{\mathsf{sp}}]
=\distro\overline K_0\cdots\overline K_{t-1}
=\mu_t^{\mathsf{av}}.
\end{equation}

Both \eqref{eq:open-loop-conditional-mean} and
\eqref{eq:open-loop-unconditional-mean} are exact identities between
(conditional) means of matrix products; being linear in the underlying
distribution, they use no concavity and are not an instance of
replacing the expectation of a nonlinear quantity by its value at the
mean. Concavity enters only now. Applying $\Ent$'s concavity to
\eqref{eq:open-loop-conditional-mean}, in the form
$\Ent(\mathbb E[\RandDistr_t^{\mathsf{st}}\mid\mathcal
A_t])\ge\mathbb E[\Ent(\RandDistr_t^{\mathsf{st}})\mid\mathcal A_t]$,
and taking expectations over $\mathcal A_t$, gives
$\Ent(\RandDistr_t^{\mathsf{sp}})\ge\mathbb
E[\Ent(\RandDistr_t^{\mathsf{st}})\mid\mathcal A_t]$ pointwise in
$\mathcal A_t$ (the left side is a function of $\mathcal A_t$ alone by
\eqref{eq:open-loop-exact-processes}), so taking a further expectation
gives $\mathbb E[\Ent(\RandDistr_t^{\mathsf{sp}})]\ge\mathbb
E[\Ent(\RandDistr_t^{\mathsf{st}})]$, the $\mathsf{sp}$ versus $\mathsf{st}$
inequality. Applying concavity to
\eqref{eq:open-loop-unconditional-mean} directly gives
$\Ent(\mu_t^{\mathsf{av}})=\Ent(\mathbb
E[\RandDistr_t^{\mathsf{sp}}])\ge\mathbb
E[\Ent(\RandDistr_t^{\mathsf{sp}})]$, the $\mathsf{av}$ versus $\mathsf{sp}$
inequality. Multiplying both by the nonnegative weights
$(1-\Discount)\Discount^t$ and summing over $t$ proves the discounted
comparison; boundedness of entropy by $\log_2n$ justifies passing from
finite partial sums to the infinite discounted sum.

Strict concavity of $\Ent$ (\cref{prop:entropy-props}) further
identifies equality at a given $t$: $\mathsf{av}$ versus $\mathsf{sp}$ is an
equality exactly when $\RandDistr_t^{\mathsf{sp}}$ is almost surely
constant, and $\mathsf{sp}$ versus $\mathsf{st}$ is an equality exactly when
$\RandDistr_t^{\mathsf{st}}$ is conditionally almost surely constant
given $\mathcal A_t$. Neither holds in general, so strictness is not
universal even in nontrivial arenas. This argument is stated only for
open-loop prescriptions because it depends on $\xi_t$ being independent
of $\RandDistr_t^{\mathsf{st}}$ given $\mathcal A_{t+1}$: under
feedback, $\sigma_t,\tau_t$ may depend on the observed
$\RandDistr_t^{\mathsf{sp}}$ or $\RandDistr_t^{\mathsf{st}}$, which in
turn depends on earlier realisations, so conditioning on a later
profile can reveal information about an earlier one, and the
independence used above need not hold.
\end{proof}

\subsection{Concavity of the averaged values and dominance}
\label{app:cmp-concavity}
For a continuous concave $f:\DistS\to\mathbb R$, write
\[
g_f^+(\mu)=\min_x\max_y f(\DistOp_{x,y}(\mu)),\qquad
g_f^-(\mu)=\max_y\min_x f(\DistOp_{x,y}(\mu)).
\]
All extrema are attained by compactness and continuity.

\begin{lemma}[Mass-weighted mixing]
\label{lem:mass-weighted-concavity}
Both $g_f^+$ and $g_f^-$ are continuous and concave. For fixed $\mu,y$,
the minimum over $x$ in $g_f^-$ is attained at a pure profile. The
minimum over $x$ in $g_f^+$ is also attained at a pure profile.
\end{lemma}
\begin{proof}
Let $\mu=\lambda\mu_1+(1-\lambda)\mu_2$ with $0<\lambda<1$. For any two
maximiser prescriptions $y_1,y_2\in\mathcal Y$, define, at every
Max-controlled state $s$ with $\mu(s)>0$,
\[
y(s)=\frac{\lambda\mu_1(s)y_1(s)+(1-\lambda)\mu_2(s)y_2(s)}{\mu(s)}.
\]
This is a nonnegative combination of two probability distributions on
the enabled actions at $s$, with total weight
$\bigl(\lambda\mu_1(s)+(1-\lambda)\mu_2(s)\bigr)/\mu(s)=1$, hence itself
a probability distribution on those actions. At a state of zero mass,
both $\mu_1(s)$ and $\mu_2(s)$ vanish; choose any enabled Dirac action
there. This defines $y\in\mathcal Y$, a single legitimate prescription,
not a strategy that observes which of $\mu_1,\mu_2$ has occurred. It
satisfies, for every $x\in\mathcal X$,
\begin{equation}
\label{eq:mass-weighted-update}
\DistOp_{x,y}(\mu)
=\lambda\DistOp_{x,y_1}(\mu_1)
+(1-\lambda)\DistOp_{x,y_2}(\mu_2).
\end{equation}
At Min states this holds because the same $x$ appears in both terms; at
Max states it holds by construction of $y$: both sides equal, at
$s'\in\States$,
$\sum_s\bigl(\lambda\mu_1(s)y_1(s)+(1-\lambda)\mu_2(s)y_2(s)\bigr)(\alpha)\Trans(s,\alpha,s')$
summed over $\alpha$, by \eqref{eq:distribution-update} and the
definition of $y$.

\emph{Concavity of $g_f^-$.} For $i=1,2$, let $y_i\in\mathcal Y$ attain
the outer maximum in $g_f^-(\mu_i)=\max_{y'}\min_xf(\DistOp_{x,y'}(\mu_i))$,
i.e.\ $\min_xf(\DistOp_{x,y_i}(\mu_i))=g_f^-(\mu_i)$, and build $y$ as
above. Concavity of $f$ and \eqref{eq:mass-weighted-update} give, for
every $x\in\mathcal X$,
\begin{align*}
f(\DistOp_{x,y}(\mu))
&\ge\lambda f(\DistOp_{x,y_1}(\mu_1))
+(1-\lambda)f(\DistOp_{x,y_2}(\mu_2))\\
&\ge\lambda\min_{x'}f(\DistOp_{x',y_1}(\mu_1))
+(1-\lambda)\min_{x'}f(\DistOp_{x',y_2}(\mu_2))\\
&=\lambda g_f^-(\mu_1)+(1-\lambda)g_f^-(\mu_2).
\end{align*}

Taking the minimum over $x\in\mathcal X$ on the left preserves this
lower bound, giving $\min_xf(\DistOp_{x,y}(\mu))\ge\lambda
g_f^-(\mu_1)+(1-\lambda)g_f^-(\mu_2)$. Since $y$ is one particular
feasible choice in $\mathcal Y$,
\[
g_f^-(\mu)=\max_{y'}\min_xf(\DistOp_{x,y'}(\mu))\ge\min_xf(\DistOp_{x,y}(\mu)),
\]
so $g_f^-(\mu)\ge\lambda g_f^-(\mu_1)+(1-\lambda)g_f^-(\mu_2)$, proving
concavity of $g_f^-$.

\emph{Concavity of $g_f^+$.} Fix $x\in\mathcal X$. For $i=1,2$, let
$y_i\in\mathcal Y$ attain $\max_{y'}f(\DistOp_{x,y'}(\mu_i))$ and build
$y$ as above; the same two displayed inequalities, with $\min_{x'}$
everywhere replaced by the fixed $x$, show that
$\mu\mapsto\max_{y'}f(\DistOp_{x,y'}(\mu))$ is concave, for this fixed
$x$. A pointwise infimum of concave functions is concave: if
$h_x(\lambda\mu_1+(1-\lambda)\mu_2)\ge\lambda h_x(\mu_1)+(1-\lambda)h_x(\mu_2)$
holds for every $x$ in an index set, then taking $\inf_x$ of both sides
preserves the inequality with $h_x$ replaced by $\inf_xh_x$ on both
occurrences. Applying this to $h_x(\mu)=\max_{y'}f(\DistOp_{x,y'}(\mu))$
shows $g_f^+=\inf_xh_x$ is concave.

\emph{Continuity.} $f$ is continuous on the compact set $\DistS$, hence
uniformly continuous, and $(\mu,x,y)\mapsto\DistOp_{x,y}(\mu)$ is
jointly continuous (\cref{prop:compact}), so
$(\mu,x,y)\mapsto f(\DistOp_{x,y}(\mu))$ is continuous on the compact
set $\DistS\times\mathcal X\times\mathcal Y$, hence itself uniformly
continuous there. Fix $\eta>0$ and let $\rho>0$ be such that
$d(\mu,\mu')<\rho$ implies
$|f(\DistOp_{x,y}(\mu))-f(\DistOp_{x,y}(\mu'))|<\eta$ for every
$x\in\mathcal X,y\in\mathcal Y$. For such $\mu,\mu'$, every term inside
the outer extrema defining $g_f^+(\mu)$ and $g_f^+(\mu')$ changes by
less than $\eta$, so, exactly as in
\eqref{eq:bellman-proof-matrix-stability}, the resulting $\min\max$
values satisfy $|g_f^+(\mu)-g_f^+(\mu')|\le\eta$; the same bound holds
for $g_f^-$. Since $\eta$ was arbitrary, both $g_f^+$ and $g_f^-$ are
(uniformly) continuous on $\DistS$.

\emph{Vertex attainment.} For fixed $\mu,y$, the map
$x\mapsto f(\DistOp_{x,y}(\mu))$ is concave on the finite product of
simplices $\mathcal X$, so its minimum is attained at an extreme point
of $\mathcal X$, i.e.\ at a pure profile: this proves the claim for
$g_f^-$, where $y$ is already fixed before the minimum over $x$ is
taken. For $g_f^+$, the map $(x,y)\mapsto f(\DistOp_{x,y}(\mu))$ is
jointly concave, since $\DistOp_{x,y}(\mu)$ is affine in the joint
action $(x,y)$ and $f$ is concave. Partial maximisation of a jointly
concave function over the convex set $\mathcal Y$ preserves concavity
in the remaining variable, so $x\mapsto\max_{y}f(\DistOp_{x,y}(\mu))$
is concave on $\mathcal X$. Its minimum over
$x\in\mathcal X$ is therefore again attained at a vertex, i.e.\ a pure
profile. Neither argument requires, or asserts, that Max has a pure
optimiser: only Min's minimisation is claimed to be vertex-attaining.
\end{proof}
\semanticvaluedominance*
\begin{proof}
Write $U_r^+,U_r^-$ for the two $r$-stage averaged values, and
$W_r^\Sem$ for the common sampled values. By the finite-horizon identification in the proof of \cref{thm:bellman} (\cref{app:bellman-proof}), $U_r^\pm=V^\pm_{\mathsf{av},r}=(\mathcal T^\pm_{\mathsf{av}})^r0$ are continuous and satisfy the averaged recursions below, with $0\le V^\pm_{\mathsf{av}}-U_r^\pm\le\Discount^r\log_2n$. The finite-horizon recursion
for $W_r^\Sem$ was proved in
\cref{thm:sp-determined}. Both start
at zero, $U_0^\pm\equiv W_0^\Sem\equiv0$, and, using the notation
$g_f^\pm$ of \cref{lem:mass-weighted-concavity}, the averaged
recursions read $U_{r+1}^{\pm}=(1-\Discount)\Ent+\Discount
g_{U_r^{\pm}}^{\pm}$. We first show by induction on $r$ that both
$U_r^\pm$ are continuous and concave. The base case $U_0^\pm\equiv0$ is
trivial. For the inductive step, if $U_r^\pm$ is continuous and
concave, \cref{lem:mass-weighted-concavity} (applied with $f=U_r^\pm$)
makes $g_{U_r^\pm}^\pm$ continuous and concave; since $\Ent$ is itself
continuous and concave (\cref{prop:entropy-props}), so is the sum
$U_{r+1}^\pm=(1-\Discount)\Ent+\Discount g_{U_r^\pm}^\pm$.

We prove simultaneously that $U_r^-\ge W_r^{\mathsf{sp}}$ and
$U_r^-\ge W_r^{\mathsf{st}}$, by induction on $r$; the base case $r=0$
is the equality $0=0$. Fix $\mu\in\DistS$ and a semantics
$\Sem\in\{\mathsf{sp},\mathsf{st}\}$, and let $\tau\in\Dist(\mathcal
P_{\max})$ be an optimal column strategy in the matrix
$M_{W_r^\Sem}^\Sem(\mu)$, i.e.\ one attaining the outer maximum in the
one-sided characterisation \eqref{eq:val-two-forms} of
$\operatorname{val}(M_{W_r^\Sem}^\Sem(\mu))$. Define $\bar y\in\mathcal
Y$ by taking, at each Max-controlled state $s$, the marginal of $\tau$
at $s$: $\bar y(s)(\beta)\defeq\sum_{b:b(s)=\beta}\tau(b)$. For every
pure row $a\in\mathcal P_{\min}$, the mean-successor identities give
\[
\DistOp_{a,\bar y}(\mu)=\sum_b\tau(b)\Phi_{a,b}(\mu)
=\sum_{b,\xi}\tau(b)P_{a,b}(\xi)\Psi_\xi(\mu).
\]
The first equality needs only the one-coordinate marginals of $\tau$,
not a product-law assumption. Put $f=U_r^-$, continuous and concave by
the previous paragraph. Recall the kernel $K_\Sem(\cdot\mid\mu,a,b)$ of
Appendix~\ref{sec:determinacy}, the law of the next distribution given
the current one and a pure-profile pair, so that
$\DistOp_{a,\bar y}(\mu)=\sum_b\tau(b)\int_{\DistS}\nu\,K_\Sem(d\nu\mid\mu,a,b)$;
and write
\[
M_{W_r^\Sem}^\Sem(\mu)_{a,b}
\defeq\int_{\DistS}W_r^\Sem(\nu)\,K_\Sem(d\nu\mid\mu,a,b)
\]
for the matrix of $r$-stage continuation values at $\mu$, matching the
notation $M_f^\Sem$ of Appendix \ref{sec:onep} at $f=W_r^\Sem$. Jensen's
inequality, applied to the concave $f$ over the mixture displayed
above, and then the induction hypothesis $f\ge W_r^\Sem$, give for
every $a\in\mathcal P_{\min}$
\begin{align}
f(\DistOp_{a,\bar y}(\mu))
&\ge\sum_b\tau(b)\int f(\nu)K_\Sem(d\nu\mid\mu,a,b)
\label{eq:svd-jensen-step}\\
&\ge\sum_b\tau(b)\int W_r^\Sem(\nu)K_\Sem(d\nu\mid\mu,a,b)
=\sum_b\tau(b)M_{W_r^\Sem}^\Sem(\mu)_{a,b}.
\notag
\end{align}
By \cref{lem:mass-weighted-concavity}, the minimum of
$x\mapsto f(\DistOp_{x,\bar y}(\mu))$ over all $x\in\mathcal X$ is
already attained at a pure row, so $\min_{a\in\mathcal
P_{\min}}f(\DistOp_{a,\bar y}(\mu))=\min_{x\in\mathcal
X}f(\DistOp_{x,\bar y}(\mu))$. Since $\bar y$ is one particular
feasible element of $\mathcal Y$,
\[
g_f^-(\mu)=\max_{y'\in\mathcal Y}\min_{x\in\mathcal X}f(\DistOp_{x,y'}(\mu))
\ge\min_{x\in\mathcal X}f(\DistOp_{x,\bar y}(\mu))
=\min_{a\in\mathcal P_{\min}}f(\DistOp_{a,\bar y}(\mu)).
\]
Taking the minimum over $a$ in \eqref{eq:svd-jensen-step} and using
that $\tau$ was chosen optimal, so that
\[
\min_a\sum_b\tau(b)M_{W_r^\Sem}^\Sem(\mu)_{a,b}=\operatorname{val}(M_{W_r^\Sem}^\Sem(\mu))
\]
exactly by \eqref{eq:val-two-forms}, chains with the previous display to
give
\[
g_f^-(\mu)\ge\min_{a\in\mathcal P_{\min}}f(\DistOp_{a,\bar y}(\mu))
\ge\operatorname{val}\bigl(M_{W_r^\Sem}^\Sem(\mu)\bigr).
\]
Multiplying by $\Discount$ and adding the common current reward
$(1-\Discount)\Ent(\mu)$ to both sides turns the left side into
$U_{r+1}^-(\mu)$, by definition of the averaged recursion with
$f=U_r^-$, and the right side into $W_{r+1}^\Sem(\mu)$, by the sampled
Bellman recursion of \cref{thm:sp-determined}; this proves
$U_{r+1}^-\ge W_{r+1}^\Sem$, completing the induction.

Uniform discounted tail estimates, \cref{thm:bellman} (proof in \cref{app:bellman-proof}) for
$U_r^+,U_r^-$ and \eqref{eq:sampled-value-limit} for $W_r^\Sem$, bound
$\|V_{\mathsf{av}}^\pm-U_r^\pm\|_\infty$ and
$\|V_\Sem-W_r^\Sem\|_\infty$ by $\Discount^r\log_2n$, so both families
converge uniformly, as $r\to\infty$, to their respective
infinite-horizon values. Letting $r\to\infty$ in $U_r^-\ge W_r^\Sem$
proves $V^-_{\mathsf{av}}\ge V_\Sem$ for $\Sem\in\{\mathsf{sp},\mathsf{st}\}$,
the second inequality of the dominance chain; a uniform
limit of continuous concave functions is again continuous and concave,
so $V^+_{\mathsf{av}}$ and $V^-_{\mathsf{av}}$ retain these properties.
The first inequality of the dominance chain,
$V^+_{\mathsf{av}}\ge V^-_{\mathsf{av}}$, is the elementary strategic
minimax inequality already noted in \cref{sec:model}.

When only Min has choices, the pure-profile reduction in
\cref{lem:pure-suffices} makes the $\mathsf{av}$ and $\mathsf{sp}$
finite-horizon recursions identical at every stage, so their limits
agree: $V_{\mathsf{av}}=V_{\mathsf{sp}}$. The dominance just proved,
specialised to $\Sem=\mathsf{st}$, then gives
$V_{\mathsf{sp}}=V_{\mathsf{av}}\ge V_{\mathsf{st}}$. The example below
establishes that no general ordering between $\mathsf{sp}$ and $\mathsf{st}$ can be
added for games with an active maximiser.
\end{proof}

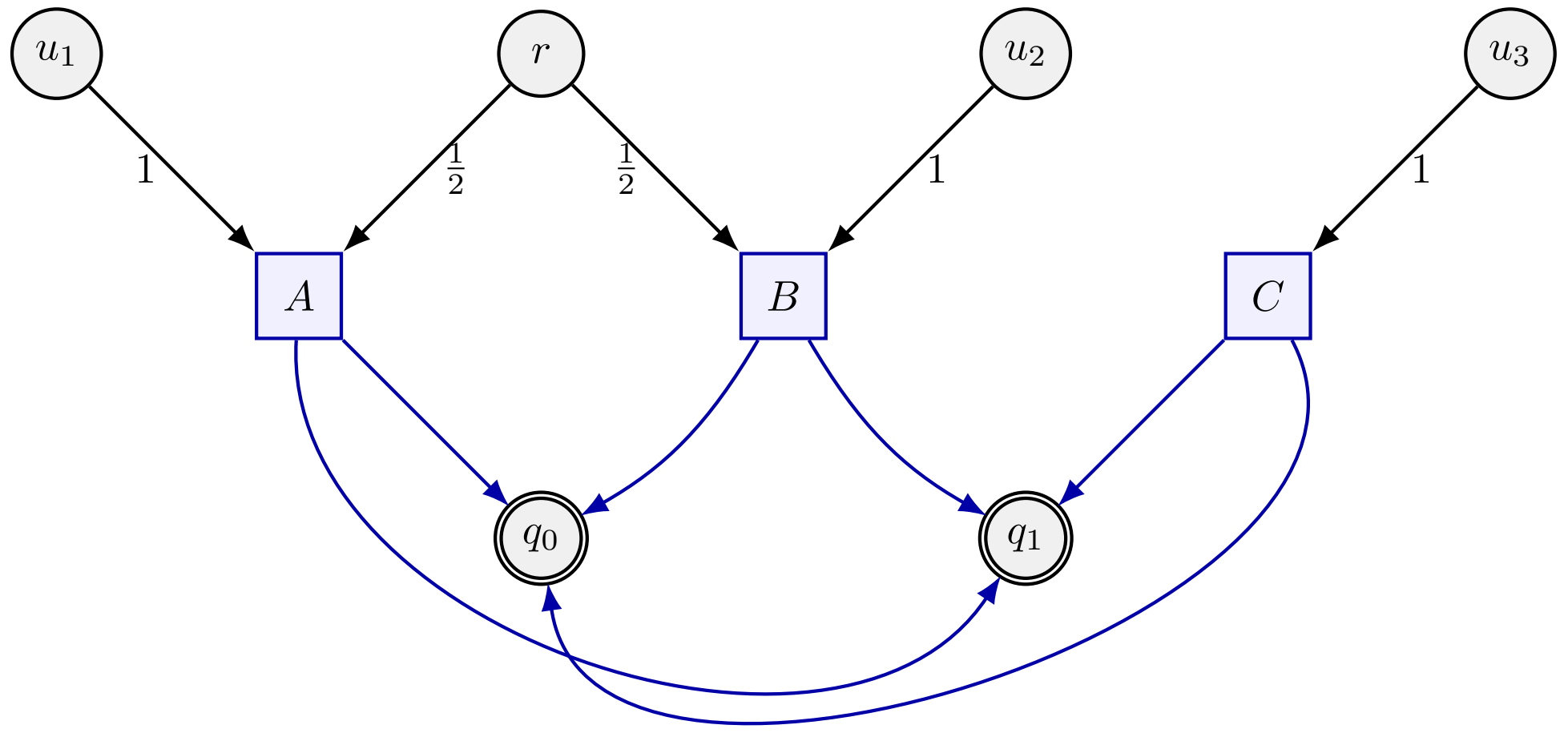
\begin{figure}[t]
\centering
\begin{tikzpicture}[>=Latex,thick,
    forced/.style={state,fill=gray!12,minimum size=7mm},
    maxstate/.style={state,rectangle,draw=blue!65!black,fill=blue!6,minimum size=7mm},
    sink/.style={state,double,fill=black!6,minimum size=7mm}]
  \node[forced]   (u1) at (-6,2)    {$u_1$};
  \node[forced]   (r)  at (-2,2)    {$r$};
  \node[forced]   (u2) at (2,2)     {$u_2$};
  \node[forced]   (u3) at (6,2)     {$u_3$};
  \node[maxstate] (A)  at (-4,0)    {$A$};
  \node[maxstate] (B)  at (0,0)     {$B$};
  \node[maxstate] (C)  at (4,0)     {$C$};
  \node[sink]     (q0) at (-2,-2)   {$q_0$};
  \node[sink]     (q1) at (2,-2)    {$q_1$};
  \draw[->,black] (u1) to node[pos=.5,left]  {$1$} (A);
  \draw[->,black] (r)  to node[pos=.5,right] {$\frac12$} (A);
  \draw[->,black] (r)  to node[pos=.5,left]  {$\frac12$} (B);
  \draw[->,black] (u2) to node[pos=.5,right] {$1$} (B);
  \draw[->,black] (u3) to node[pos=.5,right] {$1$} (C);
  \draw[->,blue!65!black] (A) to (q0);
  \draw[->,blue!65!black] (A) to[bend right=75] (q1);
  \draw[->,blue!65!black] (B) to[bend left=15] (q0);
  \draw[->,blue!65!black] (B) to[bend right=15] (q1);
  \draw[->,blue!65!black] (C) to[bend left=100] (q0);
  \draw[->,blue!65!black] (C) to (q1);
\end{tikzpicture}
\caption{The first arena of \cref{ex:sp-st-reversal}, with initial
distribution $\mu_0=\tfrac14(\delta_{u_1}+\delta_{u_2}+\delta_{u_3}+\delta_r)$,
mass $1/4$ on each of $u_1,u_2,u_3,r$ and mass $0$ elsewhere. A state is
\emph{forced} if it has exactly one enabled action, so neither player
makes any choice there; a forced transition need not be deterministic, as
witnessed by $r$, whose single enabled action nonetheless splits its mass
evenly between $A$ and $B$. Circles $u_1,u_2,u_3,r$ are forced (drawn in
black): $u_1,u_2,u_3$ route their mass deterministically to $A,B,C$
respectively, and $r$ splits its mass with probability $1/2$ each to $A$
and $B$. Rectangles $A,B,C$ are \MaxPl-states (blue), each with two
enabled actions, one routing deterministically to $q_0$ and the other to
$q_1$; this is the only place in the arena where either player exercises
a randomized choice. The double circles $q_0,q_1$ are absorbing: both of
their actions self-loop, so once mass reaches either one it stays there
for the rest of the run.}
\label{fig:sp-st-reversal}
\end{figure}

\begin{figure}[t]
\centering
\begin{tikzpicture}[>=Latex,thick,
    maxstate/.style={state,rectangle,draw=blue!65!black,fill=blue!6,minimum size=7mm},
    sink/.style={state,double,fill=black!6,minimum size=7mm}]
  \node[maxstate] (s0) at (0,0)     {$s_0$};
  \node[sink]     (q0) at (-2,-2)  {$q_0$};
  \node[sink]     (q1) at (2,-2)   {$q_1$};
  \draw[->,black] (s0) to node[pos=.5,left]  {$\frac12$} (q0);
  \draw[->,black] (s0) to node[pos=.5,right] {$\frac12$} (q1);
\end{tikzpicture}
\caption{The second arena of \cref{ex:sp-st-reversal}, giving the reverse
strict inequality $V_{\mathsf{sp}}>V_{\mathsf{st}}$. The initial
distribution is Dirac, $\mu_0=\delta_{s_0}$. In the minimal version, $s_0$
has a single enabled action, a forced fair coin flip to $q_0$ or $q_1$
(solid black edges), so $s_0$, like $q_0,q_1$, is forced or absorbing and
neither player makes any choice anywhere in this arena; the owner of
$s_0$ may then be assigned arbitrarily, since there is nothing to
optimise there. 
}
\label{fig:reverse-reversal}
\end{figure}
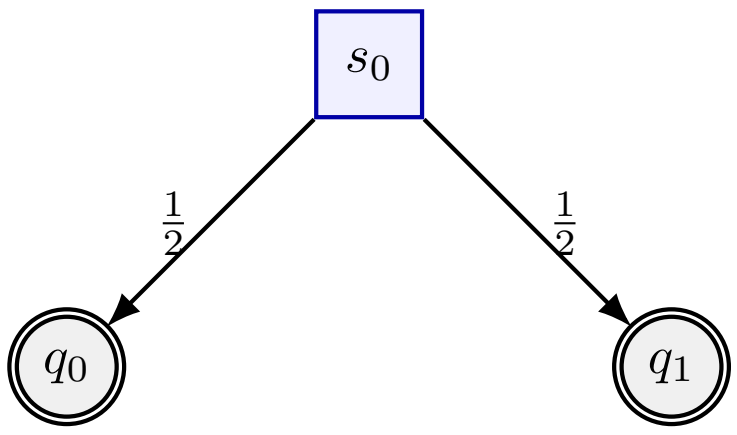

\begin{example}[Feedback reverses the comparison of $\mathsf{sp}$ and $\mathsf{st}$]
\label{ex:sp-st-reversal}
Neither inequality between $V_{\mathsf{sp}}$ and $V_{\mathsf{st}}$ holds in general. The following arena, shown in
\cref{fig:sp-st-reversal}, exhibits a strict reversal in one direction,
and a second, simpler arena, shown in \cref{fig:reverse-reversal}, gives
the reversal in the other direction.

Nine states are arranged in three layers. The top layer $u_1,u_2,u_3,r$
carries the initial mass, one quarter at each state, and every transition
out of this layer is forced, meaning the state has only one enabled
action, so it contributes no strategic choice for either player. States
$u_1,u_2,u_3$ route deterministically to $A,B,C$ respectively, and $r$
splits its mass evenly between $A$ and $B$. The middle layer $A,B,C$
belongs to the maximiser, and each of these three states has exactly two
enabled actions, one routing all of its mass deterministically to the
absorbing state $q_0$ and the other to the absorbing state $q_1$; this is
the only place in the arena where a player makes a randomized choice. The
bottom layer $q_0,q_1$ is absorbing under both actions, so once the mass
reaches there the entropy of the distribution is frozen for the rest of
the run. States with a single enabled action may be assigned to either
player without affecting any value, since there is nothing to optimise
there; call them Min-owned for concreteness, so that Min's only role in
this example is to watch the forced transitions happen.

For $\Discount=9/10$, put
$a=\Ent(3/8,3/8,1/4)$ and $b=h(3/8)$. Then
\[
V_{\mathsf{sp}}=2(1-\Discount)+\Discount(1-\Discount)a+\Discount^2b,
\quad
V_{\mathsf{st}}=2(1-\Discount)+\tfrac32\Discount(1-\Discount)+\Discount^2,
\]
Numerically $V_{\mathsf{sp}}\approx1.1136066$ and $V_{\mathsf{st}}=1.145$, respectively. So, $V_{\mathsf{st}}>V_{\mathsf{sp}}$.

In the opposite direction, the arena of
\cref{fig:reverse-reversal}, a forced fair transition from a Dirac
initial state to two absorbing sinks, gives
$V_{\mathsf{sp}}=(1-\Discount)\cdot0+\Discount\cdot1=\Discount>V_{\mathsf{st}}=0$. So neither $V_{\mathsf{sp}}\ge V_{\mathsf{st}}$ nor $V_{\mathsf{sp}}\le V_{\mathsf{st}}$ holds in general.
\end{example}
\begin{proof}[Verification]
The nine states of \cref{fig:sp-st-reversal} are
$u_1,u_2,u_3,r,A,B,C,q_0,q_1$. The first four have the forced transitions
stated above, $A,B,C$ belong to Max and each has an action to each $q_i$,
and the two $q_i$ are absorbing. All states with forced actions may
belong to Min. The initial entropy is two. Under $\mathsf{sp}$, the first
successor is deterministically $(3/8,3/8,1/4)$ on $(A,B,C)$ and has
entropy $a=11/4-(3/4)\log_2 3$. A pure profile partitions these three
masses between two bins. The possible mass at $q_1$ is
\[
0,\quad1/4,\quad3/8,\quad5/8,\quad3/4,\quad1.
\]
Its largest binary entropy is $b=h(3/8)$. A sampled mixture averages
these scalar entropies and cannot improve on their maximum. Once the
partition is chosen it is absorbing, proving the displayed $\mathsf{sp}$
value. Under $\mathsf{st}$, the first successor is either $(1/2,1/4,1/4)$ or
$(1/4,1/2,1/4)$, each with probability $1/2$. Its entropy is $3/2$.
The maximiser observes which coordinate has mass $1/2$ and routes that
coordinate alone to one sink, the other two to the other sink. The
terminal distribution is uniform on two states, attaining its maximal
possible entropy one. This proves the $\mathsf{st}$ value.

In both
models the same pure stationary feedback rule suffices: among subsets
of $\{A,B,C\}$ choose one whose current mass is nearest to $1/2$, with a
fixed lexicographic tie rule, and route it to $q_1$; this rule achieves
the optimal value under each semantics separately (they are not the same
number, only the same policy). This is a Borel rule, since it compares
finitely many continuous functions. Thus the reversal is not an artefact
of comparing unrelated policy classes.

Under $\mathsf{av}$ Max may split every current atom equally between the
sinks, so its value is $2(1-\gamma)+\gamma(1-\gamma)a+\gamma^2\approx1.1505150$, consistent
with the dominance chain.

For the reverse comparison, take the three states $s_0,q_0,q_1$ of
\cref{fig:reverse-reversal}: a Dirac initial state makes a fair
transition to two distinct absorbing sinks, and, as with the forced
states above, the owner of $s_0$ may be assigned arbitrarily since it
has only one action in this minimal version. $\mathsf{av}$ and
$\mathsf{sp}$ then have entropy one from time one onward, whereas
$\mathsf{st}$ follows a Dirac distribution forever. Thus their values
are $d,d,0$, for any $d\in(0,1)$. 

\end{proof}

\end{document}